\documentclass[3p,times]{elsarticle}
\usepackage{mathrsfs}

\usepackage{amssymb}

 \usepackage{graphicx}
 \usepackage{epstopdf}
 \usepackage{epsfig}
\usepackage[pdftex,colorlinks]{hyperref}
\usepackage{color}
 \usepackage{booktabs}

\usepackage[figuresright]{rotating}

\def\color#1{}
\begin{document}

\begin{frontmatter}


\title{Optimal low-dispersion low-dissipation  LBM schemes for computational aeroacoustics}
 \author{Hui Xu}
 \ead{xuhuixj@gmail.com or xu@lmm.jussieu.fr}
 \author{Pierre Sagaut}
 \ead{sagaut@lmm.jussieu.fr}
\address{Institut Jean le Rond d'Alembert, UMR CNRS 7190, Universit\'e Pierre et Marie Curie - Paris 6,  4 Place Jussieu case 162 Tour 55-65, 75252 Paris Cedex 05, France\fnref{label3}}





\begin{abstract}
Lattice Boltmzmann Methods (LBM) have been proved to be very
effective methods for computational  aeroacoustics (CAA), which have
been used to capture the dynamics of weak acoustic fluctuations. In
this paper, we  propose a strategy to reduce the dispersive and
disspative errors of the two-dimensional (2D) multi-relaxation-time
lattice Boltzmann method (MRT-LBM). By presenting an effective
algorithm, we obtain {\color{blue} a} uniform form of the linearized
Navier-Stokes equations corresponding to {\color{blue} the} MRT-LBM
in wave-number space. Using the matrix perturbation theory and the
 equivalent modified equation approach for finite difference
methods, we propose a class of minimization problems to optimize the
free-parameters in {\color{blue} the} MRT-LBM. We obtain this way a
dispersion-relation-preserving LBM (DRP-LBM) to circumvent the
minimized dispersion error of {\color{blue} the} MRT-LBM.
{\color{blue} The dissipation relation precision is also improved.
And the stability of {\color{blue} the} MRT-LBM with the small bulk
viscosity is guaranteed.} Von Neuman analysis of the linearized
MRT-LBM is performed to validate the optimized
dispersion/dissipation relations considering monochromatic wave
solutions. Meanwhile, dispersion and dissipation errors of
{\color{blue} the} optimized MRT-LBM are quantitatively compared
with {\color{blue} the} original MRT-LBM . Finally, some numerical
simulations are carried out to assess the new optimized MRT-LBM
schemes.
\end{abstract}


\end{frontmatter}


\section{Introduction}\label{Intro}
The lattice Boltzmann method (LBM) has emerged as a very effective
methodology for the computational modeling of a wide variety of
complex fluid flows \cite{chendoolen}. Recent researches dealing
with dispersion and dissipation relations of {\color{blue} the} LBM
have shown that {\color{blue} the} LBM possesses the required
accuracy to capture  weak acoustic pressure fluctuations
\cite{buickgreatedcampbell,simondenispierre,ricotmariesagautbailly,buickbuckleygreated}.
The analysis indicated that the simple LBS possess lower numerical
dissipation than the aeroacoustic optimized schemes of high-order
schemes for Navier-Stokes equations \cite{simondenispierre}. The
second-order accurate LBM has better dispersion capabilities than
the classical Navier-Stokes schemes with 2nd-order accuracy in space
and 3-step Runge-Kutta in time \cite{simondenispierre}. However, the
dispersion error in  {\color{blue} the} LBM is higher than that in
the finite difference method with 3rd order spatial discretization
and the 4th order time discretization, and also higher than that in
dispersion relation preserving (DRP) 6th order accurate schemes.
From the view of the numerical computations, for a given dispersion
error, {\color{blue} the} LBM is faster than the high-order schemes
for Navier-Stokes equations \cite{simondenispierre}. It has been
shown that the dispersion error can be considered as a weakness of
{\color{blue} the} LBM. For the classical lattice Boltzmann model
(BGK-LBM), it is impossible to reduce the dispersion/dissipation
errors. Meanwhile, it is reported that {\color{blue} the} original
MRT-LBM \cite{lallemandluo,dominique} and {\color{blue} the} BGK-LBM
have exactly the same dispersion error and there exists a high
dissipation of the acoustic modes for {\color{blue} the} MRT-LBM
\cite{simondenispierre}. Here, the original MRT-LBM means we use the
relaxation parameters recommended by Lallemand and Luo
\cite{lallemandluo}. Furthermore, because of a high value of the
bulk viscosity, the original MRT-LBM has a better stability compared
with {\color{blue} the} BGK-LBM \cite{simondenispierre}. So, if the
dissipation error of {\color{blue} the} MRT-LBM can be reduced and
the stability can be guaranteed, it will be a good choice to
simulate the acoustic problems using {\color{blue} the} MRT-LBM.
{\color{blue} The} MRT-LBM can be optimized thanks to the existence
of free parameters. By means of a Taylor series expansion
\cite{duboislallemand,dubois}, it is easy to establish the relation
between the linearized MRT-LBM (L-MRT-LBM) and the linearized
Navier-Stokes equations (L-NSE) with the high-order truncation. The
relations between {\color{blue} the} L-MRT-LBM and {\color{blue}
the} L-NSE offer us a way to detect the influence of free parameters
on the dispersion/dissipation relations. In the limit of linear
acoustic, von Neumann analysis is a reliable tool to recover the
dispersion and dissipation relations of {\color{blue} the} LBS.  It
is noted that this famous analysis method has been revisited and
extended \cite{simondenispierre,senguptadipankarsagaut}. Considering
plane wave solutions, the relation between the wave-number $\bf{k}$
and the wave pulsation $\omega$ is described numerically. Then, the
influence of free parameters on acoustic modes and shear modes will
be analyzed. We propose a class of optimization strategies to
minimize the dispersion/dissipation errors based on the matrix
perturbation theory and the modified equation approach, leading to
the definition of dispersion/dissipation-relation-preserving MRT-LBM
(D2RP-LBM) schemes.

 The basic idea
of {\color{blue} the} DRP-LBM is significantly different from the
idea of the classical DRP-schemes corresponding to finite difference
schemes \cite{christopherwebb}. The classical DRP-schemes were
established considering the finite difference approximation of the
first derivative $\partial f/\partial x$ at the node of an uniform
grid in wave-number space \cite{christopherwebb}. The classical DRP
finite difference approximation of the first derivative $\partial
f/\partial x$ in wave-number space is given by
\begin{equation}\label{cdrp}
\mathrm{i}\alpha\simeq\left(\frac{1}{\Delta
x}\sum_{j=-N}^{M}a_je^{\mathrm{i}\alpha j\Delta
x}\right)\widetilde{f}.
\end{equation}
The effective wave-number of Eq. (\ref{cdrp}) can be rewritten as
follows
\begin{equation}
\widetilde{\alpha}=\frac{-\mathrm{i}}{\Delta
x}\sum_{j=-N}^{M}a_je^{\mathrm{i}\alpha j\Delta x}.
\end{equation}
 In the physical spaces, the expression of Eq.
(\ref{cdrp}) is given as follows
\begin{equation}
\frac{\partial f}{\partial x}(x)\simeq\frac{1}{\Delta
x}\sum_{j=-N}^Ma_jf(x+j\Delta x).
\end{equation}
In order to minimize the dispersion error, the following integral
error $E$ is defined \cite{christopherwebb}
\begin{equation}
E=\int_{-\pi/2}^{\pi/2}|\alpha \Delta x-\widetilde{\alpha}\Delta
x|^2d(\alpha \Delta x).
\end{equation}
The classical DRP schemes only focus on proposing a best
approximation of the first-derivative $\partial f/\partial x$ on an
uniform mesh. In this paper, the proposed method to minimize the
dispersion/dissipation error focus on obtaining a best global
approximation of the exact L-NSE systems based on the recovered
L-NSE by the L-MRT-LBM in wave-number spaces. Consequently, the
resulting approximation addresses the best approaching relation
between the exact L-NSE and {\color{blue} the} recovered L-NSE, but
does not address {\color{blue}individual derivatives}. Formally, the
research in this paper is dedicated to establishing the
approximation between the following equation systems in wave-number
spaces
\begin{equation}
\partial_tW=B\cdot W\ (\mbox{exact L-NSE}),\ \partial_tW=B_{,n}\cdot W+O(\delta t^n)\ (\mbox{recovered L-NSE}),
\end{equation}
where the square matrices $B$ and $B_{,n}$ are functions of
wave-number vector $\mathbf{k}$, where $B_{,n}$ can be regarded as a
perturbation of $B$. The vector $W$ denotes the perturbed
macroscopic fluid flow quantities (density, momentum) in
{\color{blue} the} linearized-NSE systems. In the paper, in order to
reduce the dispersion error of {\color{blue} the} MRT-LBM for
zero-mean flow, the truncation error is up to $O(\delta t^5)$.
Meanwhile, in order to reduce the dissipation error of {\color{blue}
the} MRT-LBM for uniform flows, the truncation error is up to
$O(\delta t^4)$.

 Furthermore, the proposed derivation of higher-order
Taylor expansions of {\color{blue} the} MRT-LBM is very lengthy and
complicated \cite{duboislallemand}. However, this derivation still
pave a new way for establishing the relation between {\color{blue}
the} L-MRT-LBM and {\color{blue} the} L-NSE. In order to avoid using
this complex derivation, a new more effective and easy-to-use
recursive algorithm is proposed to recover {\color{blue} the} L-NSE
by {\color{blue} the} L-MRT-LBM. This recursive algorithm is
established in wavenumber space. By this algorithm, the
corresponding  linearized macroscopic equations are expressed by an
easy-to-handle matrix form. The optimization strategy is precisely
built on the basis of this matrix equation. Finally, von Neumann
analysis and numerical tests are implemented to assess the optimized
MRT-LBM.

In next section, the methodology used to establish the
transformation relation from {\color{blue} the} L-MRT-LBM to the
L-NSE is given. The optimization strategies corresponding to the
matrix equation are studied in Section 3. In Section 4, by von
Neumann analysis, the DRP-LBM schemes are analyzed in spectral
spaces. In the last section, the optimized MRT-LBM schemes are
validated by benchmark problems.

\section{Methodology for bridging from {\color{blue} the} MRT-LBM to the linearized Navier-Stokes equations }\label{Method}

In this section, the basic theory of lattice Boltzmann schemes
briefly reminded. Then, the linearized LBM is introduced and a
method to establish the relation between the linearized LBS and
{\color{blue} the} L-NSE is proposed.

\subsection{Lattice Boltzmann schemes}\label{subMRTLBM}

The evolution equations of distribution functions of the basic
lattice Boltzmann schemes are written as follows
\begin{equation}\label{blbs}
f_i(x+v_i\delta t,t+\delta
t)=f_i(x,t)+\Lambda_{ij}\left(f_j^{\rm(eq)}(x,t)-f_j(x,t)\right),\quad
0\leq i,j\leq N,
\end{equation}
where $v_i$ belongs to the discrete velocity set $\mathcal{V} $,
$f_i(x,t)$ is the discrete single particle distribution function
corresponding to $v_i$ and $f_i^{\rm(eq)}$ denotes the discrete
single particle equilibrium distribution function. $\delta t$
denotes the time step and $N+1$ is the number of discrete
velocities. $\Lambda_{ij}$ is the relation matrix. {\color{blue}
From here, the repeated index indicates the Einstein summation for 0
to N except for the special indications.} Let $\mathcal{L}\in
\mathbb{R}^d $ ($d$ denotes the spatial dimension) denote the
lattice system, the following condition is required
\cite{duboislallemand}
\begin{equation}
x+v_j\delta t\in \mathcal{L},
\end{equation}
that is to say, if $x$ is a node of the lattice, $x+v_j\delta t$ is
necessarily another node of the lattice.

For {\color{blue} the} BGK-LBM, the relaxation matrix is set as
follows
\begin{equation}\label{eq_BGK_LBM_R}
\Lambda_{ij}=s\delta_{ij},
\end{equation}
where $s$ is related to the relaxation frequency of {\color{blue}
the} BGK-LBM.

The standard MRT-LBM has the following form
\cite{duboislallemand,lallemandluo}
\begin{equation}\label{eq_conservative}
m_i=W_i=m_i^{\rm(eq)},0\leq i\leq d,
\end{equation}
and
\begin{equation}\label{eq_noconservative}
m_i(x+\delta t v_j,t+\delta
t)=m_i(x,t)+s_i\left(m_i^{\rm(eq)}(x,t)-m_i(x,t)\right),d+1\leq
i\leq N,
\end{equation}
{\color{blue} where the index $i$ doesn't indicate the summation.}
According to the work of Lallemand and Luo \cite{lallemandluo}, the
relaxation parameters in Eq. (\ref{eq_noconservative}) must satisfy
the following stability constrain
\begin{equation}
s_i\in (0,2),d+1\leq i\leq N.
\end{equation}
The relaxation matrix $\Lambda$ associated with Eqs.
(\ref{eq_conservative}) and (\ref{eq_noconservative}) is defined by
\begin{equation}
\Lambda=M^{-1}SM,
\end{equation}
where $S$ is a diagonal matrix which is related to the relaxation
parameters of {\color{blue} the} MRT-LBM.
$M=\left(M_{ij}\right)_{0\leq i\leq N,0\leq i\leq N}$ is the
transformation matrix, which satisfies the following basic
conditions \cite{lallemandluo}
\begin{equation}
M_{0j}=1,M_{\alpha j}=v_j^{\alpha},(1\leq\alpha\leq d).
\end{equation}
The macroscopic quantities are defined by
\cite{lallemandluo,duboislallemand}
\begin{equation}\label{q-macro}
m_i=M_{ij}f_j,\quad m_i^{\rm (eq)}=M_{ij}f_j^{\rm(eq)}.
\end{equation}

\subsection{{\color{blue} The} Linearized MRT-LBM and {\color{blue} the} higher-order linearized NSE}\label{subL-MRT-LBM}

 The equilibrium function $m_i^{\rm(eq)}$ ($d+1\leq i\leq
N$) is the function of conservative quantities $W_i$
($W_i=m_i^{\rm(eq)}, 0\leq i\leq d$) (we  use the same notations
as Dubois and Lallemand \cite{duboislallemand})
\begin{equation}\label{mrteq} m_i^{\rm(eq)}=G_i\left(\{W_j\}_{0\leq j\leq d}\right),d+1\leq
i \leq N.
\end{equation}

or

\begin{equation}\label{l-mrt-eq} m_i^{\rm(eq)}=G_i\left(\{W_j\}_{0\leq j\leq
d}\right)=G_{ij}W_j=G_{ij}m_j
\end{equation}

In order to implement the linear stability analysis and recover the
linearized macroscopic equations, we introduce the linearized form
of Eq. (\ref{mrteq}) around reference states
\cite{simondenispierre,duboislallemand,lallemandluo}. Using Eq.
(\ref{l-mrt-eq}), the linearized description of Eqs.
(\ref{eq_conservative}) and (\ref{eq_noconservative}) can be written
as follows

\begin{equation}\label{l-mrt-lbs}
m_i(x,t+\delta t)=M_{il}M_{lp}^{-1}\Psi_{pr}m_r(x-v_l\delta t,t),
\end{equation}
where the matrix $\Psi $ has the following form
\begin{equation}\label{mrt-psi}
\Psi=\left(\begin{array}{cc}\left(\mathrm{I}_{ij}\right)_{0\leq i\leq d,0\leq j\leq d}&0\\
\Theta&\left(\mathrm{I}_{ij}-S_{ij}\right)_{d+1\leq i\leq N,d+1\leq
j\leq N}
\end{array}\right)
\end{equation}
where $\Theta_{ij}=s_iG_{ij} $. A is matrix with the size
$(N-d-1)\times (d+1) $), $\mathrm{I}$ is an identity matrix and the
diagonal matrix $S$ is defined by
\begin{equation}\label{S_relaxation}
S={\rm
Diag}(\underbrace{0,\ldots,0}_{d+1},\underbrace{s_{d+1},\ldots,s_{N}}_{N-d-1}.
\end{equation}
In Eq. (\ref{l-mrt-lbs}), the indices $l$, $p$ and $r$ indicate
the summations from 0 to N.

In order to derive the linearized high-order equations, one assumes
that the discrete single particle distribution $f_i $ belongs to
$C^{\infty}(T\times\mathcal{L}) $ (a functional set, in which the
element possesses a sufficiently smooth property with respect to the
time domain $T$ and spatial domain $\mathcal{L} $ ). This assumption
is also used by Junk {\it et. al.} for asymptotic analysis of
{\color{blue} the} LBM \cite{junklarluo}. This regularity hypothesis
indicates that macroscopic quantities $m_i$ are  smooth ones and
that  the linearized system (\ref{l-mrt-lbs}) is well defined.

The next step consists of performing the Taylor series expansion of
the right hand of Eq. (\ref{l-mrt-lbs}),  yielding
\begin{equation}\label{expansion_1}
m_i(x,t+\delta t)=\sum_{n=0}^\infty\frac{\delta t^n}{n!}
M_{il}(-v_l^{\alpha}\partial_\alpha)^nM^{-1}_{lp}\Psi_{pr}m_r,
\end{equation}
where  $\alpha$ indicates the summation from $1$ to $d$.

 Now, we define the matrix $A^*_{,n}=\left(A^*_{ij,n}\right)_{0\leq i\leq
N,0\leq j\leq N} $ as follows
\begin{equation}
A^*_{ir,n}=\frac{1}{n!}M_{il}(-v_l^\alpha\partial_\alpha)^nM_{lp}^{-1}\Psi_{pr}.
\end{equation}
When we need to derive equivalent equations or modified equations,
it is difficult to use the matrix $A^*_{,n} $ to carry out the
calculations. In order to overcome this difficulty, we use the
differential operators in spectral space. Let us note
$\partial_\alpha=\mathrm{i}k_\alpha $, with $k_\alpha $ the
wave-number in the $\alpha$ -direction and  $\mathrm{i}^2 = -1$.
Then, in spectral
space, the matrix $A^*_{,n} $ has the following form
($A_{,n}=(A_{ij,n})_{0\leq i\leq N,0\leq j\leq N} $)
\begin{equation}
A_{ir,n}=\frac{1}{n!}M_{il}(- \mathrm{i} v_l^\alpha
k_\alpha)^nM_{lp}^{-1}\Psi_{pr}.
\end{equation}
Therefore, Eq. (\ref{expansion_1}) can be rewritten as follows

\begin{equation}\label{expansion_2}
m_i(x,t+\delta t)=\sum_{n=0}^{J-1}\delta t^n A_{ir,n}m_r+O(\delta
t^J).
\end{equation}

In order to derive the L-NSE corresponding to {\color{blue} the}
L-MRT-LBM defined by Eq. (\ref{expansion_2}), we introduce an
original   recursive algorithm. Given $m_i=W_i (0\leq i\leq d) $
(macroscopic conservative quantities), the algorithm is given as
follows

\begin{itemize}

\item {\bf Initial step}. The initial $\Phi_{,1} $ and $B_{,1} $ are given as follows
\begin{equation}
\Phi_{ij,1}=\delta_{ij} (0\leq i\leq
d),\Phi_{ij,1}=\frac{1}{s_i}\Psi_{ij} (d+1\leq i\leq N),
\end{equation}
\begin{equation}
B_{ij,1}=A_{ir,1}\Phi_{rj,1}.
\end{equation}
Let $W=\{W_i\}_{0\leq i\leq d} $ and $m=\{m_i\}_{0\leq i\leq N} $
denote the vector of the conservative quantities and the vector of
all macroscopic quantities respectively.

At the first order of $\delta t$, for all macroscopic quantities, we
have
\begin{equation}
m_i=\Phi_{ij,1}W_j+O(\delta t),0\leq i\leq N.
\end{equation}
By the matrix form, we have
\begin{equation}\label{R-IA}
m=\Phi_{,1}W+O(\delta t) .
\end{equation}
At the first-order of $\delta t$, for conservative quantities, we
have
\begin{equation}
\partial_t W_i= A_{ir,1}\Phi_{rj,1}W_j+O(\delta t).
\end{equation}
The matrix form is
\begin{equation}\label{R-IB}
\partial_t W= A_{,1}\Phi_{,1}W_j+O(\delta t).
\end{equation}

\item {\bf Recursive formula for all macroscopic quantities}.
$\Phi_{ij,n}$ can be given as follows
\begin{equation}\label{R-A}
\Phi_{ij,n}=\frac{1}{s_i}(\Psi_{ij}-\sum_{l=1}^{n-1}\frac{\delta
t^l}{l!} \Phi_{ir,n-l}B_{rj,n-l}^l+\sum_{l=1}^{n-1}\delta t^l
A_{ir,l}\Phi_{rj,n-l}),d+1\leq i\leq N
\end{equation}
and
\begin{equation}
\Phi_{ij,n}=\delta_{ij}, (0\leq i\leq d).
\end{equation}
Eliminating the higher-order term of $\delta t^{n-1}$, we have
\begin{equation}\label{R-B}
\Phi_{ij,n}=\sum_{l=0}^{n-1}\delta t^l{\rm Coeff}(\Phi_{ij,n},\delta
t,l),
\end{equation}
where ${\rm Coeff(\cdot,\cdot,\cdot)} $ is a function which extracts
the coefficients of the polynomials, for example,
$f(x)=\sum_{i=0}^na_ix^i $
\begin{equation}
{\rm Coeff}(f(x),x,i)=a_i.
\end{equation}
According to Eqs.(\ref{R-A}) and (\ref{R-B}), we have
\begin{equation}\label{R-C}
m=\Phi_{,n}W+O(\delta t^n).
\end{equation}

\item {\bf Recursive formula for conservative quantities}. $B_{ij,n}$ is
presented as follows
\begin{equation}\label{R-D}
B_{ij,n}=-\sum_{l=1}^{n-1}\frac{\delta
t^l}{(l+1)!}B_{ij,n-l}^{l+1}+\sum_{l=1}^n\delta
t^{l-1}A_{ir,l}\Phi_{rj,n+1-l}.
\end{equation}
Eliminating the higher-order term of $\delta t^{n-1}$, we have
\begin{equation}\label{R-E}
B_{ij,n}=\sum_{l=0}^{n-1}\delta t^l{\rm Coeff}(B_{ij,n},\delta t,l).
\end{equation}
Now, for the conservative quantities, we have the following equation
system
\begin{equation}\label{R-F}
\partial_tW=B_{,n}\cdot W+O(\delta t^n).
\end{equation}
Using  Eqs. (\ref{R-B}),~(\ref{R-C}),~(\ref{R-E}) and (\ref{R-F}),
we can get the coefficient matrix of the conservative quantities at
any order of $\delta t$ . Details are displayed in
\ref{AppendexDerivation}.
\end{itemize}

\subsection{Illustrating example: application to {\color{blue} a} 2D MRT-LBM}\label{L-NSE-WNS}
In {\color{blue} this} section, we  illustrate the algorithm
presented in \ref{subL-MRT-LBM} considering a 2D MRT-LBM. For the
standard 2D
 MRT-LBM, the equilibrium distribution functions are described as
 follow \cite{lallemandluo,bouzidihumiereslallemandluo}
 \begin{equation}
 m^{\rm(eq)}=\left\{\rho,j_x,j_y,-2\rho+\frac{3}{\rho}(j_x^2+j_y^2),\rho-\frac{3}{\rho}(j_x^2+j_y^2),-j_x,-j_y,\frac{1}{\rho}(j_x^2-j_y^2),\frac{1}{\rho}j_xj_y\right\},
 \end{equation}
 where $j_x$ and $j_y$ denote the x-momentum and y-momentum
 respectively, and $\rho$ represents the density
 ($W_0=m_0=\rho,W_1=m_1=j_x,W_2=m_2=j_y$).
 The corresponding matrix $\Psi $ is given by
\begin{equation}
\Psi=\left[\begin{array}{ccccccccc}
1&0&0&0&0&0&0&0&0\\
0&1&0&0&0&0&0&0&0\\
0&0&1&0&0&0&0&0&0\\
-2s_e-3(U^2+V^2)s_e&6Us_e&6Vs_e&1-s_e&0&0&0&0&0\\
s_\epsilon+3(U^2+V^2)s_\epsilon&-6 Us_\epsilon&-6 Vs_\epsilon&0&1-s_\epsilon&0&0&0&0\\
0&-s_q &0&0&0&1-s_q&0&0&0\\
0&0&-s_q &0&0&0&1-s_q&0&0\\
-(U^2-V^2)s_\nu&2U s_\nu&-2V s_\nu&0&0&0&0&1-s_\nu&0\\
-UV s_\nu&V s_\nu&U s_\nu&0&0&0&0&0&1-s_\nu\end{array}\right]
\end{equation}
The diagonal elements of the corresponding diagonal matrix $S$ are
set as follows
\begin{equation}
s_0=s_1=s_2=0,s_3=s_e,s_4=s_\epsilon,s_5=s_6=s_q,s_7=s_8=s_\nu.
\end{equation}
For the original MRT-LBM \cite{lallemandluo}, only $s_\nu$ is a free
parameter, and $s_e=1.64$, $s_\epsilon=1.54$, $s_q=1.9$. The
analogous form of $\Psi $ can be found in existing literature
\cite{lallemandluo,duboislallemand,bouzidihumiereslallemandluo}.
However, there still exist some slight differences. The derivation
of $\Psi$ can be achieved by means of the first-order Taylor series expansion with
respect to $\rho$, $j_x$ and $j_y$ at reference states. In the expression
of $\Psi$, $(U,V)$ to denote the uniform flow velocity components.

For the sake of convenience, we
introduce the following relation
\begin{equation}
\sigma_\eta=\frac{1}{s_\eta}-\frac{1}{2},
\end{equation}
where $\eta$ stands for any notations in the set
$\{e,\epsilon,q,\nu\} $.

\subsubsection{Considering the zero-mean flows $(U,V)=(0,0)$}\label{zero-mean}

Now, when the truncated error term is equal to $O(\delta t^5)$ , the
coefficient matrix $B_{,5}$ with the zero-mean flow can be described
by the summation of five matrices. The first two matrices are given
as follows, which describe the specific terms in the Navier-Stokes
equations.

The coefficient matrix associated with $\delta t^0$ is

\begin{equation}
 \mathrm{i}\cdot\left[ \begin {array}{ccc} 0&-{\it k_x}&-{\it k_y}
\\\noalign{\medskip}-\frac{1}{3}\,{\it k_x}&0&0\\\noalign{\medskip}-\frac{1}{3}\,{
\it k_y}&0&0\end {array} \right]
\end{equation}

The coefficient matrix associated with $\delta t$ is given by

\begin{equation}\label{dissipation-matrix}
 \left[ \begin {array}{ccc} 0&0&0\\
 \noalign{\medskip}0&-\frac{1}{3}\,{{\it k_x}
}^{2}\sigma_e-\frac{1}{3}\,{{\it
k_x}}^{2}\sigma_\nu-\frac{1}{3}\,{{\it k_y}}^{2}\sigma_\nu&
-\frac{1}{3}\,{\it k_x}\,{\it
k_y}\,\sigma_e\\
\noalign{\medskip}0&-\frac{1}{3}\,{\it k_x} \,{\it
k_y}\,\sigma_e&-\frac{1}{3}\,{{\it
k_x}}^{2}\sigma_\nu-\frac{1}{3}\,{{\it k_y}}^{2}
\sigma_e-\frac{1}{3}\,{{\it k_y}}^{2}\sigma_\nu\end {array} \right]
\end{equation}

At the higher-order truncated errors of $\delta t$, the coefficient
matrices are given in \ref{coef-matrices}. Compared with the results
given by Dubois and Lallemand \cite{duboislallemand}, it is shown
that the proposed algorithm yields the correct results. Meanwhile,
the higher-order terms of $\delta t$ are offered by our algorithm
explicitly.

\subsubsection{Considering the uniform flow $(U,V)$ $(U\neq 0$ or $V\neq
0)$}\label{uniform-flow} When the truncated error term is equal to
$O(\delta t^2)$ , the coefficient matrix $B_{,2}$ with the uniform
flow can be described
by the summation of two matrices.
The coefficient matrix associated with $\delta t^0$  is

\begin{equation}\label{uniform-0o}
\mathrm{i}\cdot\left[ \begin {array}{ccc} 0&-{ k_x}&-{ k_y}
\\\noalign{\medskip}-\frac{1}{3}\,{ k_x}+{ k_x}\,{U}^{2}+{ k_y}\,UV&-
2\,{ k_x}\,U-{ k_y}\,V&-{ k_y}\,U\\\noalign{\medskip}-\frac{1}{3}\,{
k_y}+{ k_y }\,{V}^{2}+{ k_x}\,UV&-{ k_x}\,V&-2\,{ k_y}\,V -{
k_x}\,U\end {array} \right]
\end{equation}
The coefficients of $\delta t$ are given in
\ref{coef-matrices-uniform}.

From the coefficient matrix, it is clear that the correct convection
terms of {\color{blue} the} L-NSE  can be given by {\color{blue}
the} L-MRT-LBM. However, the correct dissipation coefficients can
not be obtained by {\color{blue} the} L-MRT-LBM with respect to the
uniform flow, yielding the definition of a flow-dependent viscosity.
Furthermore, this dependence also becomes a source of the
non-Galilean invariance.

From \ref{zero-mean} and \ref{uniform-flow}, it is known at the
zeroth-order and the first-order of $\delta t$, the relaxation
parameters $s_\epsilon$ and $s_q$ have no influence on the recovered
L-NSE. Here, the given form of {\color{blue} the} L-NSE is generated
with respect to the small perturbations of the density $\rho$ and
the momentum quantities $(j_x,j_y)$.

\section{Optimization strategies of free parameters in {\color{blue} the} MRT-LBM}\label{optstra}

In this section, the original optimization strategies of free
parameters in {\color{blue} the} MRT-LBM are proposed based on the
matrix perturbation theory and the modified equations. The optimized
parameters will be determined in order to obtain the optimal
dispersion/dissipation relations.

\subsection{{\color{blue} The} matrix perturbation theory for {\color{blue} the} L-NSE corresponding to {\color{blue} the} L-MRT-LBM}

From Eq. (\ref{R-F}), it is known that the dispersion and
dissipation relations of {\color{blue} the} L-MRT-LBM are determined
by the matrix $B_{,n}\in C^{(1+d)\times(1+d)}$, where
$C^{(1+d)\times(1+d)}$ denotes the $(1+d)\times(1+d)$ complex matrix
set,  if the truncation error is developed up to the $n$th-order of
$\delta t$. Here,  $B$ refers to the coefficient matrix of the exact
L-NSE in wave-number space similar to the matrix $B_{,n}$ in Eq.
(\ref{R-F}). This means that for the exact L-NSE, one has the
following expression
\begin{equation}\label{L-NSE-WS}
\partial_tW=B\cdot W.
\end{equation}
 For a given $n$, if the errors terms with order higher
that $\delta t^n$ are neglected, the main deviation of dispersion
and dissipation relations between the L-MRT-LBM and the L-NSE
originates in the differences between eigenvalues of $B_{,n}$ and
those of $B$.

Now, we introduce the perturbation matrix $M_\varepsilon\in
C^{(1+d)\times(1+d)}$ defined as
\begin{equation}\label{perturbation}
B_{,n}=B+M_\varepsilon.
\end{equation}
Let $B$ have eigenvalues $\lambda_1,\ldots,\lambda_n$ and $B_{,n}$
have eigenvalues $\tilde{\lambda}_1,\ldots,\tilde{\lambda}_n$. The
spectral variation of $B_{,n}$ with respect to $B$ is
\cite{stewartsun}
\begin{equation}
{\rm SV}_{B}(B_{,n})\equiv
\max_i\min_j|\tilde{\lambda}_i-\lambda_j|.
\end{equation}
Then \cite{stewartsun},
\begin{equation}\label{approximation}
{\rm SV}_{B}(B_{,n})\leq
(\|B\|_2+\|B_{,n}\|_2)^{1-1/(d+1)}\|M_\varepsilon\|_2^{1/(d+1)},
\end{equation}
where $\|\cdot\|_2$ is the spectral norm of matrices. For all $A\in
C^{(d+1)\times(d+1)} $, $\|\cdot\|_2$ is defined by
\begin{equation}
\|A\|_2=\sqrt{\lambda_{\max}(A^HA)}=\sigma_{\max}(A),
\end{equation}
where $A^H$ denotes the conjugate transpose of $A$.
$\lambda_{\max}(A^HA)$ is the largest eigenvalue of $A^HA$ and
$\sigma_{\max}(A)$ is the largest singular value of $A$. In order to
establish the direct relation between elements and eigenvalues of
matrices explicitly, the Frobenius norm is given as follows for any
$A=(a_{ij})_{0\leq i\leq d, 0\leq j<j}\in C^{(d+1)\times(d+1)}$
\cite{HornJohnson},
\begin{equation}\label{frobenius}
\|A\|_F=\sqrt{\sum_{i=0}^d\sum_{j=0}^d|a_{ij}|^2}=\sqrt{{\rm
trace}(A^HA)}=\sqrt{\sum_{i=0}^d\sigma_i^2},
\end{equation}
where $\sigma_i$ denotes the singular values of $A$.

Furthermore, let $\sigma(B)=\{\lambda_0,\ldots,\lambda_{d+1}\}$, the
multiset of $B$'s eigenvalues, and set
\begin{equation}
\Lambda={\rm diag}(\lambda_0,\ldots,\lambda_{d+1}),\Lambda_\tau={\rm
diag}(\lambda_{\tau(0)},\ldots,\lambda_{\tau(d+1)}),
\end{equation}
where $\tau$ is a permutation of $\{1,\ldots,d+1\}$. Let
$\sigma(B_{,n})=\{\tilde{\lambda}_0,\ldots,\tilde{\lambda}_{d+1}\}$,
the multiset of $B_{,n}$'s eigenvalues, and set
\begin{equation}
\tilde{\Lambda}={\rm
diag}(\tilde{\lambda}_0,\ldots,\tilde{\lambda}_{d+1}),\tilde{\Lambda}_\tau={\rm
diag}(\tilde{\lambda}_{\tau(0)},\ldots,\tilde{\lambda}_{\tau(d+1)}),
\end{equation}
Then, there exists a permutation $\tau$ such that the following
inequality is satisfied \cite{stewartsun,hogbenbrualdi}
\begin{equation}\label{approximation2}
\|\Lambda-\tilde{\Lambda}_\tau\|_2\leq
2\left\lfloor\frac{d+1}{2}\right\rfloor
(\|B\|_2+\|B_{,n}\|_2)^{1-1/(d+1)}\|M_\varepsilon\|_2^{1/(d+1)}.
\end{equation}
Since $\|M_\varepsilon\|_2\leq \|M_\varepsilon\|_F$, the
minimization of ${\rm SV}_B(B_{,n})$ or
$\|\Lambda-\tilde{\Lambda}_\tau\|_2$ means $\|M_\varepsilon\|_F$
should be minimized to reduce both dispersion and dissipation errors
associated with the MRT-LBM schemes.

\subsection{Optimization methodology}
The following wave-number definition in Eq.(\ref{R-F}) is considered
\begin{equation}\label{wavenumber}
k_x=k\cdot{\rm cos}(\theta),k_y=k\cdot{\rm sin}(\theta).
\end{equation}
Substituting Eq. (\ref{wavenumber}) into Eq. (\ref{R-F}) and
considering the uniform flow $(U,V)$, we can get the following
formal expression for $B_{,n}$
\begin{equation}\label{B-wavenumber}
B_{,n}=\sum_{l=0}^{n-1}\delta t^l k^{l+1}b_{ij,n}^l(\theta,\Xi,U,V),
\end{equation}
where $b_{ij,n}^l$ is a function of $\theta$, $\Xi$ and $(U,V)$.
$\Xi$ denotes the following set (about extra parameters, refer to
the non-standard MRT-LBM \cite{bouzidihumiereslallemandluo})

\begin{equation}
\Xi=\{\alpha, \beta, \lambda, \sigma_e, \sigma_\epsilon, \sigma_q,
\sigma_\nu\}.
\end{equation}

For the standard MRT-LBM, the parameters $\alpha$, $\beta$ and
$\lambda$ are equal to 1, -3 and -2, respectively. In order to
handle the influences of the uniform flows, we consider the
following uniform flows
\begin{equation}
U=u_m\cdot{\rm cos}(\vartheta), V=u_m\cdot{\rm sin}(\vartheta),
\end{equation}
where $u_m$ denotes the magnitude of the uniform velocity. Now, Eq.
(\ref{B-wavenumber}) has the following form
\begin{equation}\label{B-U-wavenumber}
B_{,n}=\sum_{l=0}^{n-1}\delta t^l
k^{l+1}b_{ij,n}^{l}(\theta,\vartheta,\Xi,u_m).
\end{equation}
Furthermore, Eq. (\ref{B-U-wavenumber}) is rewritten as follows
\begin{equation}\label{B-S-U-wavenumber}
B_{ij,n}=\frac{1}{\delta t}\mathcal{B}_{ij,n}(\delta t
k,\theta,\vartheta,\Xi,u_m)=\frac{1}{\delta t}\sum_{l=1}^{n}(\delta
t k)^{l}b_{ij,n}^{l-1}(\theta,\vartheta,\Xi,u_m).
\end{equation}
For the L-NSE, there exists the similar matrix $\mathcal{B}$ defined
by $ \mathcal{B}=\delta t B$.  It is known that for the MRT-LBM, the
dispersion error corresponding to the L-MRT-LBM stems from the
odd-order coefficient matrix $M_\epsilon^O$ of $\delta t k$ in
$\mathcal{M}_\epsilon=\delta
tM_\varepsilon=\mathcal{B}_{,n}-\mathcal{B}$, and the dissipation
error comes from the even-order coefficient matrix $M_\epsilon^E$ of
$\delta t k$ in $M_\epsilon$. So, the perturbation matrix
$M_\epsilon$ can be expressed as follows
\begin{equation}
\mathcal{M}_\epsilon=M_\epsilon^O+M_\epsilon^E.
\end{equation}

\subsubsection{The zero-mean flow case}
Considering $u_m=0$, then,
$\mathcal{M}_\epsilon=\mathcal{B}_{,n}-\mathcal{B}$ is a function of
$\delta t k,\theta$ and $\Xi$. According to Sec. \ref{zero-mean},
\begin{equation}
\mathcal{M}_\epsilon=\sum_{l=3}^{n}(\delta t
k)^{l}b_{ij,n}^{l-1}(\theta,\vartheta,\Xi,u_m)=\sum_{l=3}^{n}(\delta
t k)^{l}b_{ij,n}^{l-1}(\theta,\Xi),
\end{equation}
\begin{equation}\label{zeroodd}
M_\epsilon^O=\sum_{1\leq l\leq n, 2l+1\leq n}(\delta t
k)^{2l+1}b_{ij,n}^{2l}(\theta,\Xi),
\end{equation}
\begin{equation}\label{zeroeven}
M_\epsilon^E=\sum_{1\leq l\leq n, 2l\leq n}(\delta t
k)^{2l}b_{ij,n}^{2l-1}(\theta,\Xi).
\end{equation}

According to the theory of the finite difference method (FDM)
\cite{thomas}, Eq. (\ref{R-F}) can be regarded as the modified
equation of Eq. (\ref{L-NSE-WS}). In modified equations, the higher
even-order derivatives beyond Eq. (\ref{L-NSE-WS}) cause numerical
dissipation and the higher odd-order derivatives cause numerical
dispersion \cite{thomas}.

In order to reduce the dispersion error, when $\sigma_e$ and
$\sigma_\nu$ are specified, it is proposed here to minimize the following cost function:

\begin{equation}\label{Minimization-A}
 F^o(\Xi)=\int_{0}^{\pi} \int_{0}^{2\pi}\|
M_\epsilon^O\|_F^2{\rm d}\theta{\rm d}(\delta t k).
\end{equation}

For the standard MRT-LBM, the parameters $\sigma_\epsilon$ and
$\sigma_q$ need to be determined. The corresponding
conditions are
\begin{equation}
\sigma_\epsilon\geq 0, \sigma_q\geq 0,
\end{equation}
that is to say,
\begin{equation}
s_\epsilon,s_q\in (0,2].
\end{equation}
In order to separate the kinetic modes form the modes directly
affecting hydrodynamic transport, Lallemand and Luo
\cite{lallemandluo} suggested that $s_\epsilon$ and $s_q$ should be
kept slightly larger than 1. Accordingly, it was implied that $s_\epsilon,s_q\in (1,2)$.
In this paper,
$s_\epsilon,s_q$ are taken in the range $(0,2]$.

Furthermore, the same method can be used to reduce dissipation
error. The corresponding cost function is
\begin{equation}\label{Minimization-B}
 F^e(\Xi)=\int_{0}^{\pi} \int_{0}^{2\pi}\|
M_\epsilon^E\|_F^2{\rm d}\theta{\rm d}(\delta t k).
\end{equation}

If both of dispersion and dissipation errors need to be reduced,
according to Eqs. (\ref{approximation}) and (\ref{approximation2}),
 minimizing error between hydrodynamic modes of the L-NSE and the
L-MRT-LBM is achieved by minimizing
\begin{equation}\label{Minimization-C}
 \mathcal{F}(\Xi)=\int_{0}^{\pi}
\int_{0}^{2\pi}\| \mathcal{M}_\epsilon\|_F^2{\rm d}\theta{\rm
d}(\delta t k)=\int_{0}^{\pi} \int_{0}^{2\pi}(\|
M_\epsilon^O\|_F^2+\| M_\epsilon^E\|_F^2){\rm d}\theta{\rm d}(\delta
t k).
\end{equation}

\subsubsection{The non-zero mean flow case}\label{meanopt}

Considering $u_m\neq 0$, then,
$\mathcal{M}_\epsilon=\mathcal{B}_{,n}-\mathcal{B}$ is a function of
$\delta t k,\theta,\vartheta, \Xi$ and $u_m$. It is known that it is
difficult for the non-zero mean flows to determine the values of
free parameters locally, because the optimization problems must be
solved at each lattice node. In order to avoid solving optimization
problems locally, the minimization problems will be integrated with
respect to $\vartheta$ and $u_m$. Here, it is necessary to mention
that the similar relations of Eqs. (\ref{zeroodd}) and
(\ref{zeroeven}) about $M_\epsilon^E$ and $M_\epsilon^O$ are not
satisfied for the non-zero mean flows.

In order to minimize the dispersion error, the following
cost function is introduced
\begin{equation}\label{approximation-D}
G^e(\Xi)=\int_0^{u_0}\int_0^{2\pi}\int_{0}^{\pi} \int_{0}^{2\pi}\|
M_\epsilon^E\|_F^2{\rm d}\theta{\rm d}(\delta t k){\rm d}\vartheta
{\rm d}u_m,
\end{equation}
where $u_0$ is the upper bound of lattice velocity magnitude.
Generally, $u_0$ is taken equal to 0.2. Similarly, in order to
minimize the dissipation error, we have the cost function
\begin{equation}\label{approximation-E}
G^o(\Xi)=\int_0^{u_0}\int_0^{2\pi}\int_{0}^{\pi} \int_{0}^{2\pi}\|
M_\epsilon^O\|_F^2{\rm d}\theta{\rm d}(\delta t k){\rm d}\vartheta
{\rm d}u_m.
\end{equation}
For the non-zero mean flows, the optimization problem of  minimizing
error between hydrodynamic modes of the L-NSE and the L-MRT-LBM is
associated with the following cost function
\begin{equation}\label{Minimization-F}
\mathcal{G}(\Xi)=\int_0^{u_0}\int_0^{2\pi}\int_{0}^{\pi}
\int_{0}^{2\pi}\| \mathcal{M}_\epsilon\|_F^2{\rm d}\theta{\rm
d}(\delta t k){\rm d}\vartheta {\rm d}u_m.
\end{equation}

\subsubsection{A non-zero mean flow case and separating the bulk-viscosity terms from the dissipation coefficient matrix}\label{meanoptbulk}

It is observed that when the shear viscosity is very small, the
magnitude of the bulk viscosity is very sensitive to the stability
numerically and theoretically. If the bulk viscosity is also too small, the MRT-LBM schemes will be unstable.
Meanwhile, although we adopt the optimization strategy detailed in
Sec. \ref{meanopt} and the optimized MRT-LBM appears to be more
stable than the original MRT-LBM, the stability of the obtained
MRT-LBM is still very sensitive. Furthermore, it is known that for
linear acoustic problems, the values of shear and bulk viscosity are
often very small and the dissipation effects from the shear and bulk
viscosity can nearly be neglected. In the simulations, if the bulk
viscosity is too large, the pressure fluctuations will be damped
very significantly. This over-dissipative behavior should be avoided
for aeroacoustic problems. In order to handle the low bulk viscosity
problems, we propose a new optimization strategy. For the uniform
flows, the linearized convection terms are given by the matrix
(\ref{uniform-0o}) in spectral space, and the recovered linearized
dissipation matrix is also given by a matrix in
\ref{coef-matrices-uniform}. The exact linearized dissipation
coefficient matrix with the uniform flows is given by (the simple
derivations with respect to the perturbation of $\rho$, $j_x$ and
$j_y$ are neglected)
\begin{equation}\label{dissipation-matrix-B}
 \left[ \begin {array}{ccc} 0&0&0\\
 \noalign{\medskip}0&-\frac{1}{3}\,{{\it k_x}
}^{2}\sigma_e-\frac{1}{3}\,{{\it
k_x}}^{2}\sigma_\nu-\frac{1}{3}\,{{\it k_y}}^{2}\sigma_\nu&
-\frac{1}{3}\,{\it k_x}\,{\it
k_y}\,\sigma_e\\
\noalign{\medskip}0&-\frac{1}{3}\,{\it k_x} \,{\it
k_y}\,\sigma_e&-\frac{1}{3}\,{{\it
k_x}}^{2}\sigma_\nu-\frac{1}{3}\,{{\it k_y}}^{2}
\sigma_e-\frac{1}{3}\,{{\it k_y}}^{2}\sigma_\nu\end {array} \right]
\end{equation}
It is clear that the coefficient matrix (\ref{dissipation-matrix-B})
has the same expression as the coefficient matrix
(\ref{dissipation-matrix}). The coefficient matrix corresponding to
bulk viscosity is given by
\begin{equation}\label{dissipation-matrix-Bb}
\mathscr{B}=\left[ \begin {array}{ccc} 0&0&0\\
 \noalign{\medskip}0&-\frac{1}{3}\,{{\it k_x}
}^{2}\sigma_e& -\frac{1}{3}\,{\it k_x}\,{\it
k_y}\,\sigma_e\\
\noalign{\medskip}0&-\frac{1}{3}\,{\it k_x} \,{\it
k_y}\,\sigma_e&-\frac{1}{3}\,{{\it k_y}}^{2} \sigma_e\end {array}
\right]
\end{equation}
The new perturbation matrix $\mathcal{M}_\epsilon$ is defined by
$\mathcal{M}_\epsilon=\mathcal{B}_{,n}-\mathcal{B}+\mathscr{B}$. The
matrix $\mathcal{M}_\epsilon$ possesses the information of the bulk
viscosity at the first-order of $\delta t$. The optimization
strategies are kept the same as those in Sec. \ref{meanopt}. The
parameter $\sigma_e$, which is related to the bulk viscosity, can be
taken as an user-specified or free parameter for acoustic problems.
If {\color{blue} the} MRT-LBM is considered as a high-precision
solver for the nearly incompressible flows,  $\sigma_e$ can be set
as a free parameter.

\section{von Neumann analysis of the dispersion and dissipation relations}
\label{vonN}

 In this part, the optimization strategies will be
investigated by means of von Neumann analysis. At the same time, we
will give some numerical results of the dispersion and dissipation
relations.

\subsection{Theoretical dispersion and dissipation relations for the L-MRT-LBM and the L-NSE}
In order to apply von Neumann analysis to validate the optimized
parameters, it is necessary to give the expressions of the L-MRT-LBM
in frequency-wave number space. Considering a uniform mean part
$f_i^0$ and a fluctuating part $f_i{'}$, the equilibrium
distribution function can be linearized as
\cite{simondenispierre,ricotmariesagautbailly}
\begin{equation}\label{perturb-f}
f^{\rm(eq)}_i(\{f_j^0+f_j{'}\}_{0\leq j\leq
N})=f_i^{\rm(eq),0}+\left.\frac{\partial f^{\rm(eq)}_i}{\partial
f_j}\right|_{f_j=f_j^0}\cdot f_j{'}+O(f_j'^2).
\end{equation}
Then, considering a plane wave solution of the linearized equation
\begin{equation}\label{perturb-w-f}
f_j'=A_j{\rm exp}[\mathrm{i}({\bf k}\cdot {\bf x}-\omega t)],
\end{equation}
according to Eqs. (\ref{perturb-f}), (\ref{perturb-w-f}) and
(\ref{q-macro}), we get the following eigenvalue problem for the
L-MRT-LBM in frequency-wave number space
\cite{simondenispierre,ricotmariesagautbailly,lallemandluo}
\begin{equation}
e^{-\mathrm{i}\omega}{\bf f}'=M^{\rm mrt}{\bf f}',
\end{equation}
where the matrix $M^{\rm mrt}=A^{-1}\left[I-M^{-1}SMN^{\rm
bgk}\right]$, and $N^{\rm bgk}$ is defined by
\begin{equation}
N^{\rm bgk}_{ij}=\delta_{ij}-\left.\frac{\partial
f_i^{\rm(eq)}}{f_j}\right|_{f_j=f_j^0}.
\end{equation}
For the L-NSE, the analytical acoustic modes $\omega^{\pm}$ and
shear modes $\omega^s(\bf k)$ are given by \cite{landaulifshitz}
\begin{equation}
\left\{\begin{array}{l} Re[\omega^{\pm}({\bf k})]=|{\bf k}|(\pm
c_s({\bf k})+|{\bf u}|{\bf
cos}(\widehat{\bf k\cdot u})),\\
Im[\omega^{\pm}({\bf k})]=-|{\bf
k}|^2\frac{1}{2}\left(\frac{2d-2}{d}\nu({\bf k})+\eta(\bf
k))\right),\\
Re[\omega^s(\bf k)]=|{\bf k}||{\bf u}|{\rm cos}(\widehat{\bf k\cdot u}),\\
Im[\omega^s(\bf k)]=-|{\bf k}|^2\nu({\bf k}),
\end{array}\right.
\end{equation}
where $\nu$ is the shear viscosity and $\eta $ is the bulk
viscosity.

\subsection{Optimized free parameters with the zero-mean
flows or with the non-zero mean flows}

 First, we set $n=5$ and $u_m=0$ in Eqs. (\ref{R-F}) and
(\ref{B-S-U-wavenumber}). When $\sigma_e=0.0025$ and
$\sigma_\nu=0.0025$, the analytic expressions of the problems
(\ref{Minimization-A}) and (\ref{Minimization-B}) are given in
\ref{minimization-v}. The optimized results are given in Table
\ref{tab:4-zero}.
\begin{table}
\caption{Optimized free parameters for zero-mean
flow case.\label{tab:4-zero}} \centering
\begin{tabular*}{\textwidth}{@{\extracolsep{\fill}}cccccccc}\toprule[1pt]
Groups & Methods  & $u_m$ &$\sigma_e=\sigma_\nu$ & $\sigma_\epsilon$ & $\sigma_q$ & $F^o(\Xi)$ & $\mathcal{F}(\Xi)$\\
\midrule[1pt] A & Min. (\ref{Minimization-A}) & 0 & 0.0025 & 0 & 105.468091254867 & 17.9024342612509066 & $\setminus$ \\
B & Min. (\ref{Minimization-C})&0 & 0.0025 & 0 & 105.465307838135& $\setminus$ & 17.9030645832220686\\
C & Min. (\ref{Minimization-A})&0 & 0.1
&0&26.3631592758091&17.9107477965877778
& $\setminus$\\
D & Min. (\ref{Minimization-C})&0 & 0.1 &0&26.3520430827600 &
$\setminus$ & 17.9208148202264042
\\ \bottomrule[1pt]
\end{tabular*}
\end{table}

\begin{figure}[!h]
\begin{center}
\scalebox{0.6}[0.6]{\includegraphics{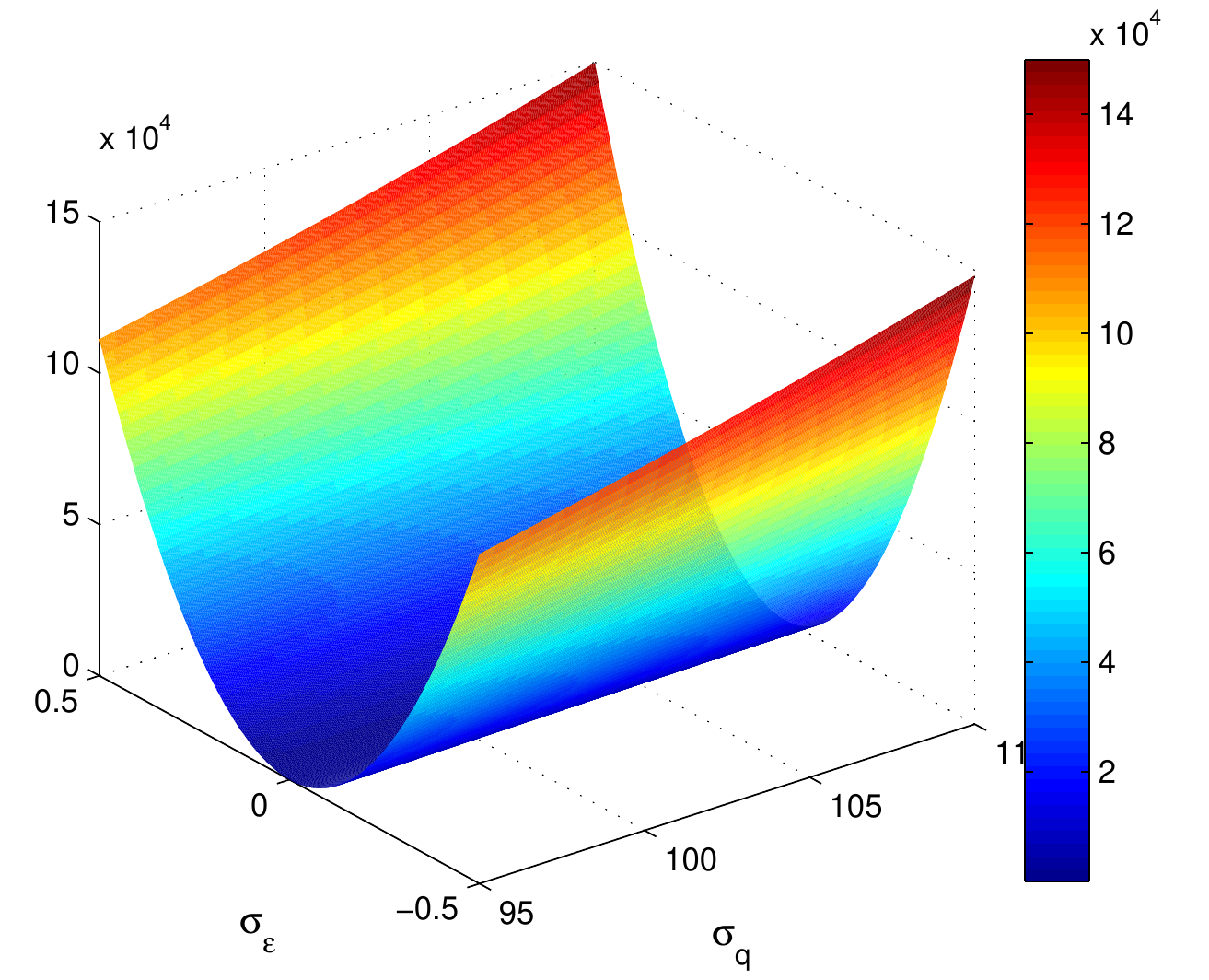}}\scalebox{0.6}[0.6]{\includegraphics{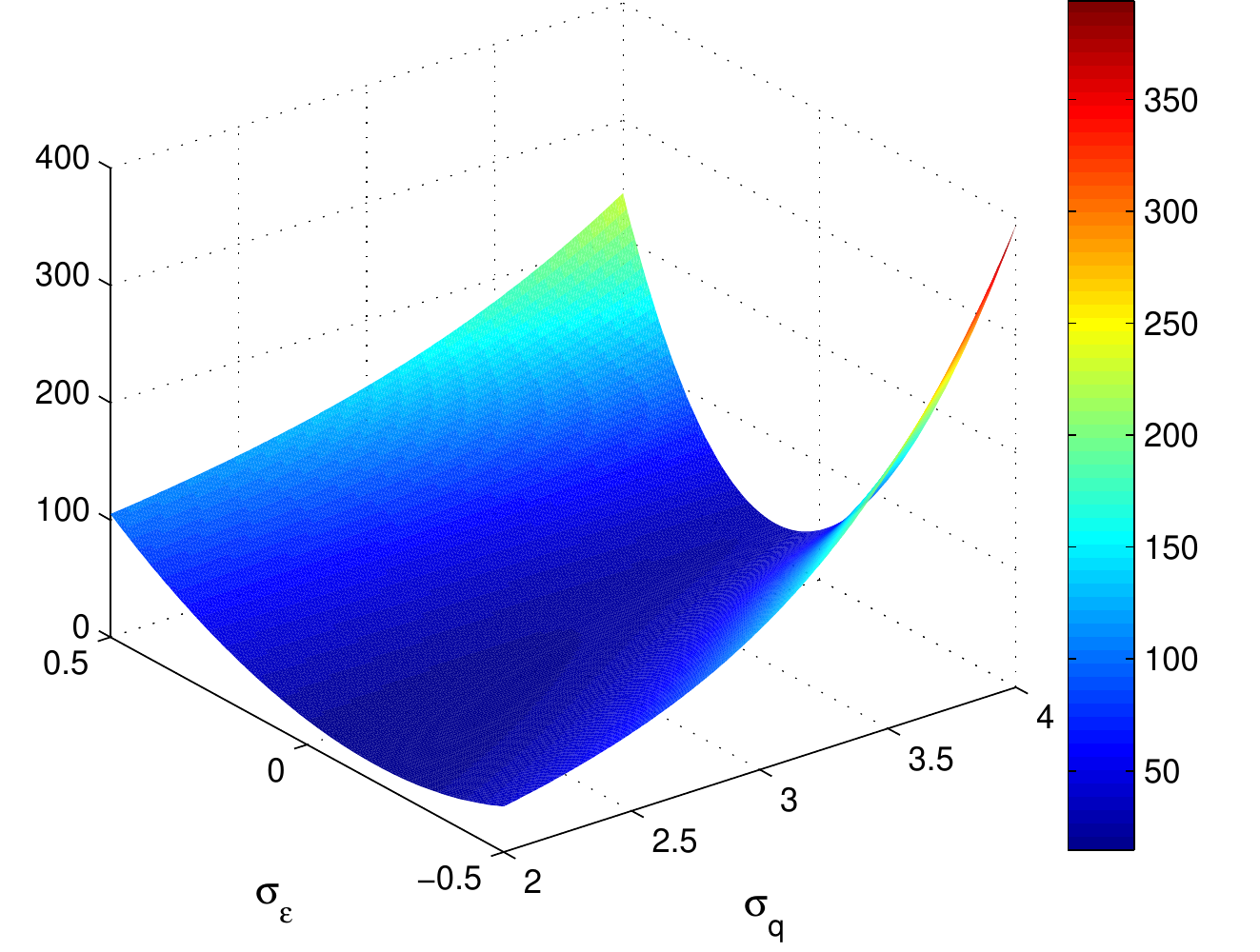}}
\caption{The 3D surfaces of the minimization function $F^o(\Xi)$:
(left) $\sigma_e=0.0025,\sigma_\nu=0.0025$ (right)
$\sigma_e=0.1,\sigma_\nu=0.1$\label{Fig:4-1}}
\end{center}
\end{figure}

It is observed
that for $n=5$ and $u_m=0$, the optimized value of
$\sigma_\epsilon$ is close to zero or equal to zero, this is to say,
the corresponding $s_\epsilon$ is close or equal to 2. When
$\sigma_e$ and $\sigma_\nu$ are attenuated, numerically,
$\sigma_\epsilon$ is equal to 0. However, the magnitude of
$\sigma_q$ increases. Furthermore, the values of $F^o(\Xi)$ and
$\mathcal{F}(\Xi)$ are around 17.9. In Fig. \ref{Fig:4-1}, the 3D
surfaces of $F^o(\Xi)$ are displayed. From the figure, it is known that
the function $F^o(\Xi)$ is convex and the extreme value can be found
along $\sigma_\epsilon=0$. According to present numerical
investigations, the value of $\sigma_\epsilon$  is always close or
equal to 0. For the zero-mean flows, when $\sigma_e$ and $\sigma_q$
are specified, we can specify $\sigma_\epsilon$ equal to 0 or
slightly larger than 0 in order to simplify the minimization
problems for practical applications.

\begin{figure}[!htbp]
\begin{center}
\scalebox{0.8}[0.8]{\includegraphics{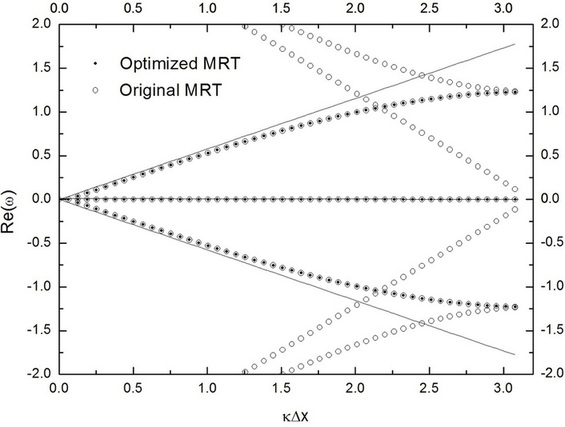}}
\scalebox{0.8}[0.8]{\includegraphics{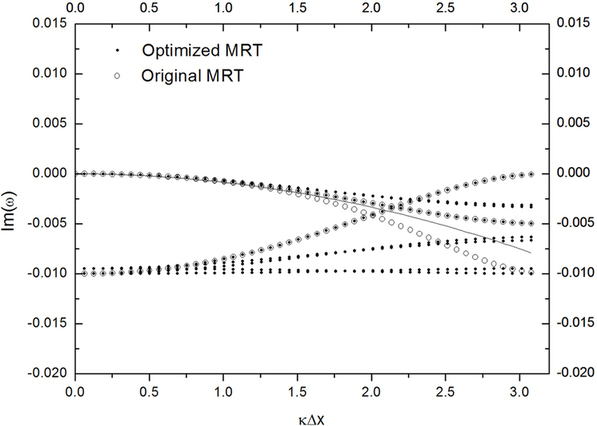}} \\
\centering{(a-1)\hspace{7cm}(a-2)} \\
\end{center}
\end{figure}
\begin{figure}[!htbp]
\begin{center}
\scalebox{0.8}[0.8]{\includegraphics{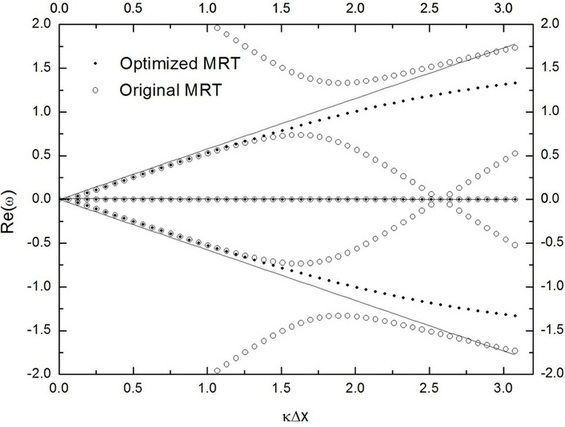}}
\scalebox{0.8}[0.8]{\includegraphics{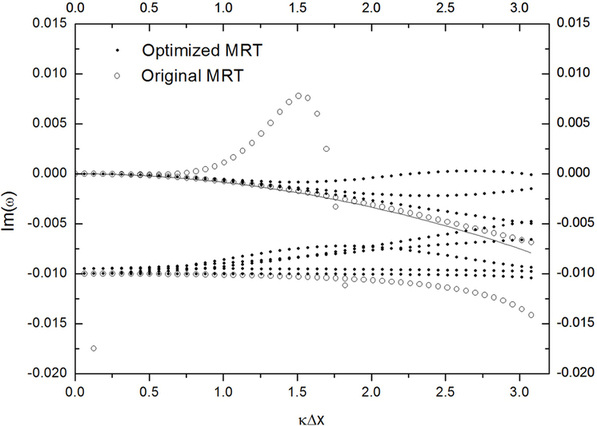}} \\
\centering{(b-1)\hspace{7cm}(b-2)}\\
\end{center}
\end{figure}
\begin{figure}[!htbp]
\begin{center}
\scalebox{0.8}[0.8]{\includegraphics{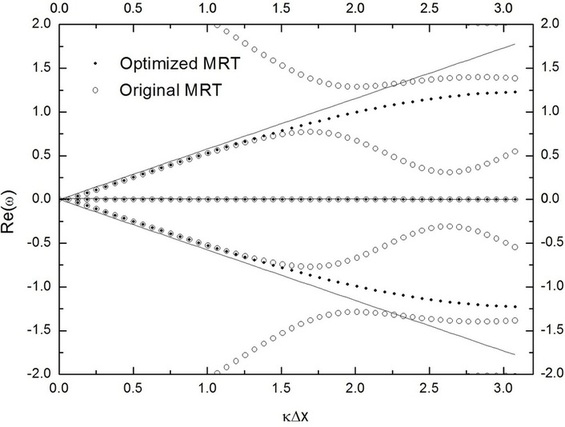}}
\scalebox{0.8}[0.8]{\includegraphics{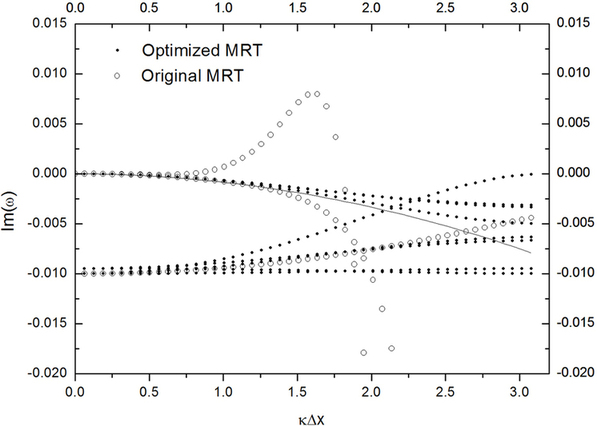}} \\
\centering{(c-1)\hspace{7cm}(c-2)} \\
\end{center}
\end{figure}
\begin{figure}[!htbp]
\begin{center}
\scalebox{0.8}[0.8]{\includegraphics{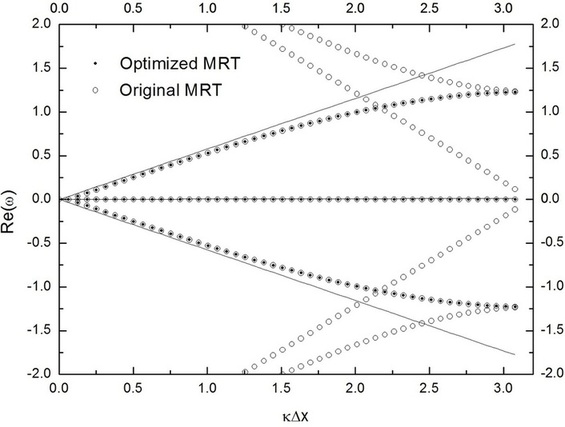}}
\scalebox{0.8}[0.8]{\includegraphics{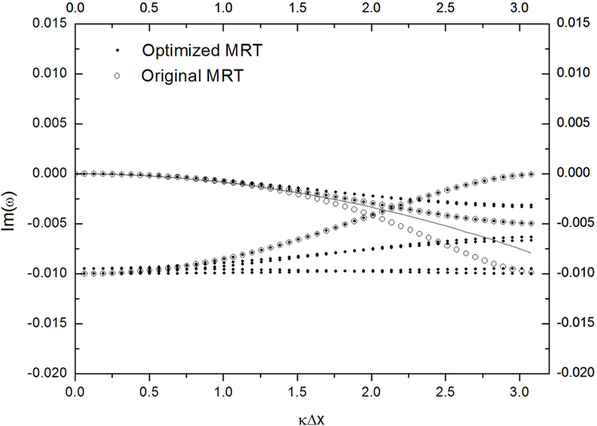}} \\
\centering{(d-1)\hspace{7cm}(d-2)} \\
\end{center}
\end{figure}

\begin{figure}[!htbp]
\begin{center}
\scalebox{0.8}[0.8]{\includegraphics{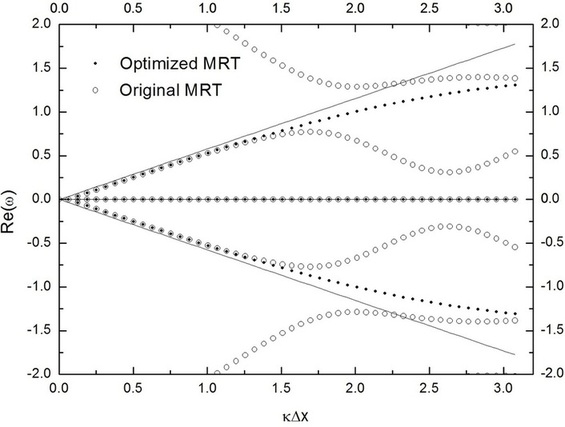}}
\scalebox{0.8}[0.8]{\includegraphics{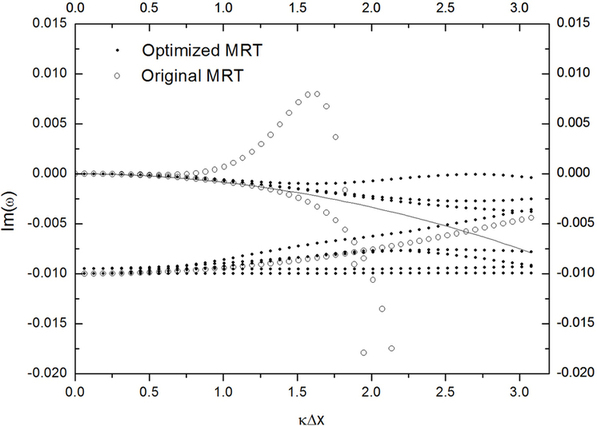}} \\
\centering{(e-1)\hspace{7cm}(e-2)} \\
\end{center}
\end{figure}
\begin{figure}[!htbp]
\begin{center}
\scalebox{0.8}[0.8]{\includegraphics{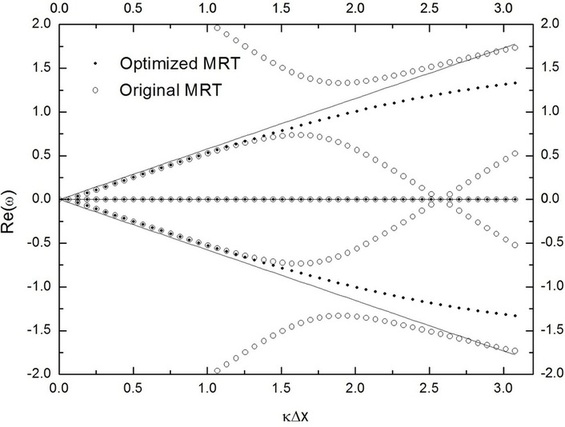}}
\scalebox{0.8}[0.8]{\includegraphics{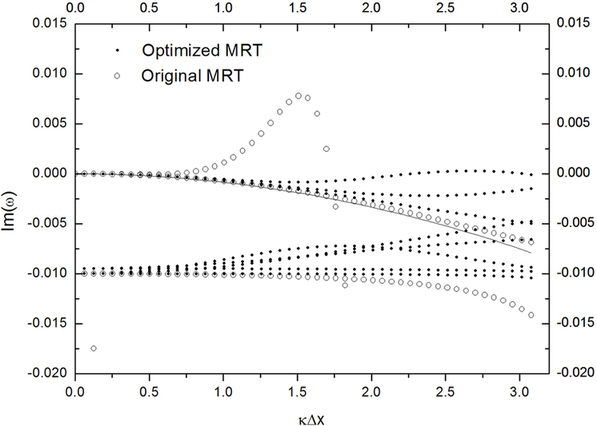}} \\
\centering{(f-1)\hspace{7cm}(f-2)} \\
\caption{Dispersion and dissipation profiles of the L-MRT-LBM based
on recommended free parameters (except $s_e$ and $s_\nu$)
\cite{lallemandluo} and optimized free parameters (group-A in Table
\ref{tab:4-zero}). The relaxation parameters $s_e$ and $s_\nu$ are
kept equal to those of the original MRT-LBM and the optimized
MRT-LBM. The (\#-1) figures display the dispersion profiles
(magnified locally) and the straight lines represent the exact
dispersion solutions. The (\#-2) figures display the dissipation
curves and the line is the expected shear mode and acoustic mode
dissipation. The angle $\theta$ between the wavenumber $\bf k$ and
the x-axis: (a) $\theta=0$;(b) $\theta=\pi/6$; (c) $\theta=\pi/4$;
(d) $\theta=\pi/2$; (e) $\theta=2\pi/3$; (f) $\theta=3\pi/4$. (The
symbol ``\#" stands for the characters a, b, c, d, e and
f.)\label{Fig:4-2}}
\end{center}
\end{figure}

\begin{figure}[!htbp]
\begin{center}
\scalebox{0.5}[0.5]{\includegraphics{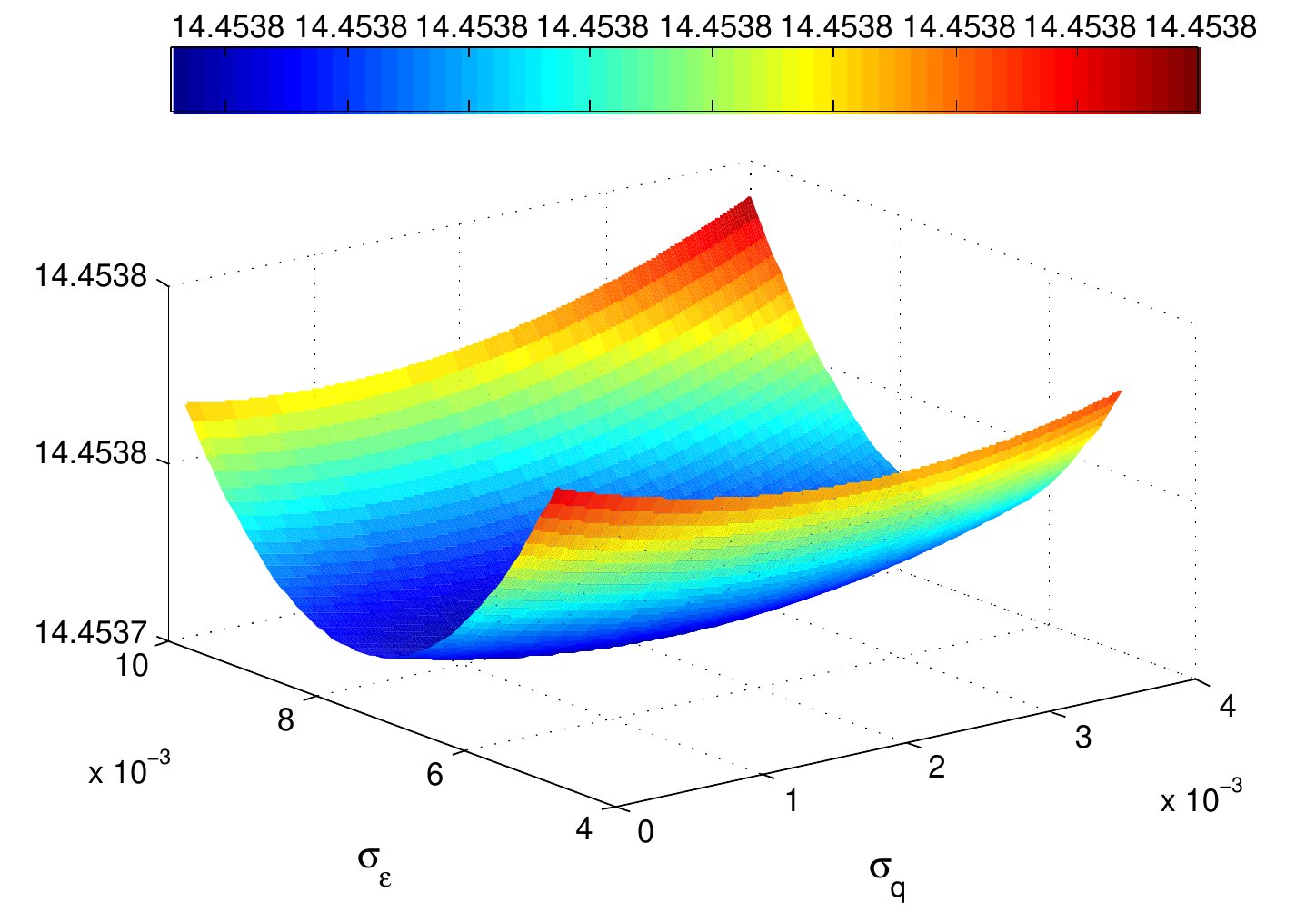}}\scalebox{0.5}[0.5]{\includegraphics{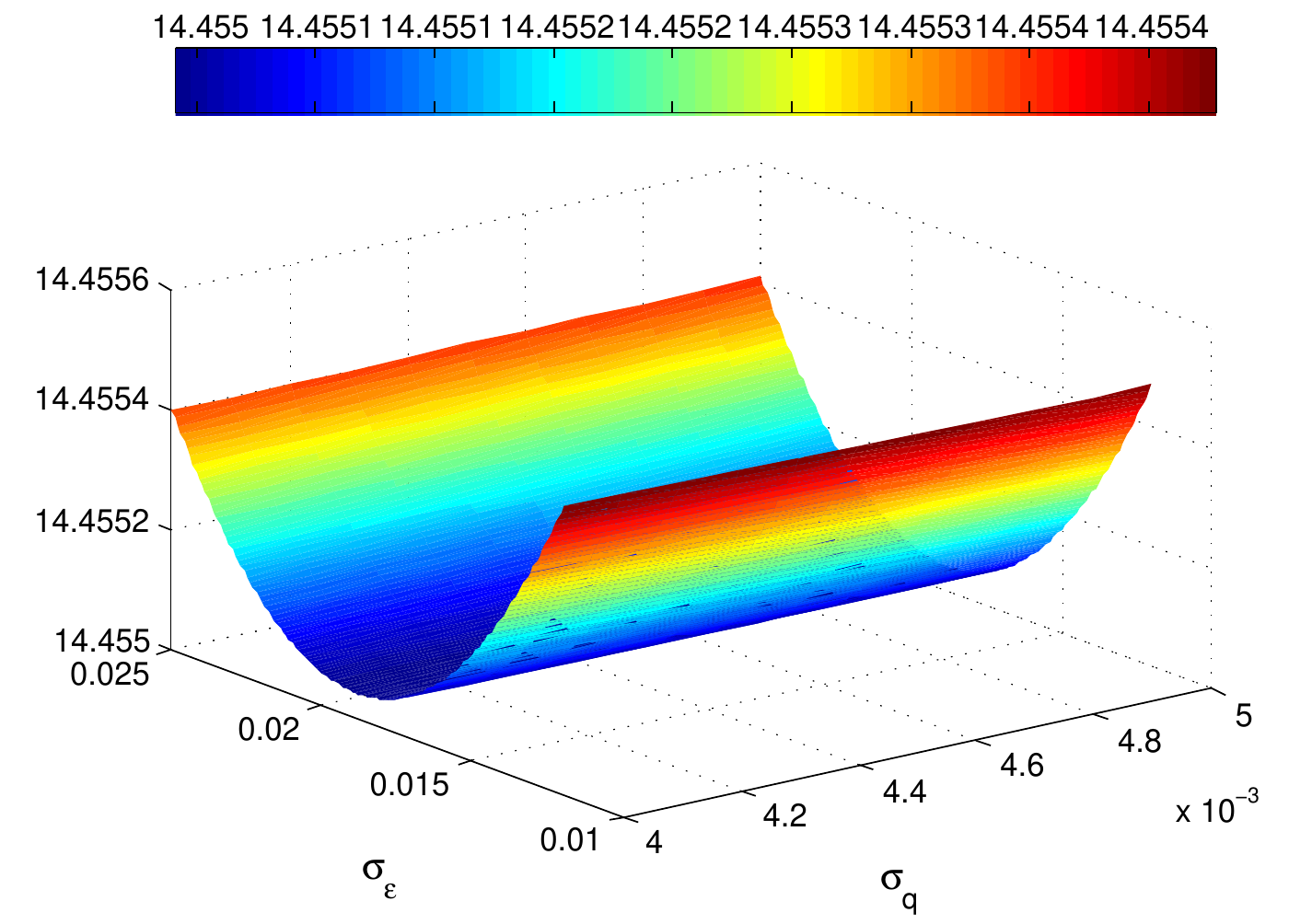}}
\caption{The 3D surfaces of the minimization function
$\mathcal{F}^o(\Xi)$: (left) $\sigma_e=0.001,\sigma_\nu=0.001$;
(right) $\sigma_e=0.0025,\sigma_\nu=0.0025$. \label{Fig:4-3}}
\end{center}
\end{figure}

The Fig.  \ref{Fig:4-2} displays the dispersion and dissipation
profiles of the L-MRT-LBM. In the original MRT-LBM, the free
parameters $\sigma_\epsilon$ and $\sigma_q$ are taken equal to the
recommended values \cite{lallemandluo} and the optimized free
parameters are given by Group A in Table \ref{tab:4-zero}. From
numerical results in Fig. \ref{Fig:4-2}, it is observed that when
the wavenumber {\bf k} is perpendicular or parallel to x-axis, the
optimized MRT-LBM has the same dispersion relations as the original
MRT-LBM. When $\theta$ is equal to other values, the optimized
MRT-LBM performs better than  the original MRT-LBM. These results
also indicate that the cross derivatives in Eq. (\ref{R-F}) are the
main source of dispersion error. For the dissipation relations, it
is observed that by the minimization problem, we can enhance the
stability of the MRT-LBM. From the profiles of the dissipation
relations, there exists one unstable mode the imaginary part ${\rm
lm}(\omega)$ of which is larger than 0, when $0.75<k\Delta x<1.75$,
$\theta\neq0$ and $\theta\neq \pi/2$.

According to the definition of $\mathcal {M}_\epsilon$ in Sec.
\ref{meanopt}, we now consider $n=4$ and $u_m=0.1$ in Eqs. (\ref{R-F})
and (\ref{B-S-U-wavenumber}). Under these conditions, the dispersion
and dissipation relations are investigated firstly. In Fig.
(\ref{Fig:4-3}), the 3D surfaces of $\mathcal{G}(\Xi)$ are shown. It
is discovered that the function $\mathcal{G}(\Xi)$ is convex. Some
optimized results for specified $\sigma_e$ and $\sigma_\nu$ are
given in Table \ref{tab:4-uniform-1}.
\begin{table}
\caption{Optimized free parameters for uniform flows ($u_m=0.1$)
based on the perturbation matrix $\mathcal{M}_\epsilon$ in Sec.
\ref{meanopt}. The parameters are obtained by the minimization
(\ref{Minimization-F})\label{tab:4-uniform-1}} \centering
\begin{tabular*}{\textwidth}{@{\extracolsep{\fill}}ccccc}\toprule[1pt]
Groups &  $\sigma_e=\sigma_\nu$ & $\sigma_\epsilon$ & $\sigma_q$ & $\mathcal{G}(\Xi)$\\
\midrule[1pt] A &  0.001 & 0.00751873323089156 & 0.00171909400064198 &14.4537555316616474\\
B &0.0025
&0.0187982349323006&0.00429903714246050&14.4550408968152340\\\bottomrule[1pt]
\end{tabular*}
\end{table}

\begin{figure}[!htbp]
\begin{center}
\scalebox{0.8}[0.8]{\includegraphics{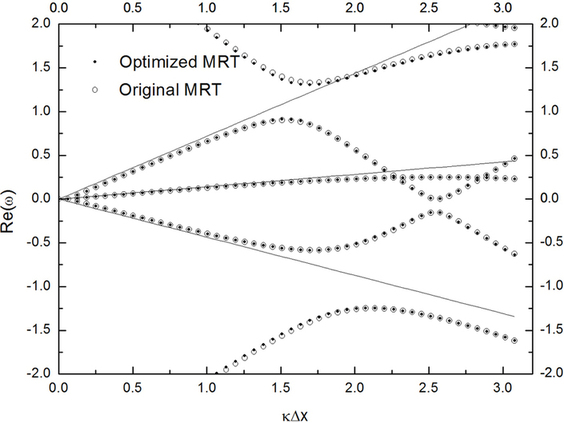}} 
\scalebox{0.8}[0.8]{\includegraphics{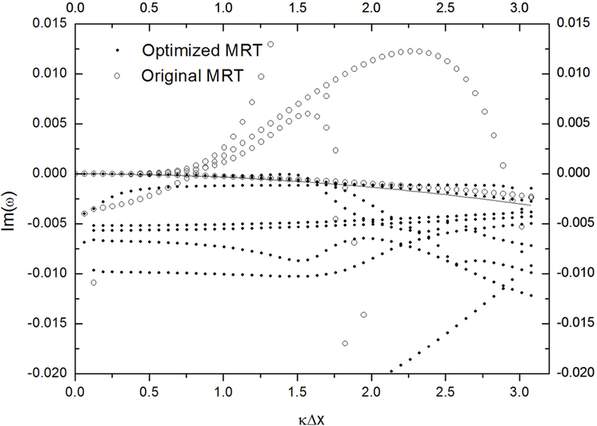}} \\
\centering{(a-1)\hspace{7cm}(a-2)} \\
\end{center}
\end{figure}
\begin{figure}[!htbp]
\begin{center}
\scalebox{0.8}[0.8]{\includegraphics{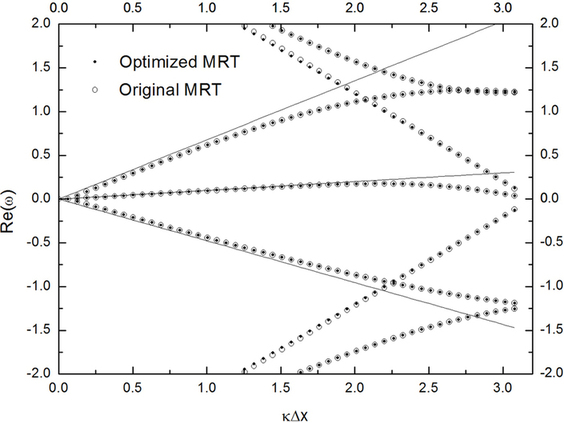}} 
\scalebox{0.8}[0.8]{\includegraphics{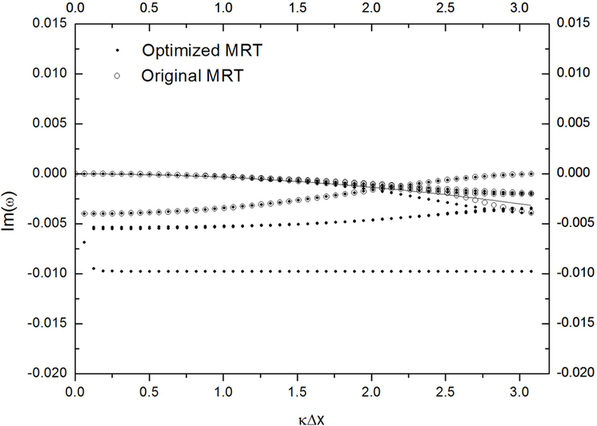}} \\
\centering{(b-1)\hspace{7cm}(b-2)} \\
\end{center}
\end{figure}
\begin{figure}[!htbp]
\begin{center}
\scalebox{0.8}[0.8]{\includegraphics{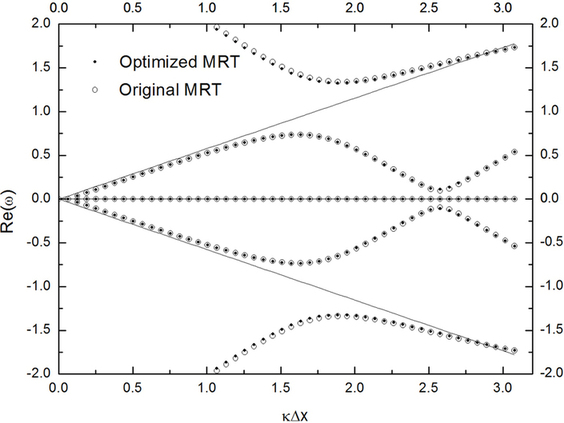}}
\scalebox{0.8}[0.8]{\includegraphics{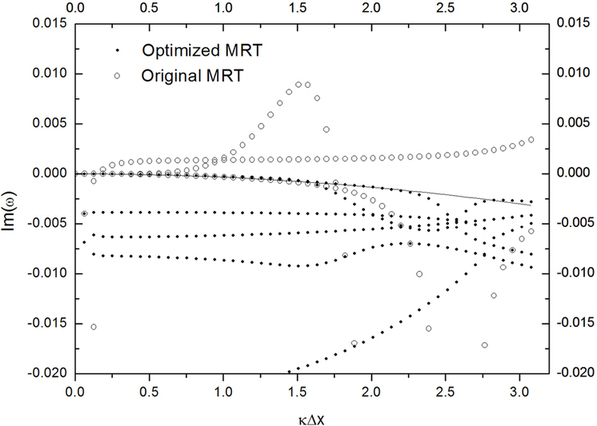}} \\
\centering{(c-1)\hspace{7cm}(c-2)} \\
\caption{Dispersion and dissipation profiles of the L-MRT-LBM based
on recommended free parameters (except $s_e$ and $s_\nu$)
\cite{lallemandluo} and optimized free parameters (group-A in Table
\ref{tab:4-uniform-1}). The relaxation parameters $s_e$ and $s_\nu$
are kept the same values for the original MRT-LBM and the optimized
MRT-LBM. The (\#-1) figures indicate the dispersion profiles
(magnified locally) and the straight lines represent the exact
dispersion solutions. The (\#-2) figures indicate the dissipation
curves and the line is the expected shear mode and acoustic mode
dissipation. The angle $\vartheta=\pi/4$, $U=0.1$ and $V=0.1$. The
angle $\widehat{\bf{k\cdot u}}$ between the wavenumber $\bf k$ and
${\bf u}$: (a) $\widehat{\bf{k\cdot u}}=0,\theta=\pi/4$;(b)
$\widehat{\bf{k\cdot u}}=\pi/4,\theta=\pi$; (c) $\widehat{\bf{k\cdot
u}}=\pi/2,\theta=3\pi/4$.  (The symbol ``\#" stands for the
characters a, b, and c.)\label{Fig:4-4}}
\end{center}
\end{figure}

In Fig.\ref{Fig:4-4}, we show the dispersion and dissipation
profiles. It is seen that the best shear mode description is given
by the optimization problem $\mathcal{G}(\Xi)$. At the same time,
the optimized MRT-LBM is more stable than the original MRT-LBM.
However, it is discovered that the dispersion error is not improved
for $n=4$ by the minimization problem (\ref{Minimization-F}). If we
want to reduce the influence of the non-zero mean flow on the
dissipation relation and avoid handling lengthy mathematical
expressions, it is suitable  to choose $n=4$ in Eq. (\ref{R-F}).
Furthermore, when $\theta=\pi$, the optimized MRT-LBM has the
similar dispersion and dissipation profiles with the original
MRT-LBM in Fig. (\ref{Fig:4-4})-b. According to authors' numerical
investigations, when the angle $\theta$ is equal to $0$, $\pi/2$ and
$\pi$, there always exist the similar dispersion and dissipation
profiles between the optimized MRT-LBM and the original MRT-LBM.
Based on the definition of $\mathcal {M}_\epsilon$ in Sec.
\ref{meanopt}, the results of numerical studies have shown that by
the optimization problems (\ref{approximation-D}),
(\ref{approximation-E}) and (\ref{Minimization-F}), the dispersion
error was not reduced when $n< 5$ in Eq. (\ref{R-F}) for the uniform
flows.

 According to the definition of $\mathcal {M}_\epsilon$ in Sec.
\ref{meanoptbulk}, we consider $n=4$ and $u_m=0.1$ in Eqs.
(\ref{R-F}) and (\ref{B-S-U-wavenumber}). For very small shear and
bulk viscosity parameters, we show some results in Table
\ref{tab:4-uniform-2}.
\begin{table}[!h]
\caption{Optimized free parameters for uniform flows ($u_m=0.1$)
based on the perturbation matrix $\mathcal{M}_\epsilon$ in Sec.
\ref{meanoptbulk}. The parameters are obtained by the minimization
(\ref{Minimization-F}).\label{tab:4-uniform-2}} \centering
\begin{tabular*}{\textwidth}{@{\extracolsep{\fill}}lllllll}\toprule[1pt]
Group& $\sigma_e$&$\sigma_\nu$ & $\sigma_\epsilon$ & $\sigma_q$ & $\mathcal{G}(\Xi)$\\
\midrule[1pt] A & 0.0025125628 & 0.00001 & 0.00947095595580122  & 0.00181622011980382 & 14.453655614637361\\
B  &0.0000025&
0.00001&0.0000311837523990053&0.00000760552251906507&14.45351071030945\\\bottomrule[1pt]
\end{tabular*}
\end{table}

\begin{figure}[!htbp]
\begin{center}
\scalebox{0.8}[0.8]{\includegraphics{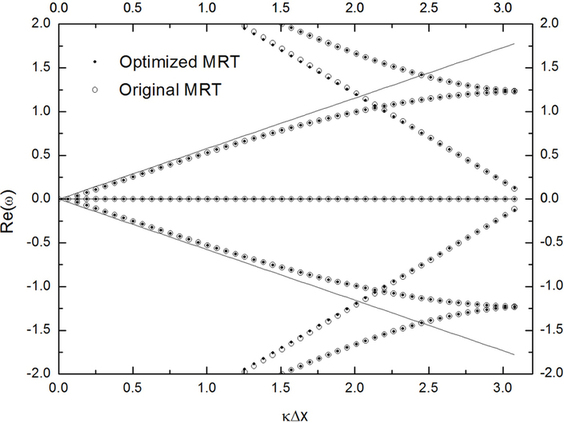}}
\scalebox{0.8}[0.8]{\includegraphics{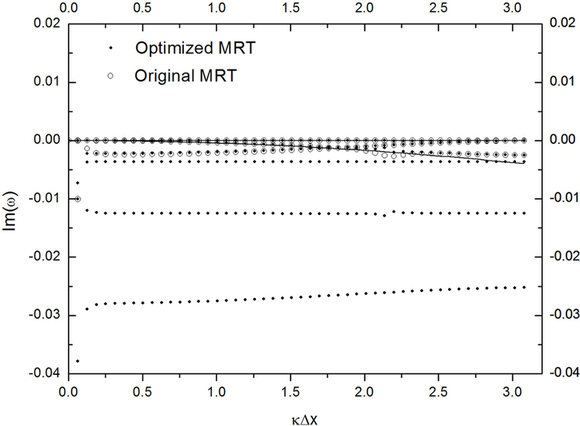}} \\
\centering{(d-1)\hspace{7cm}(d-2)} \\
\caption{Dispersion and dissipation profiles of the L-MRT-LBM based
on recommended free parameters (except $s_e$ and $s_\nu$)
\cite{lallemandluo} and optimized free parameters (group-A in Table
\ref{tab:4-uniform-2}). The relaxation parameters $s_e$ and $s_\nu$
are kept the same values for the original MRT-LBM and the optimized
MRT-LBM. The (\#-1) figures indicate the dispersion profiles
(magnified locally) and the straight lines represent the exact
dispersion solutions. The (\#-2) figures indicate the dissipation
curves and the lines are the expected shear mode and acoustic mode
dissipation. The angle $\vartheta=\pi/4$, $U=0.1$ and $V=0.0$. The
angle $\widehat{\bf{k\cdot u}}$ between the wavenumber $\bf k$ and
${\bf u}$: (a) $\widehat{\bf{k\cdot u}}=0,\theta=0$;(b)
$\widehat{\bf{k\cdot u}}=\pi/3,\theta=\pi/3$; (c)
$\widehat{\bf{k\cdot u}}=\pi/4,\theta=\pi/4$; (d)
$\widehat{\bf{k\cdot u}}=\pi/2,\theta=\pi/2$. (The symbol ``\#"
stands for the characters a, b, c and d.)\label{Fig:4-5}}
\end{center}
\end{figure}

\begin{figure}[!htbp]
\begin{center}
\scalebox{0.8}[0.8]{\includegraphics{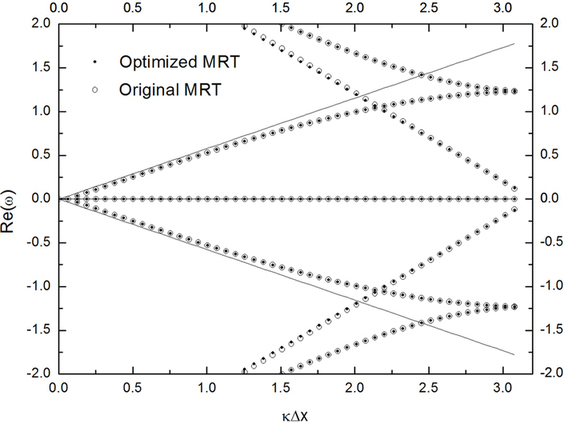}} 
\scalebox{0.8}[0.8]{\includegraphics{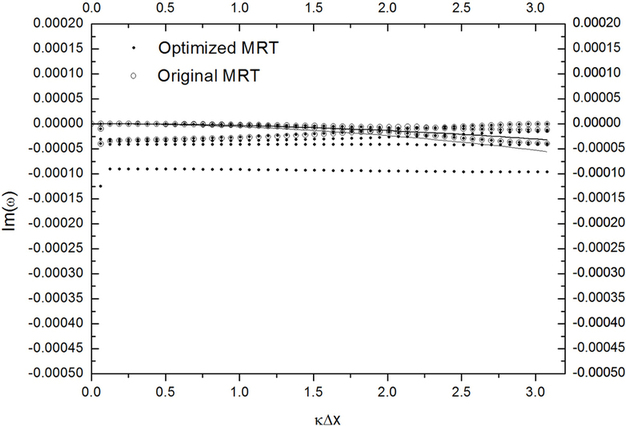}} \\
\centering{(d-1)\hspace{7cm}(d-2)} \\
\caption{Dispersion and dissipation profiles of the L-MRT-LBM based
on recommended free parameters (except $s_e$ and $s_\nu$)
\cite{lallemandluo} and optimized free parameters (group-B in Table
\ref{tab:4-uniform-2}). The relaxation parameters $s_e$ and $s_\nu$
are kept the same values for the original MRT-LBM and the optimized
MRT-LBM. The (\#-1) figures indicate the dispersion profiles
(magnified locally) and the straight lines represent the exact
dispersion solutions. The (\#-2) figures indicate the dissipation
curves and the lines are the expected shear mode and acoustic mode
dissipation. The angle $\vartheta=\pi/4$, $U=0.1$ and $V=0.0$. The
angle $\widehat{\bf{k\cdot u}}$ between the wavenumber $\bf k$ and
${\bf u}$: (a) $\widehat{\bf{k\cdot u}}=0,\theta=0$;(b)
$\widehat{\bf{k\cdot u}}=\pi/3,\theta=\pi/3$; (c)
$\widehat{\bf{k\cdot u}}=\pi/4,\theta=\pi/4$; (d)
$\widehat{\bf{k\cdot u}}=\pi/2,\theta=\pi/2$. (The symbol ``\#"
stands for the characters a, b, c and d.)\label{Fig:4-6}}
\end{center}
\end{figure}

\begin{figure}[!htbp]
\begin{center}
\scalebox{0.75}[0.75]{\includegraphics{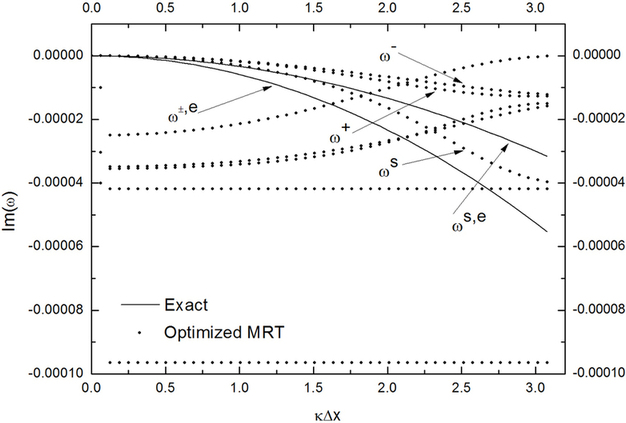}}
\scalebox{0.75}[0.75]{\includegraphics{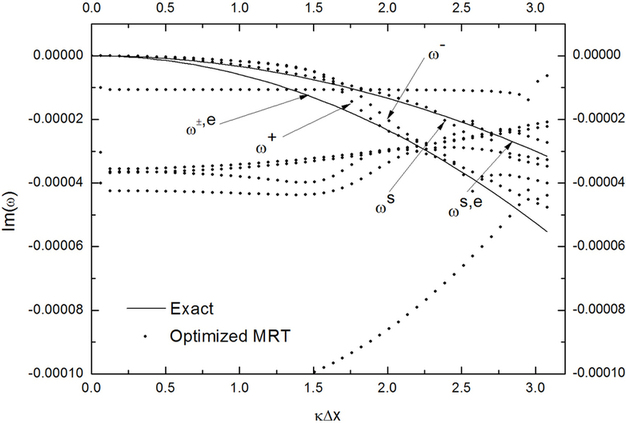}} \\
\centering{(a)\hspace{8cm}(b)} \\
\scalebox{0.75}[0.75]{\includegraphics{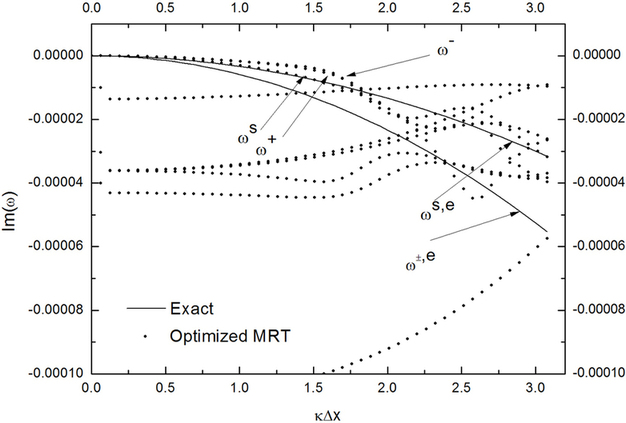}}
\scalebox{0.75}[0.75]{\includegraphics{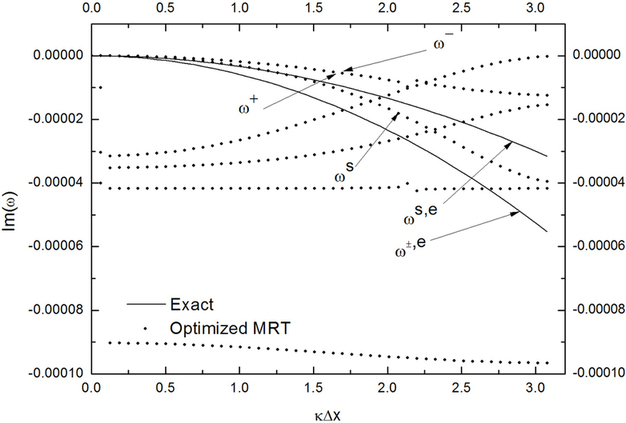}} \\
\centering{(c)\hspace{8cm}(d)}\caption{Locally-magnified dissipation profiles:
Dissipation profiles of exact solutions (acoustic mode:
$\omega^{\pm,{\rm e}}$; shear mode: $\omega^{\rm s,e}$) and
optimized free parameters (group-B in Table \ref{tab:4-uniform-2};
acoustic mode: $\omega^{\pm}$; shear mode: $\omega^{\rm s}$). The
angle $\vartheta=\pi/4$, $U=0.1$ and $V=0.0$. The angle
$\widehat{\bf{k\cdot u}}$ between the wavenumber $\bf k$ and ${\bf
u}$: (a) $\widehat{\bf{k\cdot u}}=0,\theta=0$;(b)
$\widehat{\bf{k\cdot u}}=\pi/3,\theta=\pi/3$; (c)
$\widehat{\bf{k\cdot u}}=\pi/4,\theta=\pi/4$; (d)
$\widehat{\bf{k\cdot u}}=\pi/2,\theta=\pi/2$.\label{Fig:4-7}}
\end{center}
\end{figure}

In Figs. \ref{Fig:4-5} and \ref{Fig:4-6}, the dispersion and
dissipation relations are given for both the optimized MRT-LBM and
the original MRT-LBM. It is discovered that, by the new definition
of the perturbation matrix in Sec. \ref{meanoptbulk}, the optimized
MRT-LBM is always stable for the very small viscosity and bulk
viscosity. Furthermore, when the bulk viscosity (corresponding to
Fig. \ref{Fig:4-6}) is very small, the numerical shear modes given
by the optimized MRT-LBM agree with the exact shear modes very well.
The obtained shear modes are nearly exact compared with the shear
modes given by the original MRT-LBM. In order to observe the details
of the optimized dissipation relations and the exact relations, in
Fig. \ref{Fig:4-7}, the locally-magnified dissipation relations are
given. It is clear that we observe a lower dissipation of the
acoustic modes for the optimized MRT-LBM. From Figs. \ref{Fig:4-5}
and \ref{Fig:4-6}, it is also clear that when the bulk viscosity
become smaller, the modes of the original MRT-LBM become more
unstable for all values of the angle $\theta$. These results
indicate that the original MRT-LBM is not suitable for aeroacoustic
problems, because the bulk viscosity in the original MRT-LBM can not
be chosen to be arbitrarily small. This limitation in the original
MRT-LBM means that there exists a strong dissipation of the acoustic
waves in the numerical simulations. The new definition of the
perturbation matrix $\mathcal{M}_\epsilon$ in Sec. \ref{meanoptbulk}
coupled with the optimization strategy (\ref{Minimization-F})
overcomes this drawback of the MRT-LBM under the premise of
guaranteeing the stability.

\begin{figure}[!htbp]
\begin{center}
\scalebox{0.8}[0.8]{\includegraphics{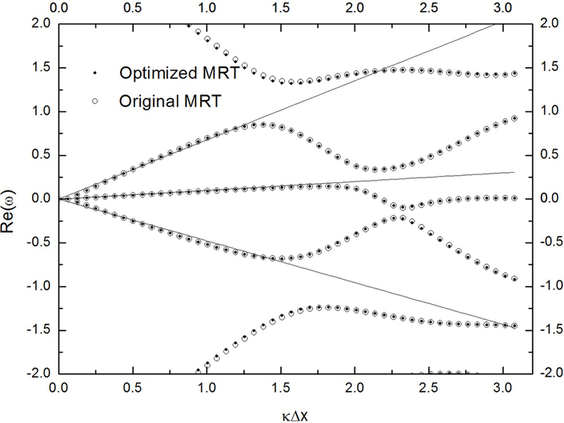}}
\scalebox{0.8}[0.8]{\includegraphics{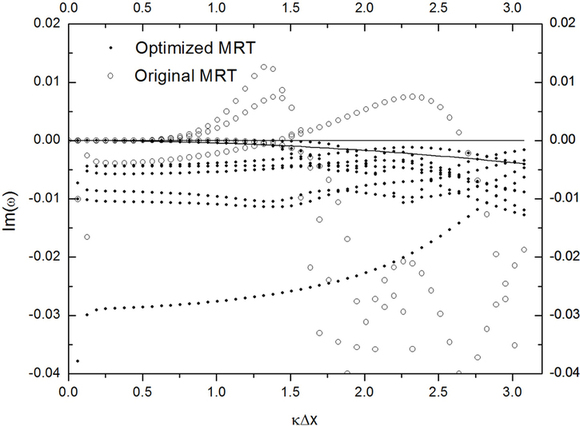}} \\
\centering{(a-1)\hspace{7cm}(a-2)} \\
\scalebox{0.8}[0.8]{\includegraphics{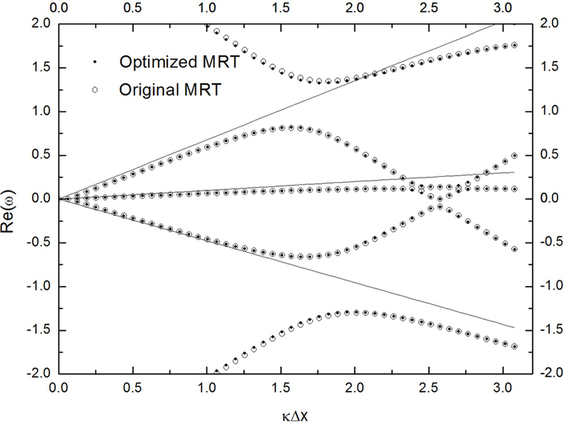}}
\scalebox{0.8}[0.8]{\includegraphics{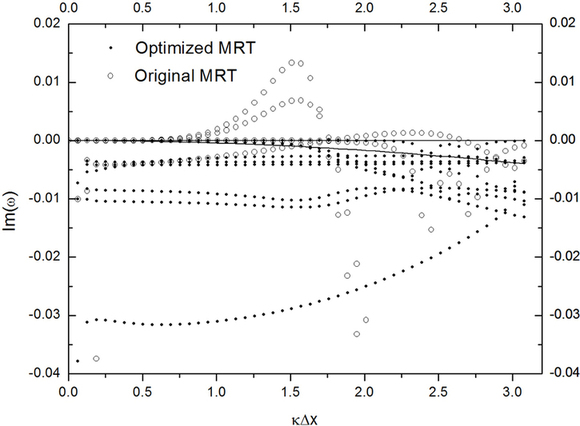}} \\
\centering{(b-1)\hspace{7cm}(b-2)} \\
\scalebox{0.8}[0.8]{\includegraphics{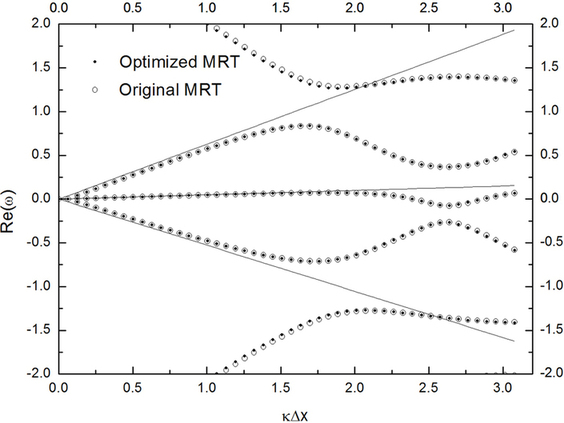}}
\scalebox{0.8}[0.8]{\includegraphics{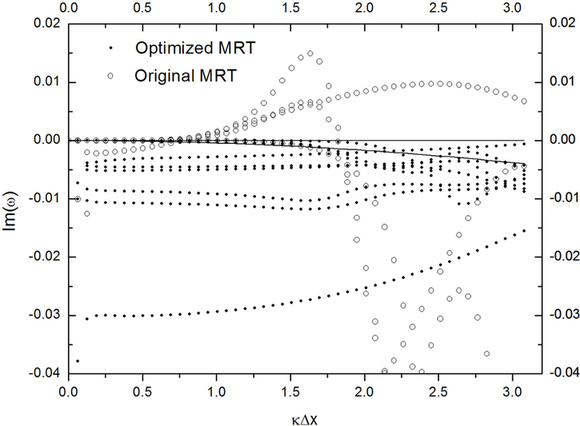}} \\
\centering{(c-1)\hspace{7cm}(c-2)}\\
\end{center}
\end{figure}

\begin{figure}[!htbp]
\begin{center}
\scalebox{0.8}[0.8]{\includegraphics{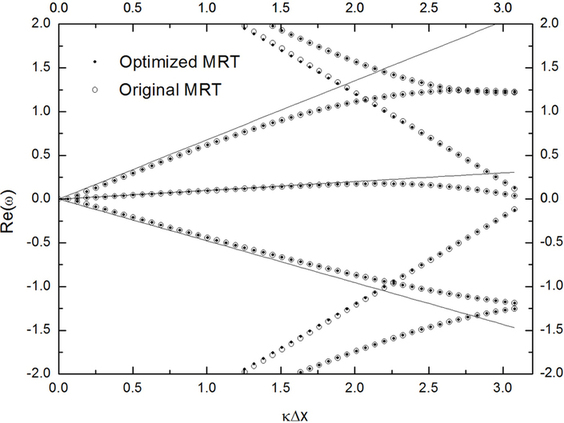}}
\scalebox{0.8}[0.8]{\includegraphics{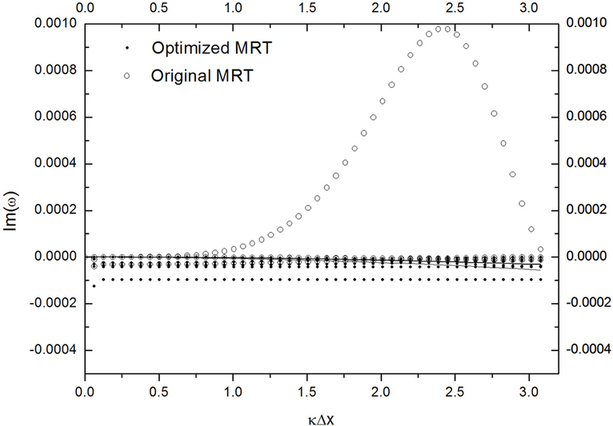}} \\
\centering{(a-1)\hspace{7cm}(a-2)} \\
\scalebox{0.8}[0.8]{\includegraphics{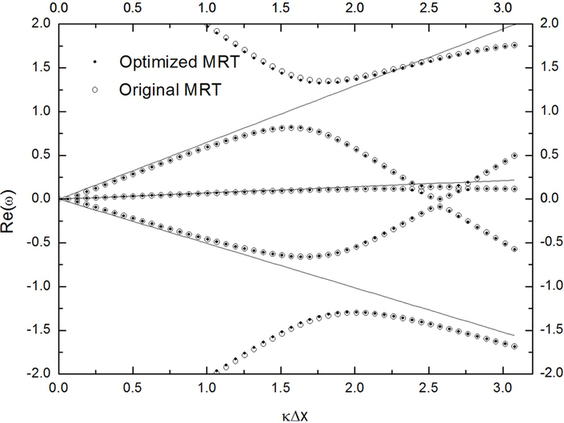}}
\scalebox{0.8}[0.8]{\includegraphics{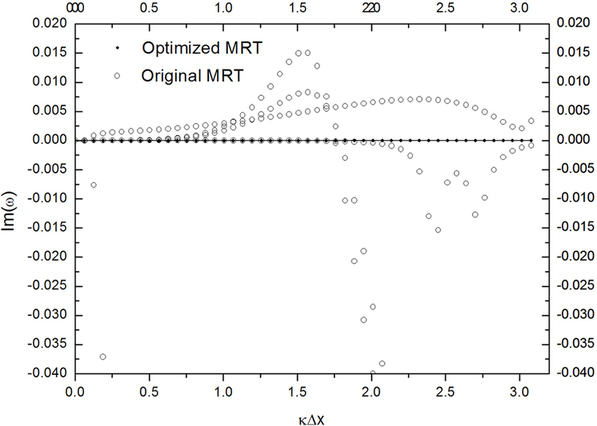}} \\
\centering{(b-1)\hspace{7cm}(b-2)} \\
\scalebox{0.8}[0.8]{\includegraphics{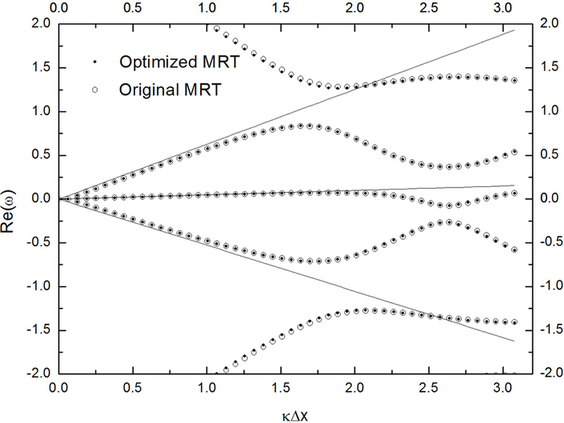}} 
\scalebox{0.8}[0.8]{\includegraphics{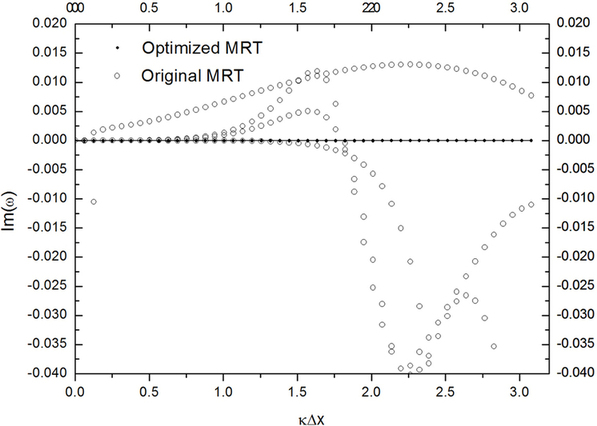}} \\
\centering{(c-1)\hspace{7cm}(c-2)} \\
\end{center}
\end{figure}

\section{Numerical simulations of acoustic problems}
In this section, the classical acoustic problems will be simulated
by optimized MRT-LBM. At the same time, some comparisons between the
D2RP MRT-LBM and the original MRT-LBM are given.
\subsection{Acoustic point source} In this part, we validate the optimized MRT-LBM by an
acoustic point which sends out a sinusoidal signal
\cite{viggen,kinsler}. The point source is set by the following
density configuration \cite{viggen}
\begin{equation}
\rho({\rm x},t)=\rho_0+\rho_s{\rm sin}\left(\frac{2\pi}{T}t\right),
\end{equation}
where $\rho_s$ is the point source amplitude, and $T$ the period of
the oscillation with respect to lattice units. In order to avoid
nonlinear wave effects, it is necessary that $\rho_s\ll \rho_0$. The
macroscopic velocity $(u,v)$ at the point source is equal to 0.

\begin{table}
\caption{Specified relaxation parameters and optimized free
relaxation parameters for zero-mean flows.\label{tab:1chp5}}
\centering
\begin{tabular*}{\textwidth}{@{\extracolsep{\fill}}lllllll}\toprule[1pt]
Groups & $s_e=s_\nu$ & $s_\epsilon$ & $s_q$ & $\nu$ &$\rho$&$\rho_s$ \\
\midrule[1pt] A & $1.99044751$ & $2$ & $0.00875438872$ &
$0.000799861$ &1&0.01
\\ B & 1.95321 &2&0.04126919093&0.00399257&1&0.01
\\ \bottomrule[1pt]
\end{tabular*}
\end{table}
\begin{table}
\caption{Specified parameters and optimized free parameters for
uniform flows based on the perturbation matrix
$\mathcal{M}_\epsilon$ in Sec. \ref{meanoptbulk}. The relaxation
parameters are obtained by the minimization
(\ref{Minimization-F})\label{tab:2chp5}} \centering
\begin{tabular*}{\textwidth}{@{\extracolsep{\fill}}lllll}\toprule[1pt]
Groups & $s_e$ & $s_\nu$ & $s_\epsilon$ & $s_q$  \\
\midrule[1pt] A &1.99& 1.999960001&1.962820428 & 1.992761413
\\ B & 1.99999 &1.999960001&1.999875273&1.999969578
\\ \bottomrule[1pt]
\end{tabular*}
\end{table}

It is known that the sound speed of the D2Q9 MRT-LBM is equal to
$c_s=1/\sqrt{3}=0.57735$. In order to avoid the effects of
boundaries, the wave propagation is limited in the computational
domain $\Omega=100\times 100$. The lattice nodes are $101\times 101$
and the acoustic point source is set in the center of computational
domain. So, if the lattice computational time $t$ is in the range
$(0,\lfloor50/c_s\rfloor]=(0,86]$, the wave will be limited in the
domain $\Omega$. Now, we will use parameters given in Table
\ref{tab:1chp5} to validate the optimized MRT-LBM with the periodic
boundary conditions.

\begin{figure}
\begin{center}
\scalebox{0.6}[0.6]{\includegraphics{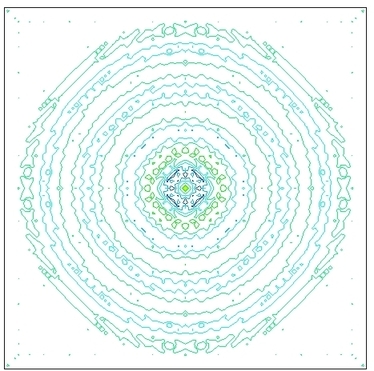}}\quad\quad
\scalebox{0.6}[0.6]{\includegraphics{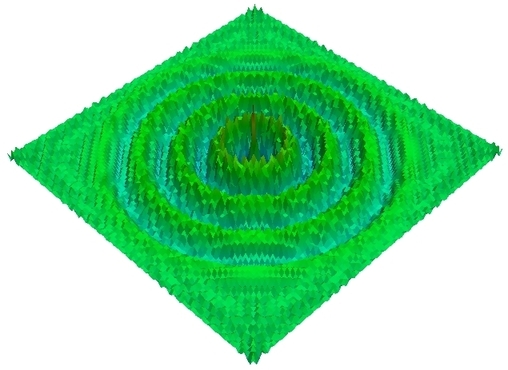}} \\
(a-1) $t=80$\hspace*{7cm} (a-2) $t=80$ \\
\scalebox{0.6}[0.6]{\includegraphics{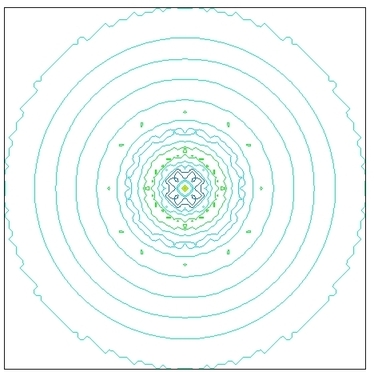}}\quad\quad
\scalebox{0.6}[0.6]{\includegraphics{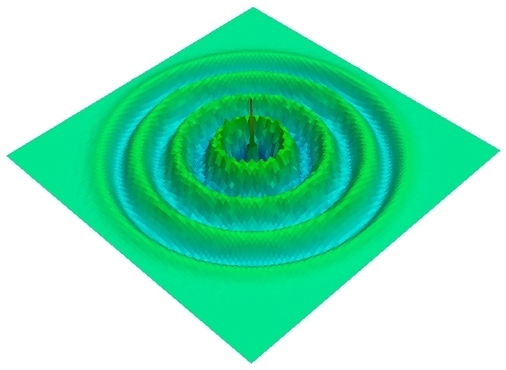}} \\
(b-1) $t=80$\hspace*{7cm} (b-2) $t=80$ \\
\scalebox{0.6}[0.6]{\includegraphics{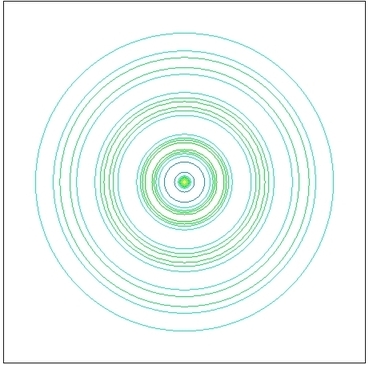}}\quad\quad
\scalebox{0.6}[0.6]{\includegraphics{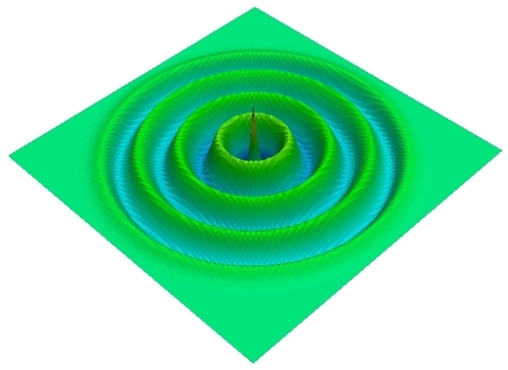}} \\
(c-1) $t=80$\hspace*{7cm} (c-2) $t=80$ \\
\caption{The density contours (left figures) and the 3D surfaces
(right figures) by the parameters of group-A in Table
\ref{tab:1chp5}: (a) the BGK-LBM; (B) the original MRT-LBM; (C) the
optimized MRT-LBM. \label{Fig:5-1}}
\end{center}
\end{figure}

\begin{figure}
\begin{center}
\scalebox{0.6}[0.6]{\includegraphics{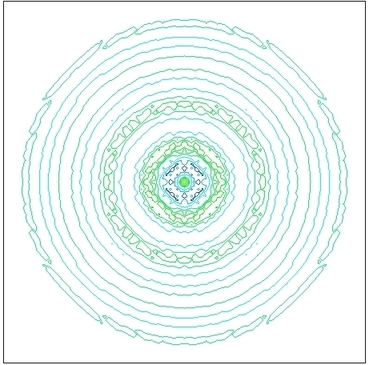}}\quad\quad
\scalebox{0.6}[0.6]{\includegraphics{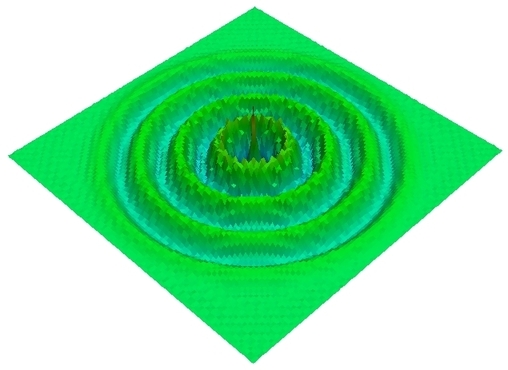}} \\
(a-1) $t=80$\hspace*{7cm} (a-2) $t=80$ \\
\scalebox{0.6}[0.6]{\includegraphics{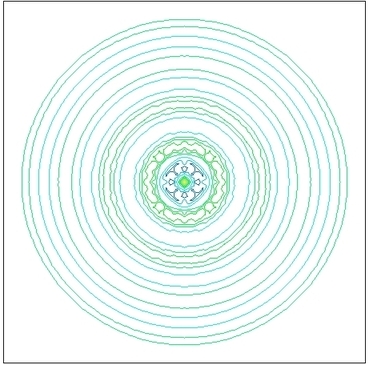}}\quad\quad
\scalebox{0.6}[0.6]{\includegraphics{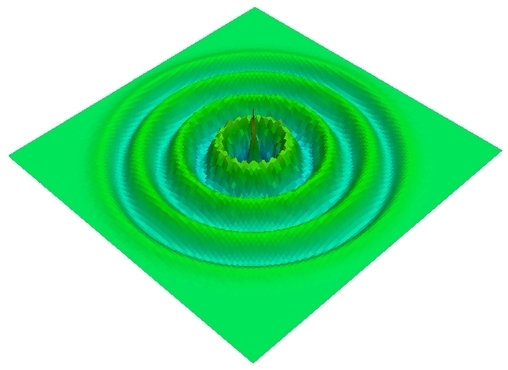}} \\
(b-1) $t=80$ \hspace*{7cm} (b-2) $t=80$ \\
\scalebox{0.6}[0.6]{\includegraphics{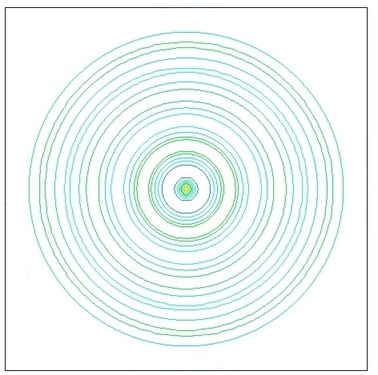}}\quad\quad
\scalebox{0.6}[0.6]{\includegraphics{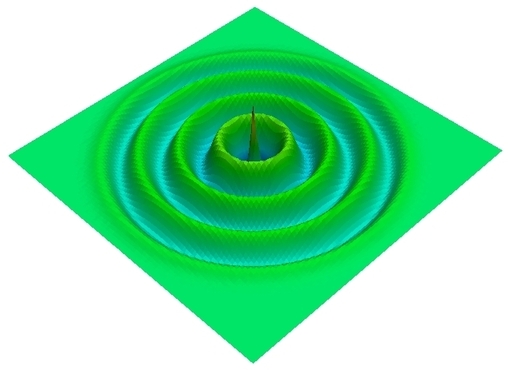}} \\
(b-1) $t=80$\hspace*{7cm} (b-2) $t=80$\\
\caption{The density contours (left figures) and the 3D surfaces
(right figures) by the parameters of group-B in Table
\ref{tab:1chp5}: (a) the BGK-LBM; (B) the original MRT-LBM; (C) the
optimized MRT-LBM.\label{Fig:5-2}}
\end{center}
\end{figure}

\begin{figure}
\begin{center}
\scalebox{0.42}[0.42]{\includegraphics{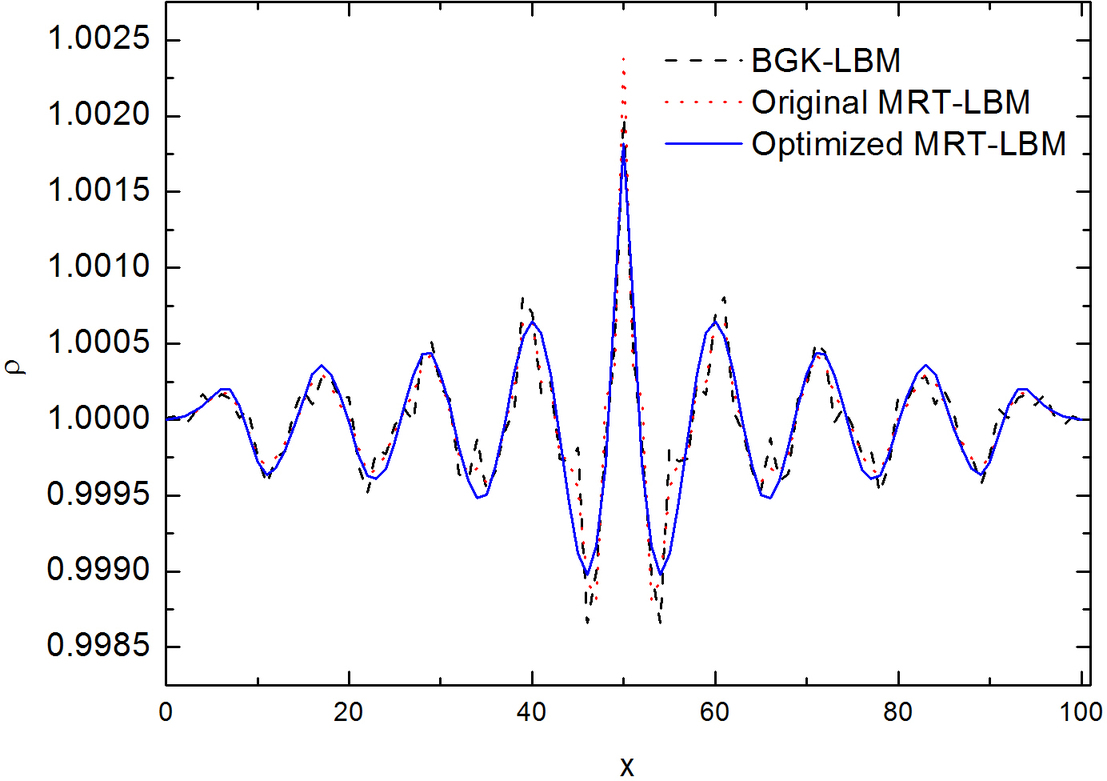}}
\scalebox{0.42}[0.42]{\includegraphics{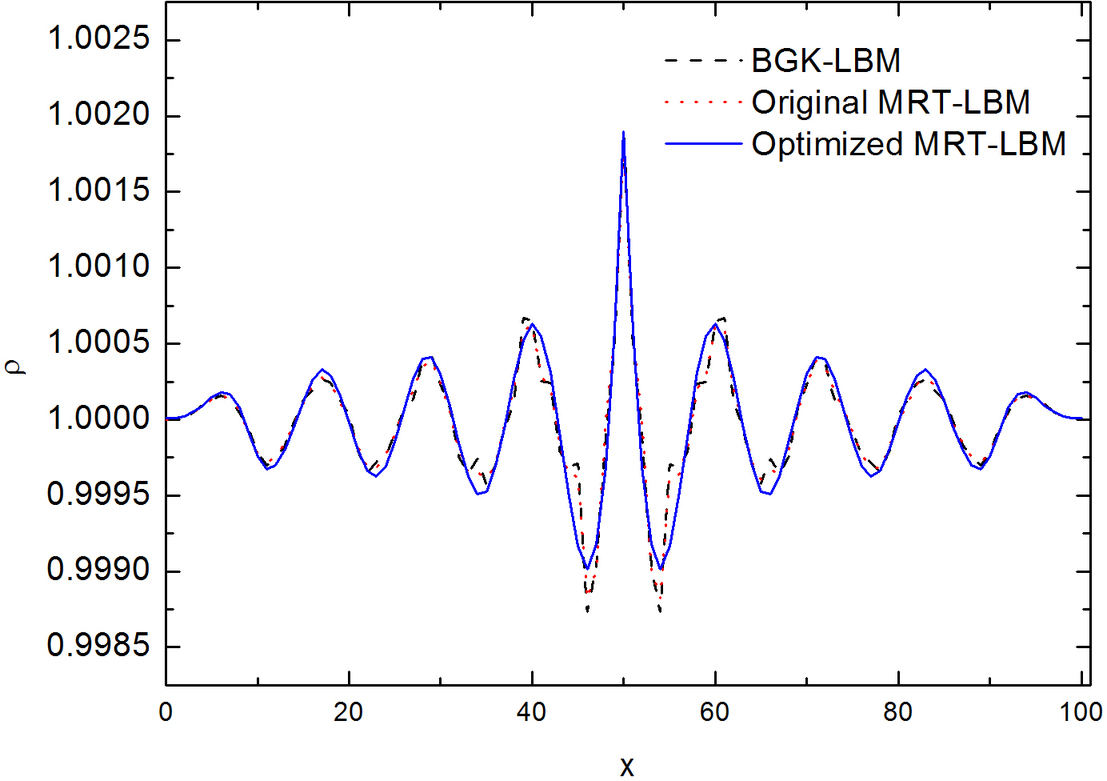}} \\
\center{\hspace{1cm}(a)\hspace{8cm}(b)}
\caption{The comparisons of cross-profiles of the density $\rho$ at
y=51 (the lattice time $t=80$): (a) Results obtained by the parameters of group-A in Table
\ref{tab:1chp5};(b) Results obtained by the parameters of group-B in Table \ref{tab:1chp5}.\label{Fig:5-4}}
\end{center}
\end{figure}

\begin{figure}
\begin{center}
\scalebox{0.42}[0.42]{\includegraphics{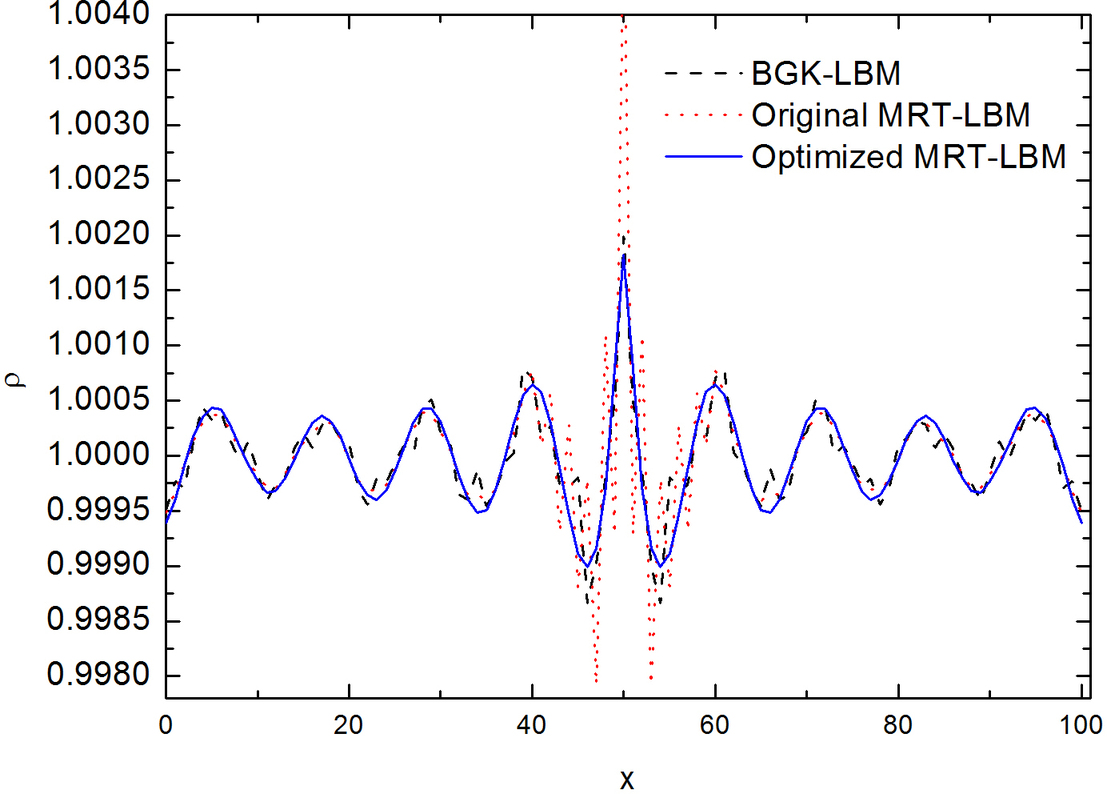}}
\scalebox{0.42}[0.42]{\includegraphics{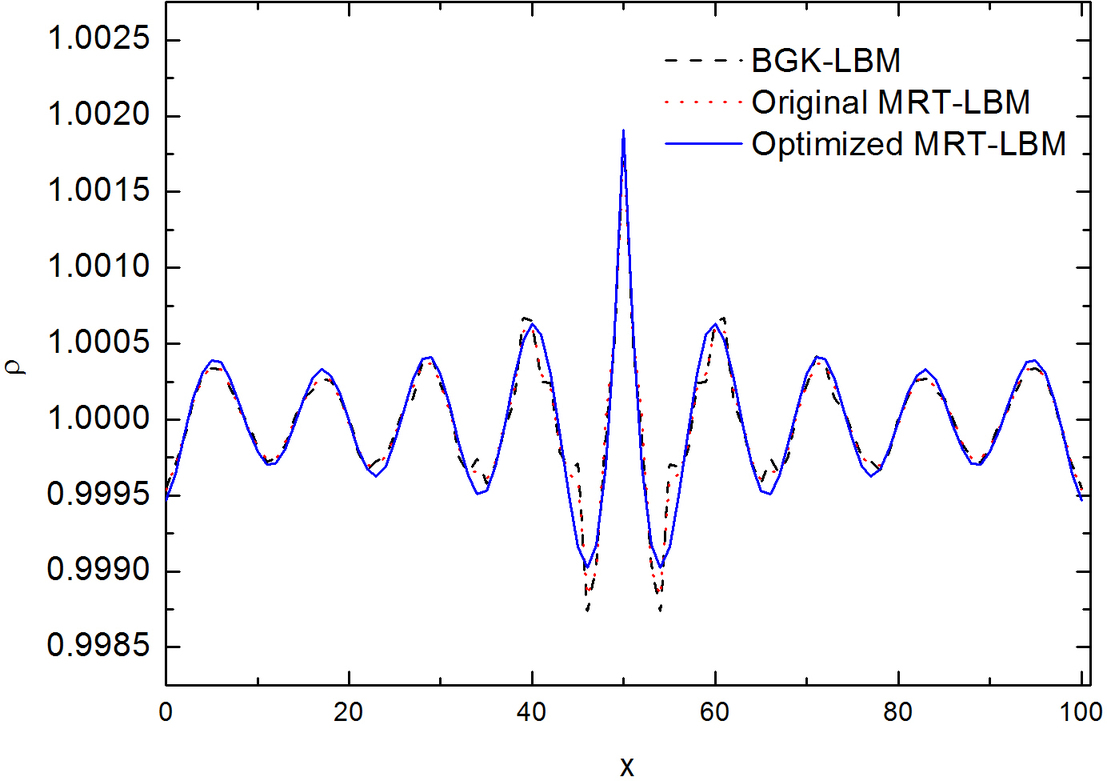}} \\
\center{\hspace{1cm}(a)\hspace{8cm}(b)}
\caption{The comparisons of cross-profiles of the density $\rho$ at
y=51 (the lattice time $t=100$):(a) Results obtained by the parameters of group-A in Table
\ref{tab:1chp5};(b) Results obtained by the parameters of group-B in Table \ref{tab:1chp5}.\label{Fig:5-5}}
\end{center}
\end{figure}

In Figs. \ref{Fig:5-1} and \ref{Fig:5-2}, the density contours and
3D surfaces are shown. Figs. \ref{Fig:5-1}(a) and \ref{Fig:5-2}(a)
show that the optimized MRT-LBM corrects the anisotropy and
annihilate the spurious fluctuations of waves. These results are
better than the BGK-LBM and the original MRT-LBM. In Figs.
\ref{Fig:5-4} and \ref{Fig:5-5}, the waves along the line $y=51$ are
given for three different methods at $t=80$ and $t=100$. It is clear
that the optimized MRT-LBM is more effective than the BGK-LBM and
the original MRT-LBM.

The numerical results demonstrate that by determining the free
parameters, we can reduce the dispersion error and the isotropy
error of the MRT-LBM.

\subsection{Acoustic pressure pulse}

\begin{figure}
\begin{center}
\scalebox{0.4}[0.4]{\includegraphics{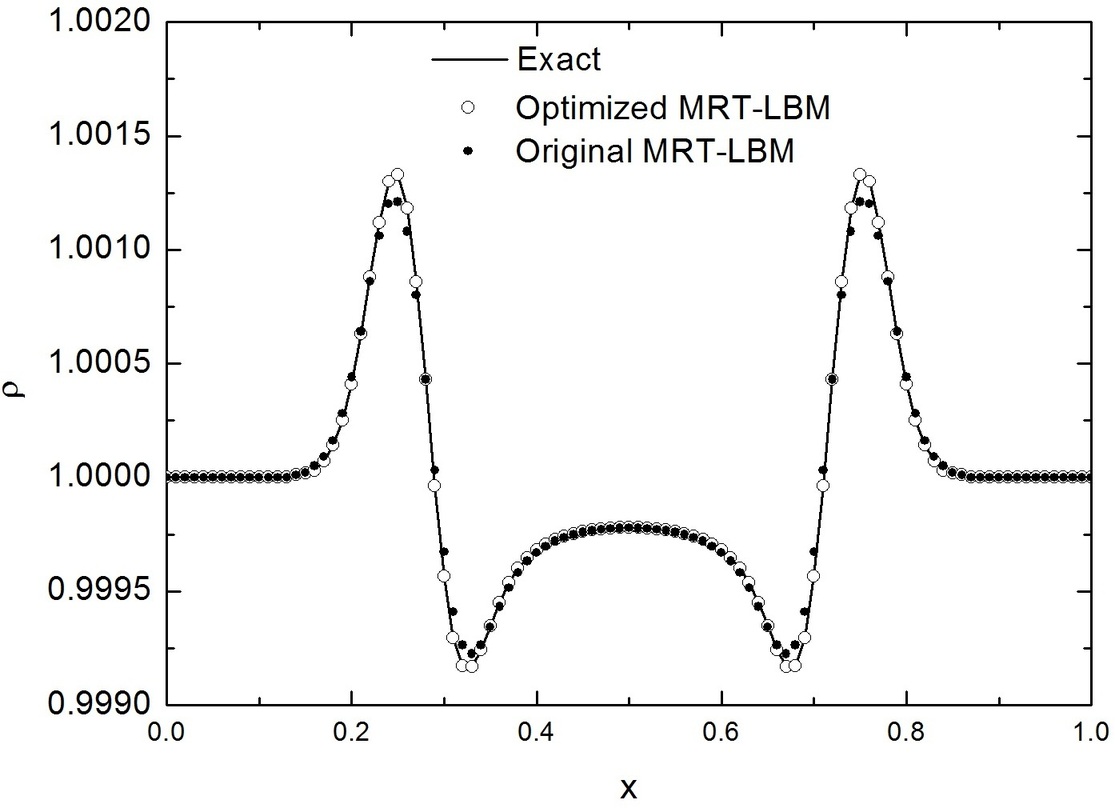}}
\scalebox{0.4}[0.4]{\includegraphics{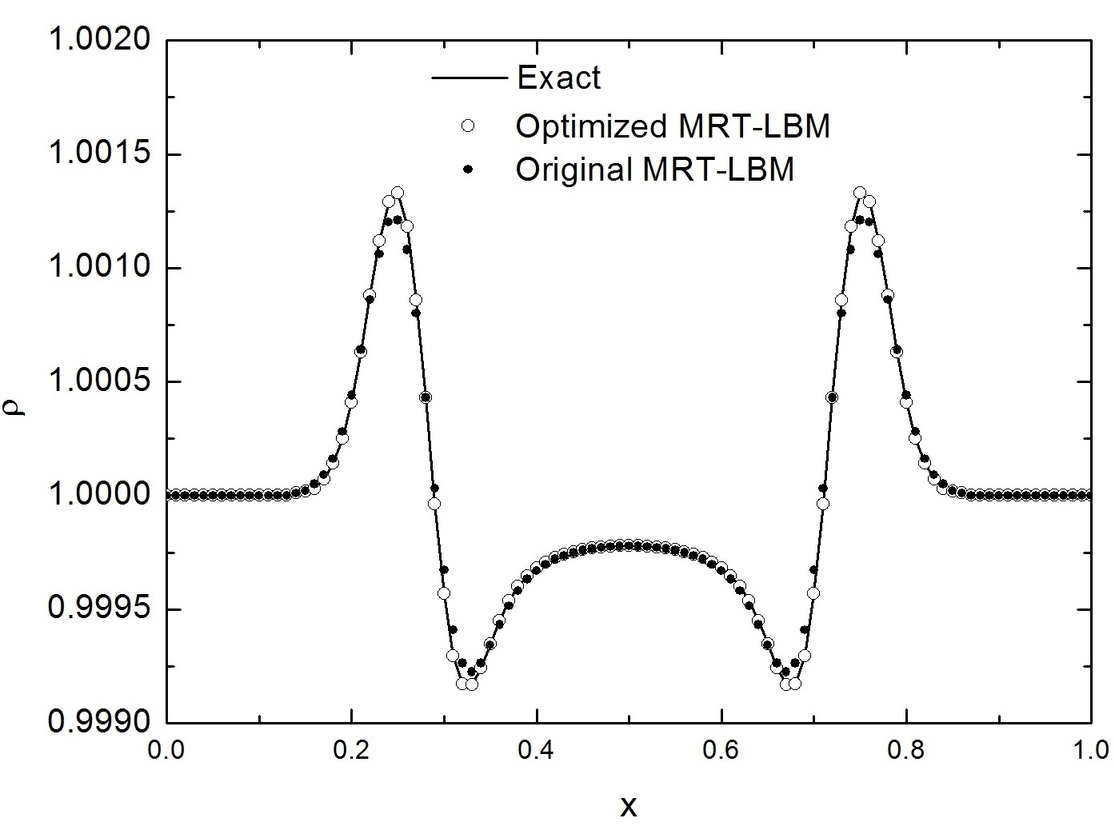}} \\
\centering{\hspace{0.5cm}(a) Horizontal density profiles at y=0.5\hspace{2cm}(b) Horizontal density profiles at y=0.5}
\caption{The comparisons of the cross profiles of the density $\rho$
at y=0.5: The horizontal velocity $U_0=0$. (a) The results of the
optimized MRT-LBM are obtained by the parameters of group-A in Table
\ref{tab:2chp5} (b) The results of the optimized MRT-LBM are
obtained by the parameters of group-B in Table \ref{tab:2chp5}. (The
computational physical time $t=0.4$)\label{Fig:5-6}}
\end{center}
\end{figure}

\begin{figure}
\begin{center}
\scalebox{0.4}[0.4]{\includegraphics{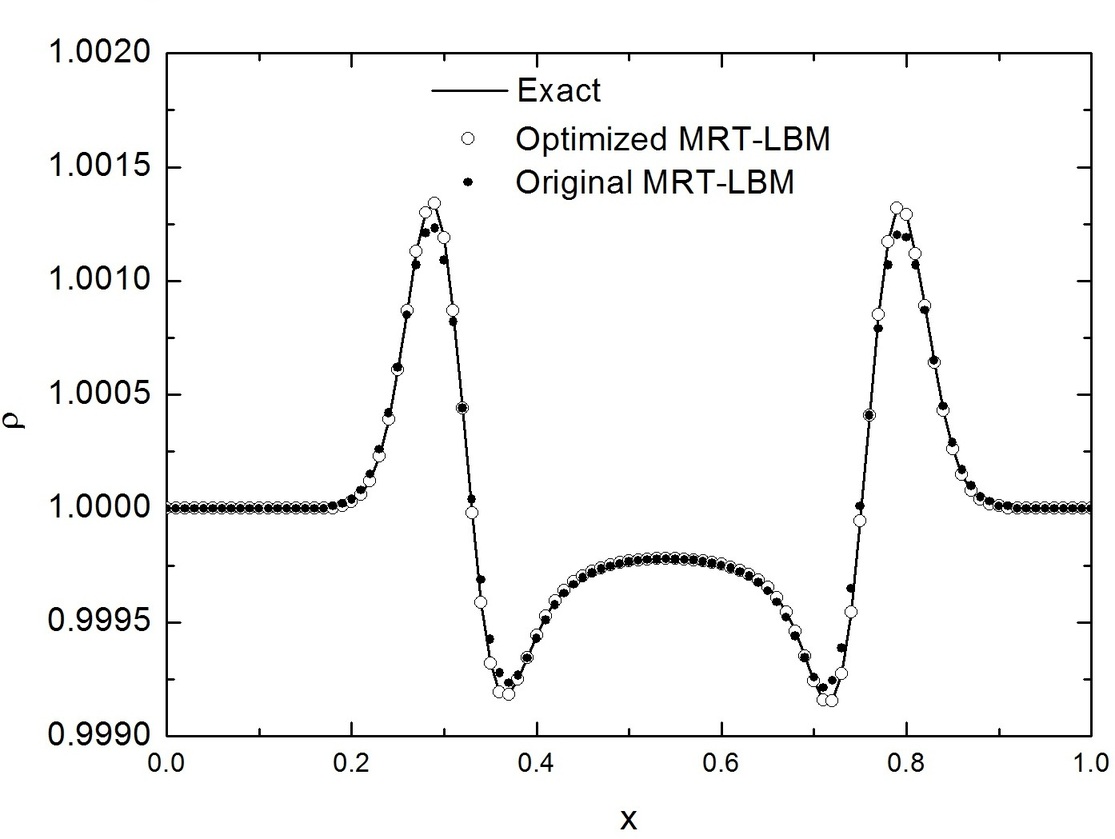}}
\scalebox{0.4}[0.4]{\includegraphics{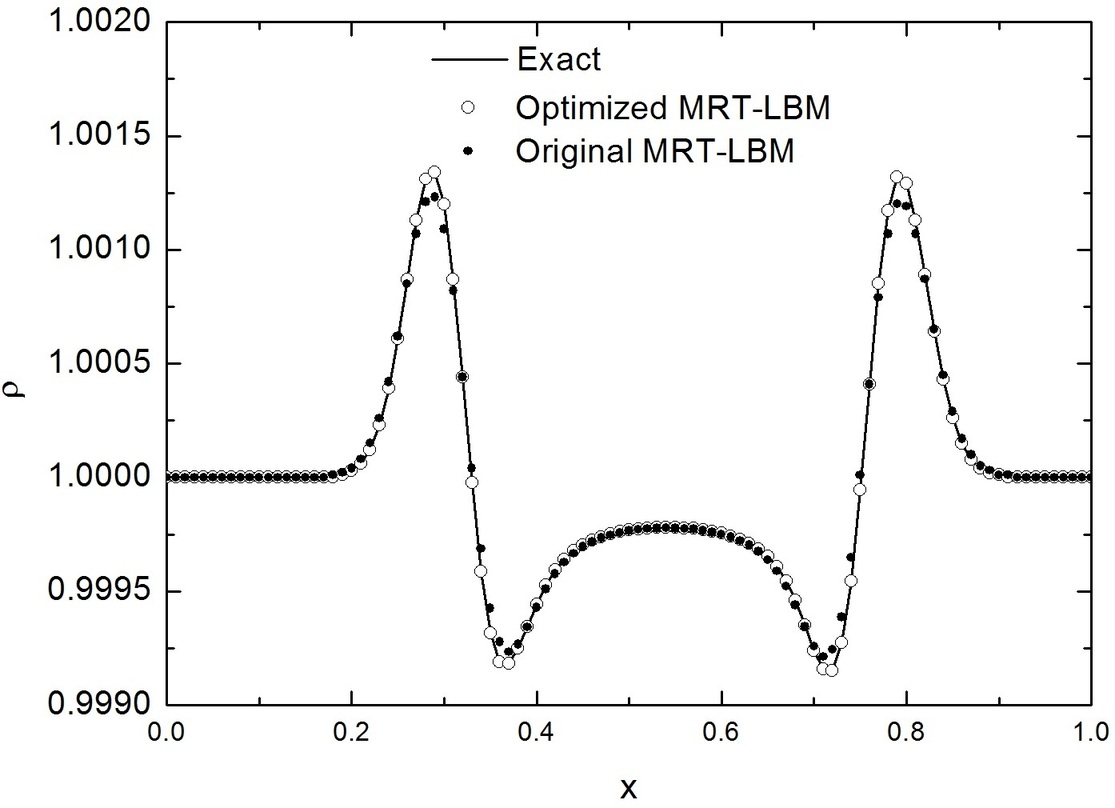}} 
\centering{\hspace{0.5cm}(a) Horizontal density profiles at y=0.5\hspace{2cm}(b) Horizontal density profiles at y=0.5}
\caption{The comparisons of the cross profiles of the density $\rho$
at y=0.5: The horizontal velocity $U_0=0.1$. (a) The results of the
optimized MRT-LBM are obtained by the parameters of group-A in Table
\ref{tab:2chp5} (b) The results of the optimized MRT-LBM are
obtained by the parameters of group-B in Table \ref{tab:2chp5}. (The
computational physical time $t=0.4$)\label{Fig:5-7}}
\end{center}
\end{figure}

\begin{figure}
\begin{center}
\scalebox{0.4}[0.4]{\includegraphics{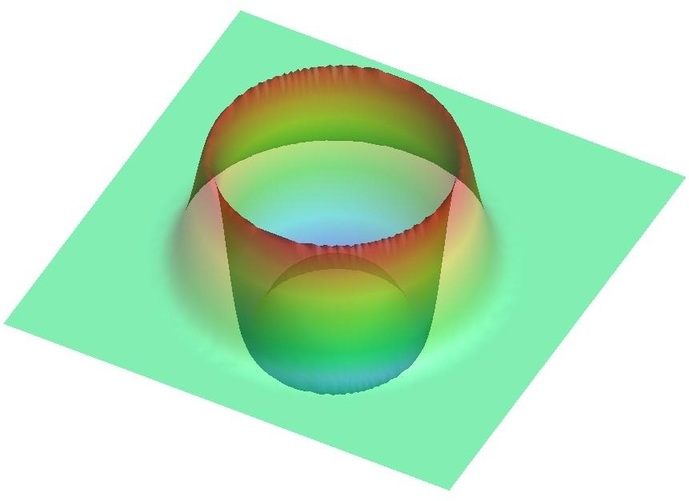}}\scalebox{0.4}[0.4]{\includegraphics{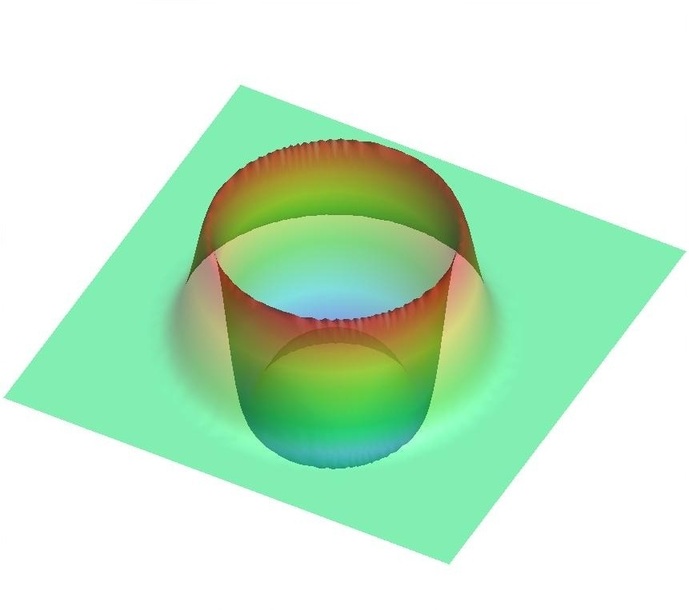}}\\
{\small (a) Density distribution by the optimized MRT-LBM \hspace*{1cm}(b) Density profiles by the original MRT-LBM.}\\
\caption{The comparisons of density distribution with the horizontal
velocity $U_0=0.1$. (a) The results of the optimized MRT-LBM are
obtained by the parameters of group-A in Table \ref{tab:2chp5} (b)
The results of the original MRT-LBM.  (The computational physical
time $t=0.4$)\label{Fig:5-8}}
\end{center}
\end{figure}

\begin{figure}
\begin{center}
\scalebox{0.5}[0.5]{\includegraphics{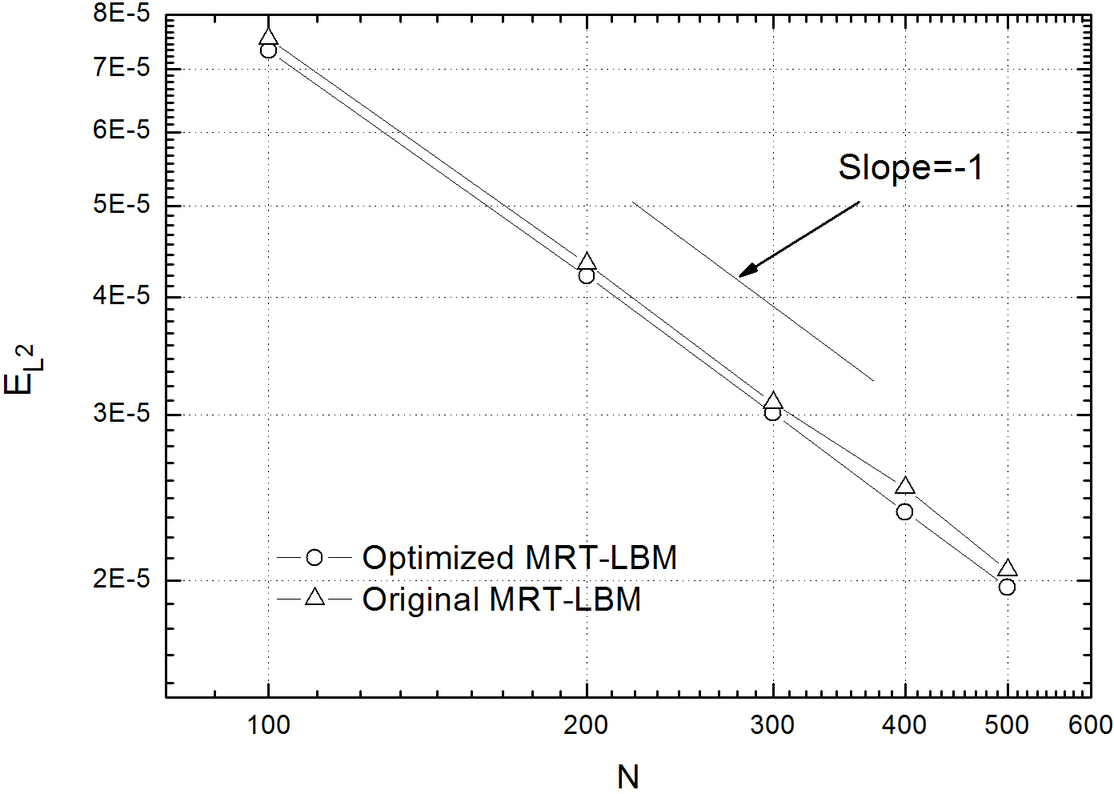}}\\
\caption{The comparisons of relative errors of the density $\rho$ in
the log-log coordinate. The results of the optimized MRT-LBM are
obtained by the parameters of group-A in Table \ref{tab:2chp5}  (The
computational physical time $t=0.4$)\label{Fig:5-9}}
\end{center}
\end{figure}

In this part, the quality of the optimized MRT-LBM is assessed
considering the acoustic pressure pulse problem. These comparisons
provide an evidence for the accuracy of computed solutions. In order
to compare the results with the exact solution and neglect the
effects of the dissipation, the very small values of shear and bulk
viscosity are chosen for a MRT-LBM. The initial perturbation is
given by a Gaussian density distribution at the center of the domain
at $t=0$ \cite{christopherwebb}
\begin{equation}
\left\{\begin{array}{rl} \rho(x,y,0)&=\rho_0+\varepsilon
\mathrm{exp}(-\alpha\eta^2),\\
u(x,y,0)&=U_0,\\
v(x,y,0)&=0,
\end{array}\right.
\end{equation}
where $\alpha$ is related to the half-width Gaussian , $b$, by
$\alpha=\mathrm{ln}2/b^2$. $\eta$ is defined by
$\eta=\left[(x-U_0t)^2+y^2\right]$, which is equal to the radial
coordinate at $t=0$. $\varepsilon$ is the density pulse amplitude.
The analytical solution of the problem for the pressure and density
can be given by a zero-order Bessel function $J_0$
\cite{christopherwebb}
\begin{equation}\label{exactpules}
\rho(x,y,t)=\rho_0+\frac{\varepsilon}{2\alpha}\int_0^{\infty}\mathrm{exp}(-\xi^2/(4\alpha))\mathrm{cos}(c_s\xi
t)J_0(\xi\eta)\xi \mathrm{d}\xi.
\end{equation}
It is noted that Eq. (\ref{exactpules}) does not include the
dissipation introduced by the viscosity of the fluid. So, in order
to implement the computation, the influence of viscosity on pressure
wave must be minimized by the small magnitude of viscosity. The
computational domain is $[0,1]\times[0,1]$. The parameters
$\varepsilon$ and $b$ are equal to 0.01 and 0.04, respectively. In
Figs. \ref{Fig:5-6} and \ref{Fig:5-7}, we show the horizontal
density profiles at y=0.5. The simulation physical time $t=0.4$.
From these figures, it is clear that the results by the optimized
MRT-LBM are superior to the results by the original MRT-LBM.
Obviously, the amplitudes of crests and troughs are damped by the
bulk viscosity in the original MRT-LBM. It is demonstrated that by
the proposed strategy (\ref{Minimization-F}) and the new definition
of $\mathcal {M}_\epsilon$ in Sec. \ref{meanoptbulk}, the
dissipation influence from the bulk viscosity on pressure waves can
be reduced to a negligible level. Meanwhile, the bulk viscosity can
be attenuated. In Fig. \ref{Fig:5-8}, the figures of density
distribution are given. In order to test the accuracy, the following
$L^2$-norm relative error is defined
\begin{equation}
E_{L^2}=\sqrt{\frac{\sum_{i=1}^{N_{\rm nodes}}{(\rho_i^{\rm
th}-\rho_i^{\rm num})^2}}{\sum_{i=1}^{N_{\rm nodes}}(\rho_i^{\rm
th})^2}},
\end{equation}
where $\rho_i^{\rm th}$ and $\rho_i^{\rm num}$ denote the
theoretical and numerical solutions, respectively.
 In Fig. \ref{Fig:5-9}, the $L^2$-norm relative errors of
density are given based on lattice nodes in a log-log coordinate.
From this figure, it is discovered that the convergence orders of
pressure pulse for the optimized MRT-LBM and the original MRT-LBM
are very close to 1. However, the result given by the optimized
MRT-LBM are better than that given by the original MRT-LBM.

\section{Conclusion}
In this paper, we have proposed several numerical strategies to
reduce the dispersion/dissipation errors (regarded the optimized
MRT-LBM as {\color{blue} the} D2PR-LBM). We also gave an easy-to-use
algorithm to derive linearized Navier-Stokes with the high-order
truncation errors starting from the linearized MRT-LBM. Von Neumann
analysis of the isothermal linearized MRT-LBM and the linearized
BGK-LBM has been made to investigate the dispersion and dissipation
relations. The von Neumann stability analysis also shows that by
optimizing free parameters, the acoustic modes are decoupled from
the shear modes and other modes with respect to zero-mean flows when
the truncation error is up to $O(\delta t^5)$. For uniform flows, it
is discovered that when the truncation error is up to $O(\delta
t^4)$, by optimization strategies, the dissipation error can only be
reduced and the influences from mean flows on dissipation relations
are also reduced. The stability of the MRT-LBM is enhanced.
Especially, when the shear viscosity and bulk viscosity are very
small, the optimized dissipation relation is nearly exact. The
optimized MRT-LBM can annihilate the spurious waves and the
isotropic error suffered by the original MRT-LBM and reduce the
over-damping influence of the bulk viscosity on pressure waves.
Numerical simulations of acoustic problems demonstrated that for
acoustic problems, the optimized MRT-LBM is more effective than the
original MRT-LBM.

\section*{Acknowledgement} This work was supported by the FUI project LaBS (Lattice Boltzmann Solver, http://www.labs-project.org).
 Dr. Orestis Malaspinas is warmly acknowledged for useful
 discussions. We appreciate the referee's comments to this manuscript.

\appendix
\section{The Taylor expansion strategy of the L-MRT-LBM in wave-number
spaces}\label{AppendexDerivation}
\noindent In this part, we show the derivation details from the L-MRT-LBM to the L-NSE and the truncation error is up to $O(\delta t^4)$.\\
(1) When $J=1$ in Eq. (\ref{expansion_2}), we have
\begin{equation}\label{J-1A}
W_i=m_i,0\leq i\leq d,
\end{equation}
\begin{equation}\label{J-1B}
m_i=\frac{1}{s_i}\Psi_{ij}W_j+O(\delta t),d<i\leq N.
\end{equation}
We rewrite Eqs. (\ref{J-1A}) and (\ref{J-1B}) as a uniform
expression
\begin{equation}\label{J-1U}
m_i=\Phi_{ij,1}W_j+O(\delta t),0\leq i\leq N,
\end{equation}
where
\begin{equation}
\Phi_{ij,1}=\delta_{ij} (0\leq i\leq
d),\Phi_{ij,1}=\frac{1}{s_i}\Psi_{ij} (d+1\leq i\leq N).
\end{equation}
(2) When $J=2$ in Eq. (\ref{expansion_2}), we have
\begin{equation}\label{J-2}
m_i+\delta t\partial_t m_i=A_{ir,0}m_r+\delta t A_{ir,1}m_r+O(\delta
t ^2).
\end{equation}
When $0\leq i\leq d $, we have
\begin{equation}\label{J-2a}
W_i+\delta t\partial_t W_i=A_{ir,0}m_r+\delta t A_{ir,1}m_r+O(\delta
t ^2).
\end{equation}
Because $A_{ir,0}=\delta_{ir} $, we have
\begin{equation}\label{J-2b}
W_i+\delta t\partial_t W_i=\delta_{ir}m_r+ \delta t
A_{ir,1}m_r+O(\delta t ^2).
\end{equation}
That is,
\begin{equation}\label{J-2c}
\partial_t W_i= A_{ir,1}m_r+O(\delta t).
\end{equation}
According to Eq. (\ref{J-1U}), we have
\begin{equation}\label{J-2d}
\partial_t W_i= A_{ir,1}\Phi_{rj,1}W_j+O(\delta t).
\end{equation}
Let
\begin{equation}
B_{ij,1}=A_{ir,1}\Phi_{rj,1}.
\end{equation}
Then, we obtain
\begin{equation}\label{J-2dd}
\partial_t W_i=B_{ij,1}W_j+O(\delta t).
\end{equation}
When $d+1\leq i\leq N $, we have
\begin{equation}\label{J-2e}
m_i+\delta t\partial_t m_i=A_{ir,0}m_r+\delta t A_{ir,1}m_r+O(\delta
t ^2).
\end{equation}
It is known that when $d+1\leq i\leq N $, $A_{ir,0} $ is defined by
\cite{duboislallemand}
\begin{equation}\label{J-2f}
A_{ir,0}m_r=(\delta_{ir}-S_{ir})m_r-\Psi_{ij}W_j.
\end{equation}
So, we have (the combination of Eqs. (\ref{J-2f}) and (\ref{J-1B}))
\begin{equation}
m_i+\delta t
\Phi_{ij,1}\partial_tW_j=(\delta_{ir}-S_{ir})m_r-\Psi_{ij}W_j+\delta
t A_{ir,1}\Phi_{rj,1}W_j+O(\delta t ^2),
\end{equation}
\begin{equation}
m_i=\frac{1}{s_i}(-\Psi_{ij}-\delta t \Phi_{ik,1}B_{kj,1}+\delta t
A_{ir,1}\Phi_{rj,1})W_j+O(\delta t ^2).
\end{equation}
Now, introducing $\Phi_{ij,2} $ as follows
\begin{equation}
\Phi_{ij,2}=\delta_{ij} (0\leq i\leq d),
\end{equation}
we have
\begin{equation}
\Phi_{ij,2}=\frac{1}{s_i}(-\delta
t\Phi_{ik,1}B_{kj,1}+\Psi_{ij}+\delta t A_{ir,1}\Phi_{rj,1}),
(d+1\leq i\leq N).
\end{equation}
So, we obtain
\begin{equation}\label{J-2g}
m_{i}=\Phi_{ij,2}W_j+O(\delta t^2).
\end{equation}
(3) When $J=3$ in Eq. (\ref{expansion_2}), we have
\begin{equation}\label{J-3a}
m_i+\delta t\partial_t m_i+\frac{\delta t^2}{2!}\partial_t
m_i=A_{ir,0}m_r+\delta t A_{ir,1}m_r+\delta t^2 A_{ir,2}m_r+O(\delta
t ^3).
\end{equation}
When $0\leq i\leq d $, we have
\begin{equation}\label{J-3b}
\partial_t W_i+\frac{\delta t}{2!}\partial_t W_i=A_{ir,1}m_r+\delta t A_{ir,2}m_r+O(\delta t
^2).
\end{equation}
By Eqs. (\ref{J-2dd}), (\ref{J-1U}) and (\ref{J-2g}), we get
\begin{equation}\label{J-3c}
\partial_t W_i=-\frac{\delta t}{2!}B_{ij,1} W_j+A_{ir,1}\Phi_{rj,2}W_j+\delta t A_{ir,2}\Phi_{rj,1}W_j+O(\delta t^2).
\end{equation}
Let
\begin{equation}
B_{ij,2}=-\frac{\delta t}{2!}B_{ij,1}
W_j+A_{ir,1}\Phi_{rj,2}W_j+\delta t A_{ir,2}\Phi_{rj,1},
\end{equation}
we get
\begin{equation}\label{J-3d}
\partial_t W_i=B_{ij,2}W_j+O(\delta t^2)
\end{equation}
When $d+1\leq i\leq N $, by Eqs.(\ref{J-2g}), (\ref{J-1U}) and
(\ref{J-3d}), we have
\begin{equation}\label{J-3e}
m_i=\frac{1}{s_i}(-\delta t\Phi_{ik,2}B_{kj,2}-\frac{\delta
t^2}{2!}\Phi_{ik,1}B_{kj,1}^2+\Psi_{ij}+\delta t
A_{ir,1}\Phi_{rj,2}+\frac{\delta t^2}{2!}
A_{ir,2}\Phi_{rj,1})W_j+O(\delta t ^3).
\end{equation}
Now, introducing $\Phi_{ij,3} $, we get
$$
\Phi_{ij,3}=\delta_{ij} (0\leq i\leq d),
$$
and for $ (d+1\leq i\leq N) $
\begin{equation}\label{J-3g}
\Phi_{ij,3}=\frac{1}{s_i}(-\delta t\Phi_{ik,2}B_{kj,2}-\frac{\delta
t^2}{2!}\Phi_{ik,1}B_{kj,1}^2+\Psi_{ij}+\delta t
A_{ir,1}\Phi_{rj,2}+\delta t^2 A_{ir,2}\Phi_{rj,1}).
\end{equation}
In order to restrict the truncated error of Eq. (\ref{J-3g}) equal
to $O(\delta t^3)$, we rewrite Eq. (\ref{J-3g}) as follows
$$
\Phi_{ij,3}=\sum_{l=0}^{2}\delta t^l{\rm Coeff}(\Phi_{ij,3},\delta
t,l).
$$
So, we have
\begin{equation}\label{J-3U}
m_i=\Phi_{ij,3}W_j+O(\delta t^3).
\end{equation}
(4) When $J=4$ in Eq. (\ref{expansion_2}), we have
\begin{equation}\label{J-4a}
m_i+\delta t\partial_t m_i+\frac{\delta t^2}{2!}\partial_t^2
m_i+\frac{\delta t^3}{3!}\partial_t^3 m_i= A_{ir,0}m_r+\delta t
A_{ir,1}m_r+\delta t^2A_{ir,2}m_r+\delta t^3 A_{ir,3}m_r+O(\delta t
^4).
\end{equation}
When $0\leq i\leq d $, we have
\begin{equation}\label{J-4b}
\partial_t W_i+\frac{\delta t}{2!}\partial_t^2 W_i+\frac{\delta t^2}{3!}\partial_t^3 W_i=
 A_{ir,1}m_r+\delta t A_{ir,2}m_r+\delta t^2 A_{ir,3}m_r+O(\delta t
 ^3).
\end{equation}
By Eqs. (\ref{J-3U}), (\ref{J-3d}) and (\ref{J-2dd}), we get the
R.H.S of Eq. (\ref{J-4b}),
\begin{equation}\label{J-4c}
{\rm R.H.S}=A_{ir,1}\Phi_{rj,3}W_j+\delta t
A_{ir,2}\Phi_{rj,2}W_j+\delta t^2 A_{ir,3}B_{rj,1}W_j+O(\delta t^3).
\end{equation}
By Eqs. (\ref{J-3d}) and (\ref{J-2dd}),  we get the L.H.S of Eq.
(\ref{J-4b}),
\begin{equation}
{\rm L.H.S}=\partial_t W_i+\frac{\delta
t}{2!}B_{ij,2}^2W_j+\frac{\delta t^2}{3!}B_{ij,1}^3W_j.
\end{equation}
So, we have
\begin{equation}
\partial_t W_i=(-\frac{\delta t}{2!}B_{ij,2}^2-\frac{\delta t^2}{3!}B_{ij,1}^3+A_{ir,1}\Phi_{rj,3}+\delta t A_{ir,2}\Phi_{rj,2}+\delta t^2 A_{ir,3}\Phi_{rj,1})W_j+O(\delta
t^3).
\end{equation}
Let
\begin{equation}\label{J-4d} B_{ij,3}=-\frac{\delta
t}{2!}B_{ij,2}^2-\frac{\delta
t^2}{3!}B_{ij,1}^3+A_{ir,1}\Phi_{rj,3}+\delta t
A_{ir,2}\Phi_{rj,2}+\delta t^2 A_{ir,3}\Phi_{rj,1},
\end{equation}
we restrict the truncated error of Eq. (\ref{J-4d}) equal to
$O(\delta t^3)$ and get
\begin{equation}
B_{ij,3}=\sum_{l=0}^{2}\delta t^l{\rm Coeff}(B_{ij,3},\delta t,l).
\end{equation}
Then, we have
\begin{equation}\label{J-4dd}
\partial_t W_i=B_{ij,3}W_j+O(\delta t^3).
\end{equation}
When $d+1\leq i\leq N $, we have
\begin{equation}\label{J-4e}
m_i+\delta t\partial_t m_i+\frac{\delta t^2}{2!}\partial_t^2
m_i+\frac{\delta t^3}{3!}\partial_t^3 m_i= A_{ir,0}m_r+\delta t
A_{ir,1}m_r+\delta t^2A_{ir,2}m_r+\delta t^3 A_{ir,3}m_r+O(\delta t
^4).
\end{equation}
By Eqs. (\ref{J-1U}), (\ref{J-2dd}), (\ref{J-2g}), (\ref{J-3d}),
(\ref{J-3U}) and (\ref{J-4dd}), we get the L.H.S of Eq. (\ref{J-4a})
\begin{equation}
{\rm L.H.S}=m_i+\delta t\Phi_{ir,3}B_{rj,3} W_j+\frac{\delta
t^2}{2!}\Phi_{ir,2}B_{rj,2}^2 W_j+\frac{\delta
t^3}{3!}\Phi_{ir,1}B_{rj,1}^3 W_j.
\end{equation}
By Eqs. (\ref{J-4dd}), (\ref{J-3d}) and (\ref{J-1U}), we gain the
R.H.S of Eq. (\ref{J-4a})
\begin{equation}
{\rm R.H.S}=(1-s_i)m_i+\Psi_{ij}W_j+\delta t
A_{ir,1}\Phi_{rj,3}W_j+\delta t^2A_{ir,2}\Phi_{rj,2}W_j+\delta t^3
A_{ir,3}\Phi_{rj,1}W_j+O(\delta t ^4).
\end{equation}
So, we have
\begin{equation}
m_i=\frac{1}{s_i}(-\delta t\Phi_{ir,3}B_{rj,3}-\frac{\delta
t^2}{2!}\Phi_{ir,2}B_{rj,2}^2-\frac{\delta
t^3}{3!}\Phi_{ir,1}B_{rj,1}^3+\Psi_{ij}+\delta t
A_{ir,1}\Phi_{rj,3}+\delta t^2A_{ir,2}\Phi_{rj,2}+\delta t^3
A_{ir,3}\Phi_{rj,1})W_j+O(\delta t ^4).
\end{equation}
Let
$$
\Phi_{ij,4}=\delta_{ij} (0\leq i\leq d),
$$
and for $d+1\leq i\leq N$
\begin{equation}\label{J-4f}
\Phi_{ij,4}=\frac{1}{s_i}(-\delta t\Phi_{ir,3}B_{rj,3}-\frac{\delta
t^2}{2!}\Phi_{ir,2}B_{rj,2}^2-\frac{\delta
t^3}{3!}\Phi_{ir,1}B_{rj,1}^3 +\Psi_{ij}+\delta t
A_{ir,1}\Phi_{rj,3}+\delta t^2A_{ir,2}\Phi_{rj,2}+\delta t^3
A_{ir,3}\Phi_{rj,1}),
\end{equation}
and we restrict the truncated error of Eq. (\ref{J-4f}) equal to
$O(\delta t^4)$ and get
$$
\Phi_{ij,4}=\sum_{l=0}^{3}\delta t^l{\rm Coeff}(\Phi_{ij,4},\delta
t,l).
$$
So, we have
\begin{equation}\label{J-4U}
m_j=\Phi_{ij,4}W_j+O(\delta t^4).
\end{equation}
\section{The coefficient matrices of the higher-order L-NSE with the zero-mean flow}\label{coef-matrices}
\noindent (A) The coefficients of $\delta t^2$  are given by the
following matrix
\begin{equation}
 \mathrm{i}\cdot\left[ \begin {array}{ccc} 0&-\frac{1}{18}\,{ k_x}\, \left( {{ k_x}}^{2}+
{{ k_y}}^{2} \right) &-\frac{1}{18}\,{ k_y}\, \left( {{ k_x}}^{2}+{{
k_y}}^{2} \right) \\\noalign{\medskip}-\frac{1}{27}\,{ k_x}\, \left(
-{{ k_x}}^{2}-{{ k_y}}^{2}+3\,{{ k_y}}^{2}{\sigma_e}^{2}+3\,{{
k_x}}^ {2}{\sigma_e}^{2}+3\,{{ k_y}}^{2}{\sigma_\nu}^{2}+3\,{{
k_x}}^{2}{ \sigma_\nu}^{2} \right)
&0&0\\\noalign{\medskip}-\frac{1}{27}\,{ k_y}\,
 \left( -{{ k_x}}^{2}-{{ k_y}}^{2}+3\,{{ k_y}}^{2}{\sigma_e}^{2}
+3\,{{ k_x}}^{2}{\sigma_e}^{2}+3\,{{ k_y}}^{2}{\sigma_\nu}^{2}+3\,{{
 k_x}}^{2}{\sigma_\nu}^{2} \right) &0&0\end {array} \right]
\end{equation}
(B) The coefficients of $\delta t^3$  are given by the following
matrix
\begin{equation}{\rm Coeff}(B_{,5}[1,1],\delta
t^3)={\frac{1}{108}}\,{{ k_y}}^{4}\sigma_e+{\frac{1}{54}}\,{{
k_x}}^{ 2}{{ k_y}}^{2}\sigma_e+{\frac{1}{108}}\,{{
k_x}}^{4}\sigma_e+{ \frac{1}{108}}\,{{
k_y}}^{4}\sigma_\nu+{\frac{1}{108}}\,{{ k_x}}^{
4}\sigma_\nu+{\frac{1}{54}}\,{{ k_x}}^{2}{{ k_y}}^{2}\sigma_\nu
\end{equation}
\begin{equation}
{\rm Coeff}(B_{,5}[1,2],\delta t^3)=0,{\rm Coeff}(B_{,5}[1,3],\delta
t^3)=0,{\rm Coeff}(B_{,5}[2,1],\delta t^3)=0
\end{equation}
\begin{equation}
\begin{array}{r}
{\rm Coeff}(B_{,5}[2,2],\delta t^3)=-\frac{5}{9}\,{{ k_x}}^{2}{{
k_y}}^{2}\sigma_\nu\,\sigma _\epsilon \,\sigma_q +\frac{1}{9}\,{{
k_x}}^{4}\sigma _\epsilon \,\sigma_q\,\sigma_\nu+{\frac{5}{18 }}\,{{
k_x}}^{2}{{ k_y}}^{2}{\sigma_\nu}^{2}\sigma_q-1/9\,{{ k_x}} ^{2}{{
k_y}}^{2}{\sigma_\nu}^{2}\sigma_e-{\frac{5}{54}}\,{{ k_x}}^{ 2}{{
k_y}}^{2}\sigma_\nu-\\[2mm]
{\frac{13}{108}}\,{{ k_x}}^{2}{{ k_y}}^{ 2}\sigma_e-\frac{1}{9}\,{{
k_x}}^{4}\sigma_\nu\,{\sigma_e}^{2}+\frac{1}{6}\,{{ k_x}}
^{4}{\sigma_e}^{2}\sigma_q-{\frac{1}{108}}\,{{ k_x}}^{4}\sigma_\nu-{
\frac{1}{108}}\,{{ k_x}}^{4}\sigma_e-\frac{1}{9}\,{{
k_x}}^{4}\sigma_e\, \sigma _\epsilon \,\sigma_q-\frac{1}{9}\,{{
k_x}}^{2}{{ k_y}}^{2}\sigma_e\,
\sigma _\epsilon \,\sigma_q-\\[2mm]
\frac{1}{36}\,{{ k_y}}^{4}\sigma_\nu-\frac{1}{9}\,{{ k_y}
}^{4}{\sigma_\nu}^{3}+{\frac{10}{9}}\,{{ k_x}}^{2}{{ k_y}}^{2}
\sigma_\nu\,\sigma_e\,\sigma_q- \frac{2}{9}\,{{
k_x}}^{4}\sigma_\nu\,\sigma_e\, \sigma_q+\frac{1}{6}\,{{ k_x}}^{2}{{
k_y}}^{2}{\sigma_e}^{2}\sigma_q-\frac{1}{9}\, {{ k_x}}^{2}{{
k_y}}^{2}{\sigma_e}^{2}\sigma_\nu+\frac{1}{18}\,{{ k_x}}^{4
}{\sigma_\nu}^{2}\sigma_q-\\[2mm]
\frac{1}{9}\,{{ k_x}}^{4}{\sigma_\nu}^{2}\sigma_e-\frac{1}{9} \,{{
k_x}}^{2}{{ k_y}}^{2}{\sigma_\nu}^{3}+\frac{2}{9}\,{{ k_y}}^{4}{
\sigma_\nu}^{2}\sigma_q+\frac{1}{18}\,{{ k_x}}^{2}{{ k_y}}^{2}\sigma
_\epsilon
\end{array}
\end{equation}
\begin{equation}
\begin{array}{r}
{\rm Coeff}(B_{,5}[2,3],\delta t^3)=\frac{1}{6}\,{ k_x}\,{{
k_y}}^{3}{\sigma_\nu}^{2}\sigma_q-\frac{5}{9}\,{ k_x}\,{{
 k_y}}^{3}\sigma_\nu\,\sigma _\epsilon \,\sigma_q-\frac{1}{9}\,{ k_x}\,{{
k_y}}^{3}\sigma_e\,\sigma _\epsilon \,\sigma_q+\frac{1}{6}\,{{
k_x}}^{3}{ k_y}\,{\sigma_\nu}^{2}\sigma_q+\frac{1}{9}\,{{ k_x}}^{3}{
k_y}\,\sigma _\epsilon \,\sigma_q\,\sigma_\nu-\\[2mm]
\frac{1}{9}\,{{ k_x}}^{3}{ k_y}\,\sigma_e\, \sigma _\epsilon
\,\sigma_q+\frac{1}{6}\,{ k_x}\,{{ k_y}}^{3}{\sigma_e}^{2}
\sigma_q+{\frac {7}{9}}\,{ k_x}\,{{ k_y}}^{3}\sigma_\nu\,\sigma_e\,
\sigma_q+\frac{1}{6}\,{{ k_x}}^{3}{ k_y}\,{\sigma_e}^{2}\sigma_q+
\frac{1}{9}\,{{
 k_x}}^{3}{ k_y}\,\sigma_e\,\sigma_\nu\,\sigma_q-{\frac {5}{54}}\,{
 k_x}\,{{ k_y}}^{3}\sigma_e-\\[2mm]
 {\frac {7}{108}}\,{ k_x}\,{{ k_y}}
^{3}\sigma_\nu-{\frac {7}{108}}\,{{ k_x}}^{3}{
k_y}\,\sigma_\nu-\frac{1}{27}\, {{ k_x}}^{3}{
k_y}\,\sigma_e-\frac{1}{9}\,{ k_x}\,{{ k_y}}^{3}{\sigma
e}^{2}\sigma_\nu+\frac{1}{9}\,{{ k_x}}^{3}{ k_y}\,{\sigma_\nu}^{3}+
\frac{1}{9}\,{ k_x}\,{{ k_y}}^{3}{\sigma_\nu}^{3}-\frac{1}{9}\,{{
k_x}}^{3}{ k_y}\,{ \sigma_\nu}^{2}\sigma_e+\\[2mm]
\frac{1}{18}\,{ k_x}\,{{ k_y}}^{3}\sigma _\epsilon - \frac{1}{9}\,{{
k_x}}^{3}{ k_y}\,\sigma_\nu\,{\sigma_e}^{2}-\frac{1}{9}\,{ k_x}\,{{
 k_y}}^{3}{\sigma_\nu}^{2}\sigma_e
 \end{array}
\end{equation}
\begin{equation}
{\rm Coeff}(B_{,5}[3,1],\delta t^3)=0
\end{equation}
\begin{equation}
\begin{array}{r}
{\rm Coeff}(B_{,5}[3,2],\delta t^3)=\frac{1}{18}\,{{ k_x}}^{3}{
k_y}\,\sigma_\epsilon -\frac{5}{9}\,{{ k_x}}^{3}{
 k_y}\,\sigma_\epsilon \,\sigma_q\,\sigma_\nu+\frac{1}{9}\,{ k_x}\,{{ k_y}
}^{3}\sigma_\nu\,\sigma_\epsilon \,\sigma_q-\frac{1}{9}\,{ k_x}\,{{
k_y}}^{3 }\sigma_e\,\sigma_\epsilon \,\sigma_q-\frac{1}{9}\,{{
k_x}}^{3}{ k_y}\, \sigma_e\,\sigma_\epsilon
\,\sigma_q+\\[2mm]
\frac{1}{6}\,{{ k_x}}^{3}{ k_y}\,{ \sigma_\nu}^{2}\sigma_q-
\frac{1}{9}\,{{ k_x}}^{3}{ k_y}\,\sigma_\nu\,{\sigma_e
}^{2}-\frac{1}{27}\,{ k_x}\,{{ k_y}}^{3}\sigma_e-{\frac {7}{108}}\,{
k_x }\,{{ k_y}}^{3}\sigma_\nu-{\frac {7}{108}}\,{{ k_x}}^{3}{ k_y}\,
\sigma_\nu-{\frac {5}{54}}\,{{ k_x}}^{3}{
k_y}\,\sigma_e+\frac{1}{9}\,{ k_x}\,{{
k_y}}^{3}{\sigma_\nu}^{3}+\\[2mm]{\frac {7}{9}}\,{{ k_x}}^{3}{
k_y}\,\sigma_e\,\sigma_\nu\,\sigma_q+ \frac{1}{9}\,{ k_x}\,{{
k_y}}^{3}\sigma v\,\sigma_e\,\sigma_q-\frac{1}{9}\,{{ k_x}}^{3}{
k_y}\,{\sigma_\nu}^{2} \sigma_e+\frac{1}{6}\,{ k_x}\,{{
k_y}}^{3}{\sigma_\nu}^{2}\sigma_q-\frac{1}{9}\,{
 k_x}\,{{ k_y}}^{3}{\sigma_e}^{2}\sigma_\nu+\frac{1}{6}\,{{ k_x}}^{3}{
k_y}\,{\sigma_e}^{2}\sigma_q+\\[2mm]
\frac{1}{9}\,{{ k_x}}^{3}{ k_y}\,{\sigma_\nu}^{3 }-\frac{1}{9}\,{
k_x}\,{{ k_y}}^{3}{\sigma_\nu}^{2}\sigma_e+\frac{1}{6}\,{ k_x}\,{ {
k_y}}^{3}{\sigma_e}^{2}\sigma_q
\end{array}
\end{equation}
\begin{equation}
\begin{array}{r}
{\rm Coeff}(B_{,5}[3,3],\delta t^3)=-\frac{5}{9}\,{{ k_x}}^{2}{{
k_y}}^{2}\sigma_\nu\,\sigma_\epsilon \,\sigma_q +{\frac {5}{18}}\,{{
k_x}}^{2}{{ k_y}}^{2}{\sigma_\nu}^{2}\sigma_q-\frac{1}{9}\,{{
k_x}}^{2}{{ k_y}}^{2}{\sigma_\nu}^{2}\sigma_e-{\frac {5}{54}} \,{{
k_x}}^{2}{{ k_y}}^{2}\sigma_\nu-{\frac {13}{108}}\,{{ k_x}}^{ 2}{{
k_y}}^{2}\sigma_e-\\[2mm]
\frac{1}{36}\,{{ k_x}}^{4}\sigma_\nu-1/9\,{{ k_x}}^ {2}{{
k_y}}^{2}\sigma_e\,\sigma_\epsilon \,\sigma_q-{\frac {1}{108}} \,{{
k_y}}^{4}\sigma_\nu+\frac{1}{9}\,{{
k_y}}^{4}\sigma_\nu\,\sigma_\epsilon \,\sigma_q+{\frac {10}{9}}\,{{
k_x}}^{2}{{ k_y}}^{2}\sigma_\nu\, \sigma_e\,\sigma_q+\frac{1}{6}\,{{
k_y}}^{4}{\sigma_e}^{2}\sigma_q-\\[2mm]
\frac{1}{9}\,{{
 k_y}}^{4}{\sigma_e}^{2}\sigma_\nu-\frac{1}{9}\,{{ k_y}}^{4}{\sigma_\nu}^{2}
\sigma_e-\frac{1}{9}\,{{ k_x}}^{4}{\sigma_\nu}^{3}-{\frac
{1}{108}}\,{{ k_y} }^{4}\sigma_e+\frac{1}{6}\,{{ k_x}}^{2}{{
k_y}}^{2}{\sigma_e}^{2}\sigma_q- \frac{1}{9}\,{{ k_x}}^{2}{{
k_y}}^{2}{\sigma_e}^{2}\sigma_\nu+\frac{2}{9}\,{{ k_x}
}^{4}{\sigma_\nu}^{2}\sigma_q-\\[2mm]
\frac{1}{9}\,{{ k_x}}^{2}{{ k_y}}^{2}{\sigma_\nu
}^{3}+\frac{1}{18}\,{{
k_y}}^{4}{\sigma_\nu}^{2}\sigma_q-\frac{2}{9}\,{{ k_y}}^{4}
\sigma_\nu\,\sigma_e\,\sigma_q-\frac{1}{9}\,{{
k_y}}^{4}\sigma_e\,\sigma \epsilon \,\sigma_q+\frac{1}{18}\,{{
k_x}}^{2}{{ k_y}}^{2}\sigma_\epsilon
\end{array}
\end{equation}
(C) The coefficient matrix of $\delta t^4$  are given by the
following matrix
\begin{equation}
{\rm Coeff}(B_{,5}[1,1],\delta t^4)=0
\end{equation}
\begin{equation}
\begin{array}{r}
{\rm Coeff}(B_{,5}[1,2],\delta
t^4)=-{\frac{1}{540}}\,\mathrm{i}\cdot{ k_x}\, \left( 20\,{{
k_x}}^{2}{{ k_y}}^{2} \sigma_\nu\,\sigma_e+10\,{{
k_x}}^{4}\sigma_\nu\,\sigma_e-90\,{{ k_y}}
^{2}\sigma_e\,\sigma_q\,{{ k_x}}^{2}+10\,{{ k_y}}^{4}\sigma_\nu\,
\sigma_e+4\,{{ k_x}}^{4}+\right.\\[2mm]
\left.7\,{{ k_y}}^{4}-30\,{{ k_y}}^{4}\sigma_\nu\,\sigma_q+60\,{{
k_x}}^{2}\sigma_\epsilon \,\sigma_q\,{{ k_y}}^{ 2}+14\,{{
k_x}}^{2}{{ k_y}}^{2}-30\,{{ k_x}}^{2}{{ k_y}}^{2}
\sigma_\nu\,\sigma_q \right)
\end{array}
\end{equation}
\begin{equation}
\begin{array}{r}
{\rm Coeff}(B_{,5}[1,3],\delta
t^4)=-{\frac{1}{540}}\,\mathrm{i}\cdot{ k_y}\, \left( 60\,{{
k_x}}^{2}\sigma_\epsilon \,\sigma_q\,{{ k_y}}^{2}+14\,{{ k_x}}^{2}{{
k_y}}^{2}- 90\,{{ k_y}}^{2}\sigma_e\,\sigma_q\,{{ k_x}}^{2}+20\,{{
k_x}}^{2 }{{ k_y}}^{2}\sigma_\nu\,\sigma_e+ 7\,{{
k_x}}^{4}-\right.\\[2mm]
\left.30\,{{ k_x}}^{2}{{ k_y}}^{2} \sigma_\nu\,\sigma_q-30\,{{
k_x}}^{4}\sigma_\nu\, \sigma_q+10\,{{
k_x}}^{4}\sigma_\nu\,\sigma_e+10\,{{ k_y}}^{4}
\sigma_\nu\,\sigma_e+4\,{{ k_y}}^{4} \right)
\end{array}
\end{equation}
\begin{equation}
{\rm Coeff}(B_{,5}[2,1],\delta t^4)
=\mathrm{i}\cdot(\zeta_{4,1}k_xk_y^4+\zeta_{2,3}k_x^3k_y^2+\zeta_{0,5}k_y^5)
\end{equation}
\begin{equation}
{\rm Coeff}(B_{,5}[2,2],\delta t^4)=0,{\rm Coeff}(B_{,5}[2,3],\delta
t^4)=0
\end{equation}
\begin{equation}
{\rm Coeff}(B_{,5}[3,1],\delta t^4)
=\mathrm{i}\cdot(\zeta_{4,1}k_x^4k_y+\zeta_{2,3}k_x^2k_y^3+\zeta_{0,5}k_x^5)
\end{equation}
\begin{equation}
\begin{array}{r}
\zeta_{4,1}={\frac{1}{54}}{\sigma_\epsilon }^{2}-{\frac{19}{648}}{
\sigma_e}^{2}-\frac{1}{27}{\sigma_\nu}^{4}-{\frac{5}{27}}\sigma_\epsilon
{\sigma_q}^{2}\sigma_\nu+{\frac{7}{27}}{\sigma_\nu}^{3}\sigma_q+\frac{1}{9}{
\sigma_e}^{3}\sigma_q+\frac{1}{18}{\sigma_e}^{2}{\sigma_q}^{2}-{\frac{2}{27
}}{\sigma_e}^{2}{\sigma_\nu}^{2}-\frac{1}{27}\sigma_\nu{\sigma_e}^{3}-
\frac{1}{12}\sigma_e\sigma_q+\\[2mm]
{\frac{7}{54}}{\sigma_\nu}^{2}{\sigma_q}^{2}+{
\frac{7}{27}}\sigma_e\sigma_q{\sigma_\nu}^{2}+{\frac{7}{27}}
\sigma_e{\sigma_q}^{2}\sigma_\nu+{\frac{5}{108}}\sigma_\epsilon
\sigma_\nu+{\frac{1}{108}}\sigma_e\sigma_\epsilon -{\frac{17}{324}
}\sigma_\nu\sigma_e-{\frac{7}{108}}\sigma_\nu\sigma_q+ \frac{1}{18}
\sigma_\epsilon \sigma_q-{\frac{5}{27}}\sigma_\epsilon \sigma_q
{\sigma_\nu}^{2}-\\[2mm]
{\frac{5}{27}}{\sigma_\epsilon }^{2}\sigma_\nu
\sigma_q-\frac{1}{27}{\sigma_\epsilon
}^{2}\sigma_q\sigma_e-\frac{1}{27}\sigma_\epsilon
\sigma_q{\sigma_e}^{2}+{\frac{7}{27}}\sigma_\nu\sigma
q{\sigma_e}^{2}-\frac{1}{27}{\sigma_q}^{2}\sigma_\epsilon
\sigma_e-\frac{1}{27}
{\sigma_e}^{4}+{\frac{13}{1620}}-{\frac{31}{648}}{\sigma_\nu}^{2}-
\frac{1}{27}\sigma_e{\sigma_\nu}^{3} \end{array}
\end{equation}
\begin{equation}
\begin{array}{r}\zeta_{2,3}=
{\frac{13}{810}}+{\frac{1}{54}}{\sigma _\epsilon }^{2}-{
\frac{4}{81}}{\sigma_e}^{2}-{\frac{2}{27}}{\sigma_\nu}^{4}-{\frac
{4}{27}}\sigma _\epsilon {\sigma_q}^{2}\sigma_\nu+{\frac{8}{27}}{
\sigma_\nu}^{3}\sigma_q+\frac{2}{9}{\sigma_e}^{3}\sigma_q+\frac{1}{9}{\sigma_e}^{2}{
\sigma_q}^{2}-{\frac{4}{27}}{\sigma_e}^{2}{\sigma_\nu}^{2}-{\frac{2}
{27}}\sigma_\nu{\sigma_e}^{3}- \\[2mm]
{\frac{13}{108}}\sigma_e\sigma_q+
{\frac{4}{27}}{\sigma_\nu}^{2}{\sigma_q}^{2}+{\frac{11}{27}}
\sigma_e\sigma_q{\sigma_\nu}^{2}+{\frac{11}{27}}\sigma_e{
\sigma_q}^{2}\sigma_\nu+\frac{1}{27}\sigma _\epsilon
\sigma_\nu+{\frac{1}{54}} \sigma_e\sigma _\epsilon
-{\frac{25}{324}}\sigma_\nu\sigma_e-{
\frac{13}{108}}\sigma_\nu\sigma_q+\frac{1}{18}\sigma _\epsilon
\sigma_q- \\[2mm]
{\frac{4}{27}}\sigma _\epsilon \sigma_q{\sigma_\nu}^{2}- {\frac{4}
{27}}{\sigma _\epsilon }^{2}\sigma_\nu\sigma_q-{\frac{2}{27}}{
\sigma _\epsilon }^{2}\sigma_q\sigma_e-{\frac{2}{27}}\sigma
_\epsilon \sigma_q{\sigma_e}^{2}+{\frac{11}{27}}\sigma_\nu
\sigma_q{\sigma_e}^{2}-{\frac{2}{27}}{\sigma_q}^{2}\sigma _\epsilon
\sigma_e-{\frac{2}{27}}{\sigma_e}^{4}-{\frac{19}{324}}
{\sigma_\nu}^{2}-{\frac{2}{27}}\sigma_e{\sigma_\nu}^{3}
\end{array}
\end{equation}
\begin{equation}
\begin{array}{r}\zeta_{0,5}=-\frac{1}{27}\,\sigma_e\,{\sigma_\nu}^{3}+{\frac
{1}{54}}\,{\sigma_\nu}^{2}{ \sigma_q}^{2}+\frac{1}{27}\,\sigma
_\epsilon \,{\sigma_q}^{2}\sigma_\nu-{\frac {2
}{27}}\,{\sigma_e}^{2}{\sigma_\nu}^{2}+\frac{1}{27}\,{\sigma
_\epsilon }^{2}
\sigma_\nu\,\sigma_q+1/9\,{\sigma_e}^{3}\sigma_q-\frac{1}{27}\,{\sigma_e}^{4}+
\frac{1}{27}\,{\sigma_\nu}^{3}\sigma_q-\\[2mm]\frac{1}{27}\,{\sigma_\nu}^{4}+
1/18\,{\sigma_e}^{2}{ \sigma_q}^{2}-{\frac
{2}{27}}\,\sigma_e\,\sigma_q\,{\sigma_\nu}^{2}-{ \frac
{1}{108}}\,\sigma _\epsilon \,\sigma_\nu+{\frac {1}{108}}\,\sigma_e
\,\sigma _\epsilon -{\frac
{2}{27}}\,\sigma_e\,{\sigma_q}^{2}\sigma_\nu+ \frac{1}{27}\,\sigma
_\epsilon \,\sigma_q\,{\sigma_\nu}^{2}-\frac{1}{27}\,\sigma
_\epsilon
\,\sigma_q\,{\sigma_e}^{2}-\\[2mm]\frac{1}{27}\,{\sigma_q}^{2}\sigma
_\epsilon \, \sigma_e-
 {\frac {1}{648}}\,{\sigma_e}^{2}-{\frac
{2}{27}}\,\sigma_\nu\,
\sigma_q\,{\sigma_e}^{2}-\frac{1}{27}\,\sigma_\nu\,{\sigma_e}^{3}+{\frac
{1}{ 1620}}-\frac{1}{27}\,{\sigma _\epsilon
}^{2}\sigma_q\,\sigma_e+{\frac {5}{648}} \,{\sigma_\nu}^{2}+{\frac
{5}{162}}\,\sigma_\nu\,\sigma_e
\end{array}
\end{equation}
\begin{equation}
{\rm Coeff}(B_{,5}[3,2],\delta t^4)=0,{\rm Coeff}(B_{,5}[3,3],\delta
t^4)=0
\end{equation}
\section{The coefficient matrices of the higher-order L-NSE with the uniform flow}\label{coef-matrices-uniform}
\begin{equation}
{\rm Coeff}(B_{,2}[1,1],\delta t)=0,{\rm Coeff}(B_{,2}[1,2],\delta
t)=0,{\rm Coeff}(B_{,2}[1,3],\delta t)=0
\end{equation}
\begin{equation}
\begin{array}{r}
{\rm Coeff}(B_{,2}[2,1],\delta
t)=-2\,{{k_y}}^{2}U{V}^{2}\sigma_\nu+\frac{1}{3}\,{{k_y}}^{2}U\sigma_\nu-3\,{
k_x}\,{k_y}\,{U}^{2}V\sigma_\nu+{{k_x}}^{2}U{V}^{2}\sigma_\nu+{
k_x}\,{k_y}\,{V}^{3}\sigma_\nu+ \frac{1}{3}\,{{k_x}}^{2}U\sigma_\nu-\\[2mm]{{
k_x}}^{2}{U}^{3}\sigma_\nu-{{k_x}}^{2}U{V}^{2}\sigma_e+\frac{1}{3}\,{{k_x}}
^{2}U\sigma_e-{{k_x}}^{2}{U}^{3}\sigma_e-{k_x}\,{k_y}\,{U}^{2
}V\sigma_e+\frac{1}{3}\,{k_x}\,{k_y}\,V\sigma_e-{k_x}\,{k_y}\,{V}^
{3}\sigma_e
\end{array}
\end{equation}
\begin{equation}
\begin{array}{r}
{\rm Coeff}(B_{,2}[2,2],\delta
t)=-\frac{1}{3}\,{{k_x}}^{2}\sigma_\nu+{{k_y}}^{2}{V}^{2}\sigma_\nu-\frac{1}{3}\,{{
k_y}}^{2}\sigma_\nu-\frac{1}{2}\,{{k_x}}^{2}{V}^{2}\sigma_\nu+\frac{3}{2}\,{{k_x}}^{2
}{U}^{2}\sigma_\nu+3\,{k_x}\,{k_y}\,UV\sigma_\nu+\\[2mm]{k_x}\,{k_y}
\,UV\sigma_e+\frac{1}{2}\,{{k_x}}^{2}{V}^{2}\sigma_e-\frac{1}{3}\,{{k_x}}^{2}
\sigma_e+\frac{3}{2}\,{{k_x}}^{2}{U}^{2}\sigma_e
\end{array}
\end{equation}
\begin{equation}
\begin{array}{r}
{\rm Coeff}(B_{,2}[2,3],\delta
t)=2\,{{k_y}}^{2}UV\sigma_\nu+\frac{3}{2}\,{k_x}\,{k_y}\,{U}^{2}\sigma_\nu-\frac{3}{2}\,{k_x}\,{k_y}\,{V}^{2}\sigma_\nu-{{k_x}}^{2}UV\sigma_\nu+\frac{1}{2}\,
{k_x}\,{k_y}\,{U}^{2}\sigma_e+{{k_x}}^{2}UV\sigma_e-\\[2mm]\frac{1}{3}\,{
k_x}\,{k_y}\,\sigma_e+\frac{3}{2}\,{k_x}\,{k_y}\,{V}^{2}\sigma_e
\end{array}
\end{equation}

\begin{equation}
\begin{array}{r} {\rm Coeff}(B_{,2}[3,1],\delta t)=-3\,{k_y}\,{k_x}\,U{V}^{2}\sigma_\nu-{{k_y}}^{2}{V}^{3}\sigma_\nu
+\frac{1}{3}\,{{k_x}}^{2}V\sigma_\nu+{{k_y}}^{2}{U}^{2}V\sigma_\nu+{k_y}
\,{k_x}\,{U}^{3}\sigma_\nu+\frac{1}{3}\,{{k_y}}^{2}V\sigma_\nu-\\[2mm]
2\,{{k_x}}
^{2}V{U}^{2}\sigma_\nu-{{k_y}}^{2}{V}^{3}\sigma_e-{k_y}\,{k_x}
\,U{V}^{2}\sigma_e+\frac{1}{3}\,{{k_y}}^{2}V\sigma_e-{k_y}\,{k_x}\,{U
}^{3}\sigma_e+\frac{1}{3}\,{k_y}\,{k_x}\,U\sigma_e-{{k_y}}^{2}{U}^{2}
V\sigma_e
\end{array}
\end{equation}
\begin{equation}
\begin{array}{r}
{\rm Coeff}(B_{,2}[3,2],\delta
t)=-{{k_y}}^{2}UV\sigma_\nu-\frac{3}{2}\,{k_x}\,{k_y}\,{U}^{2}\sigma_\nu+\frac{3}{2}
\,{k_x}\,{k_y}\,{V}^{2}\sigma_\nu+2\,{{k_x}}^{2}UV\sigma_\nu-\frac{1}{3}
\,{k_x}\,{k_y}\,\sigma_e+ \frac{3}{2}\,{k_x}\,{k_y}\,{U}^{2}\sigma_e+\\[2mm]
\frac{1}{2}\,{k_x}\,{k_y}\,{V}^{2}\sigma_e+{{k_y}}^{2}UV\sigma_e
\end{array}
\end{equation}
\begin{equation}
\begin{array}{r}
{\rm Coeff}(B_{,2}[3,3],\delta
t)=-\frac{1}{3}\,{{k_x}}^{2}\sigma_\nu+3\,{k_x}\,{k_y}\,UV\sigma_\nu-\frac{1}{3}\,{{
k_y}}^{2}\sigma_\nu+{{k_x}}^{2}{U}^{2}\sigma_\nu-\frac{1}{2}\,{{k_y}}^{2}
{U}^{2}\sigma_\nu+\frac{3}{2}\,{{k_y}}^{2}{V}^{2}\sigma_\nu-\\[2mm] \frac{1}{3}\,{{k_y}}^{2}
\sigma_e+{k_x}\,{k_y}\,UV\sigma_e+\frac{3}{2}\,{{k_y}}^{2}{V}^{2}
\sigma_e+\frac{1}{2}\,{{k_y}}^{2}{U}^{2}\sigma_e
\end{array}
\end{equation}
\section{The optimized values of free parameters}\label{minimization-v}
Considering $n=5$ and $u_m=0$ in Eqs. (\ref{R-F}) and
(\ref{B-S-U-wavenumber}), $\sigma_e=0.0025$ and $\sigma_\nu=0.0025$,
the analytic expressions of the problems (\ref{Minimization-A}) and
(\ref{Minimization-B}) are given by
\begin{equation}
\begin{array}{rl}
F^o(\Xi)= & - 0.9276312550\,\sigma_q+ 0.1201533868\,\sigma_\epsilon
+ 7.203744088 \,{\sigma_\epsilon }^{4}+ 129.3388220\,\sigma_\epsilon
\,\sigma_q-
 0.4322246453\,\sigma_q\,{\sigma_\epsilon }^{4}+\\[2mm]
 & 0.01296673935\,{
\sigma_q}^{3}{\sigma_\epsilon }^{3}-
0.00007699001495\,{\sigma_q}^{4} \sigma_\epsilon +
113.4657225\,{\sigma_\epsilon }^{2}{\sigma_q}^{2}+
 43.21814228\,\sigma_q\,{\sigma_\epsilon }^{3}+ \\[2mm]
 & 0.006483369681\,{
\sigma_q}^{2}{\sigma_\epsilon }^{4}+ 0.01836896999\,{\sigma_q}^{3}
\sigma_\epsilon - 2.155399630\,{\sigma_q}^{2}\sigma_\epsilon -
 1.728866164\,{\sigma_q}^{2}{\sigma_\epsilon }^{3}+ \\[2mm]
 & 0.006483369681\,{
\sigma_q}^{4}{\sigma_\epsilon }^{2}- 0.5121252003\,\sigma_q\,{\sigma_\epsilon }^{2}- 1.296718509\,{\sigma_q}^{3}{\sigma_\epsilon }^{2}+\\[2mm]
  & 0.1080561613\,{\sigma_\epsilon }^{3}+  0.009393468450\,{\sigma_q}^{2}+
 16.02085679\,{\sigma_\epsilon }^{2}- 0.00006398720339\,{\sigma_q}^{3}
+ 57.80191535+ \\[2mm]
&0.0000002304635316\,{\sigma_q}^{4}
\end{array}
\end{equation}
\begin{equation}
\begin{array}{rl}
 F^e(\Xi)= & 0.007225929590\,{\sigma_\epsilon }^{2}{\sigma_q}^{2}+ 8.028810655\,{
\sigma_\epsilon }^{2}+ 0.0000002866536304\,{\sigma_q}^{2}-
 0.1773049091\,\sigma_\epsilon -\\[2mm]
 & 0.00003525193584\,\sigma_q-
 0.4817286393\,\sigma_q\,{\sigma_\epsilon }^{2}- 0.00008580791386\,{
\sigma_q}^{2}\sigma_\epsilon +\\[2mm]
& 0.008179411071\,\sigma_\epsilon \, \sigma_q+ 0.001159746366\\[2mm]
\end{array}
\end{equation}



\bibliographystyle{elsarticle-num}



\end{document}